\documentclass[apj,twocolumn,twocolappendix,floatfix]{openjournal}

\DeclareRobustCommand{\VAN}[3]{#2}
\let\VANthebibliography\thebibliography
\def\thebibliography{\DeclareRobustCommand{\VAN}[3]{##3}\VANthebibliography}

\usepackage{amsmath}	
\usepackage{amssymb}	
\usepackage{booktabs}
\usepackage{lipsum}
\usepackage[T1]{fontenc}
\usepackage{graphicx}	

\usepackage[breaklinks,colorlinks,citecolor=blue,urlcolor=blue,linkcolor=blue]{hyperref}
\usepackage{cleveref}
\usepackage{multirow}
\usepackage{newtxtext,newtxmath}
\usepackage{orcidlink}
\usepackage{pifont}
\usepackage[nobottomtitles]{titlesec}
\usepackage{natbib}
\usepackage{xparse}
\usepackage{ifthen}
\titleformat{\section}{\filcenter\MakeUppercase}{\thesection.}{0.5em}{}
\usepackage{graphicx}
\usepackage{savesym}
\savesymbol{tablenum}
\usepackage{siunitx}
\restoresymbol{SIX}{tablenum}
\usepackage{needspace}
\usepackage{overpic}
\usepackage{xcolor}
\usepackage{lipsum}

\begin{document}

\title{AT2021YKY: A FAST-RISING OPTICAL TRANSIENT WITH EVOLVING BROAD HYDROGEN EMISSION CONSISTENT WITH AN AMBIGUOUS NUCLEAR TRANSIENT}

\author{\vspace{-1.3cm}
Paarmita Pandey\,\orcidlink{0009-0003-6803-2420}$^{1,2}$,
Jason~T.~Hinkle\,\orcidlink{0000-0001-9668-2920}$^{3,4,5,a}$,
Christopher~S.~Kochanek\,\orcidlink{0000-0002-1790-3148}$^{1,2}$,
Michael~A.~Tucker\,\orcidlink{0000-0002-2471-8442}$^{1,2}$,
Mark~T.~Reynolds\,\orcidlink{0000-0003-1621-9392}$^{1,6}$,
Katie~Auchettl\,\orcidlink{0000-0002-4449-9152}$^{7,8}$,
C.~Ashall\,\orcidlink{0000-0002-5221-7557}$^{5}$,
Dhvanil~D.~Desai\,\orcidlink{0000-0002-2164-859X}$^{5}$,
Aaron Do\,\orcidlink{0000-0003-3429-7845}$^{9}$,
Willem~B.~Hoogendam\,\orcidlink{0000-0003-3953-9532}$^{5,b}$,
M.~E.~Huber\,\orcidlink{0000-0003-1059-9603}$^{5}$,
T.~de~Jaeger \,\orcidlink{0000-0001-6069-1139}$^{5}$,
Thomas~B.~Lowe\,\orcidlink{0000-0002-9438-3617}$^{5}$,
Anna~V.~Payne\,\orcidlink{0000-0003-3490-3243}$^{10}$,
Benjamin~J.~Shappee\,\orcidlink{0000-0003-4631-1149}$^{5}$,
Todd~A.~Thompson\,\orcidlink{0000-0003-2377-9574}$^{1,2,11}$,
Daniel~R.~Wilkins\, \orcidlink{orcid:0000-0002-4794-5998}$^{1,2}$\\
}

\affiliation{
$^{1}$Department of Astronomy, The Ohio State University, 140 W. 18th Ave., Columbus, OH 43210, USA\\
$^{2}$Center for Cosmology and Astroparticle Physics, The Ohio State University, 191 W. Woodruff Ave., Columbus, OH 43210, USA\\
$^{3}$Department of Astronomy, University of Illinois Urbana-Champaign, 1002 West Green Street, Urbana, IL 61801, USA\\
$^{4}$NSF-Simons AI Institute for the Sky (SkAI), 172 E. Chestnut St., Chicago, IL 60611, USA \\
$^{5}$Institute for Astronomy, University of Hawai`i at Manoa, 2680 Woodlawn Dr., Hawai`i, HI 96822, USA \\
$^{6}$Department of Astronomy, University of Michigan, 1085 S. University Ave., Ann Arbor, MI 48109, USA\\
$^{7}$School of Physics, The University of Melbourne, Parkville, VIC 3010, Australia.\\
$^{8}$Department of Astronomy and Astrophysics, University of California, Santa Cruz, CA 95064, USA\\
$^{9}$Institute of Astronomy and Kavli Institute for Cosmology, University of Cambridge, Madingley Road, Cambridge CB3 0HA, UK\\
$^{10}$Space Telescope Science Institute, 3700 San Martin Drive, Baltimore, MD 21218, USA\\
$^{11}$Department of Physics, The Ohio State University, 191 W. Woodruff Ave, Columbus, OH 43210, USA\\
$^{a}$NHFP Einstein Fellow \\
$^{b}$NSF Graduate Research Fellow
}

\begin{abstract}
Nuclear transients are powerful probes of supermassive black hole properties, offering insight into black hole mass, accretion physics, and the structure of galactic nuclei. Among these, a growing class of events cannot be classified as either tidal disruption events (TDEs) or active galactic nuclei (AGN) flares, and their physical origins remain poorly understood. We present a multi-wavelength photometric and spectroscopic analysis of AT2021yky (ZTF21abzciqh), an ambiguous nuclear transient (ANT) at a redshift of $z = 0.076$. AT2021yky reached a peak bolometric luminosity of $L_{\rm peak} = (4.1 \pm 1.1) \times 10^{43}~\mathrm{erg~s^{-1}}$, with a rise-time of $18.2 \pm 0.7$ days. The early-time UV/optical emission is well described by a blackbody with a temperature of $T \simeq 1.4 \times 10^{4}$~K, cooler than most optically selected TDEs. No X-ray emission from the transient is detected, with a $3\sigma$ limit of $L_X \lesssim 3.4 \times 10^{41}$~erg~s$^{-1}$ near peak. Spectroscopic observations reveal a largely featureless blue continuum with broad (FWHM$\sim 11,000$ km s$^{-1}$) H$\alpha$ emission line that appears around 20$-$40 days post-peak. The host-galaxy emission-line ratios indicate the presence of an AGN, though the absence of optical or mid-IR variability and a non-AGN mid-IR color suggest it is weak. AT2021yky exhibits a rapid rise time comparable to that of luminous fast blue optical transients (LFBOTs), while its decay timescale and late-time broad H$\alpha$ emission resemble those observed in TDEs. However, its cooler blackbody temperature and the absence of He~II and Balmer emission lines other than H$\alpha$ instead favour its classification as an ANT.

\end{abstract}

\maketitle

\section{Introduction}
The advent of wide-field, high-cadence optical transient surveys has fundamentally transformed our understanding of the time-domain sky. These include the All-Sky Automated Survey for Supernovae \citep[ASAS-SN;][]{shappee14, kochanek17}, the Zwicky Transient Facility \citep[ZTF;][]{bellm19, Graham2019}, the Asteroid Terrestrial-impact Last Alert System \citep[ATLAS;][]{tonry18}, the Panoramic Survey Telescope and Rapid Response System \citep[Pan-STARRS;][]{chambers16} and The Young Supernova Experiment \citep[YSE;][]{Jones_2021}. These surveys have dramatically increased the discovery rate of known transient classes and, by probing a wide range of timescales and luminosities, revealed entirely new phenomena.

Among the most energetic and prominent of these discoveries are the transient events associated with supermassive black holes (SMBHs; $10^6$--$10^9~M_\odot$), which are ubiquitous in galaxy centres. Such events provide a rare and direct means of measuring black hole masses and spins, extending black hole demographics to galaxies where dynamical mass measurements and reverberation mapping are not possible.

One class of such SMBH transients arises from actively accreting SMBHs, or Active Galactic Nuclei (AGNs). The stochastic variability of AGNs across all wavelengths and timescales was almost immediately recognized after their discovery \citep[e.g.,][]{Sandage64}. In the optical, the variability is stochastic and can be reasonably described by a damped random walk (DRW) process of modest amplitude \citep{Kelly2009, Kozlowski2010, MacLeod2010}, with evidence for deviations from this model on short timescales \citep[e.g.,][]{Zu2011, Kasliwal2015, Kozlowski2016}. The DRW model parameters, such as amplitude and characteristic timescale, are correlated with AGN physical properties such as luminosity, black hole mass, and accretion rate \citep[e.g.,][]{Kelly2009, Kozlowski2010, MacLeod2010, Burke2021, Tarrant2025}. The physical origin of the optical variability remains an open question, with evidence for contributions from both reverberation, where the outer disk varies in response to illumination from the inner disk, and intrinsic local variability within the disk itself \citep[e.g.,][]{blandford82, edelson15, fausnaugh16, neustadt20}.

In recent years, more dramatic episodes of variability have been observed from SMBHs. One category is changing-look AGNs (CL-AGNs), which are characterized by rapid changes in their optical and/or X-ray emissions on timescales of months to years. Their optical variability is typically attributed to a change in the accretion state, whereas the X-ray variability is largely due to changes in the obscuration (e.g., \citealt{bianchi05,shappee14,macleod16,trakhtenbrot19a, Mcloed2019, sheng2017, Ricci2023, Guo2025}). In the optical regime, these changes manifest as the appearance or disappearance of broad emission lines, causing the source to transition between Seyfert Type 1 and Type 2 classifications or vice versa (e.g., \citealt{macleod16, temple2023}). Changing-look low-ionization nuclear emission regions (CL-LINERs) are nuclei previously characterized by weak or negligible activity that undergo a rapid transition into a luminous, AGN-like accretion state. Such turn-on events offer a rare opportunity to observe the ignition of SMBH accretion in real time and provide direct observational constraints on the physical conditions that trigger and sustain enhanced accretion onto otherwise quiescent black holes (e.g., \citealt{gezari17, yan19, frederick19}).

In contrast to the persistent accretion that powers AGNs, SMBHs can also be revealed through Tidal Disruption Events (TDEs), which occur when a star passes within the tidal radius of a SMBH and is disrupted by tidal forces \citep{rees88, phinney89, evans89}. A portion of the disrupted stellar debris becomes gravitationally bound and falls back onto the black hole to power a luminous, short-lived accretion-powered flare. These events are usually dominated by UV and optical emission, but some can also show X-ray and radio emission. The transient fades over months to years \citep{ holoien14-14ae,  holoien16-14li, auchettl17, holoien19b, holoien19c,  holoien20, Gezari2021, hinkle21a, hammerstein23, Yao2023, Mockler2023}. Optical spectroscopic observations reveal considerable differences among TDEs, particularly in the occurrence, intensity, and width of their emission features \citep[e.g.,][]{arcavi14, hung17, leloudas19, holoien20, nicholl20}. The optical spectra are generally characterized by prominent hydrogen and/or helium emission lines, with some events additionally exhibiting oxygen emission associated with Bowen fluorescence  \citep[e.g.,][]{leloudas19-18pg, vanvelzen20b}.  Several physical mechanisms have been proposed for the spectral diversity, including variations in photoionization conditions \citep[e.g.,][]{gaskell14, roth16, kara18}, differences in the chemical composition of the disrupted star \citep[e.g.,][]{kochanek16b, Mockler_2021}, disruption of stripped helium stars \citep{Gezari2021}, differences between partial and total disruptions \citep{nicholl20}, and ionization conditions resulting from the contraction of the photospheric radius during the event \citep{Charalampopoulos_2022}. Despite these theoretical developments, a comprehensive framework explaining the full range of observed spectroscopic properties has yet to be established. %

The prompt identification of TDEs through modern time-domain surveys is particularly important for the study of the emerging subclass known as faint and fast (FaF) TDEs \citep{Charalampopoulos23, Hoogendam_2024}. These transients are characterized by peak luminosities that are typically an order of magnitude lower than in typical TDEs. These TDEs exhibit both more rapid rises to maximum brightness and faster post-peak declines. Some examples of this class are iPTF16fnl \citep{blagorodnova17, brown18}, AT2019qiz \citep{nicholl20}, AT2020neh \citep{Angus22}, and AT2020wey \citep{Charalampopoulos23}.

\begin{figure*}[t]
    \begin{center}
\includegraphics[width=\textwidth]{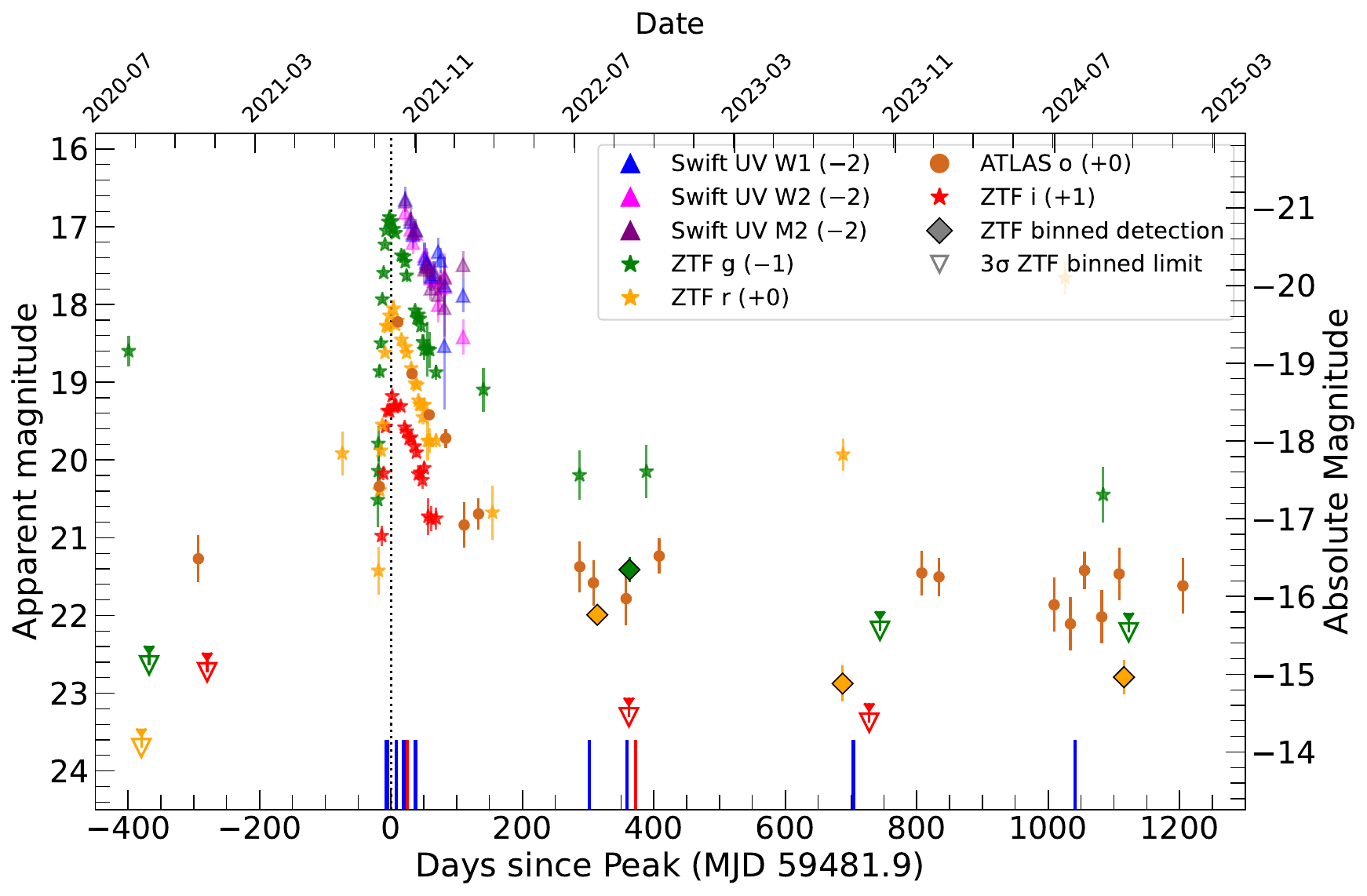}
       \caption{Host-subtracted UV and optical light curves of AT2021yky, showing the Swift UV bands (blue, pink, and purple triangles), ZTF $g$ (green star), $r$ (yellow star), $i$ (red star), and ATLAS $o$ (orange circle). The photometry spans from roughly 400 days before peak to roughly 1200 days after peak in the observer frame. The diamonds (inverted triangles) show binned detections ($3\sigma$ upper limits) from the ZTF seasonal stacking. We detect residual emission in the ZTF $g$ and $r$ bands, consistent with the late-time ATLAS $o$ band flux. The blue (red) bars along the time axis show the epochs of the optical (near-IR) spectra. The black dotted line marks the peak time (MJD = 59481.9) used as the time reference. All data are corrected for Galactic extinction and are in the AB magnitude system. The light curves have been offset for visual clarity. } 
       \label{fig:all_lc}
    \end{center}
\end{figure*}

\begin{figure*}[t]
    \begin{center}
\includegraphics[width=20cm]{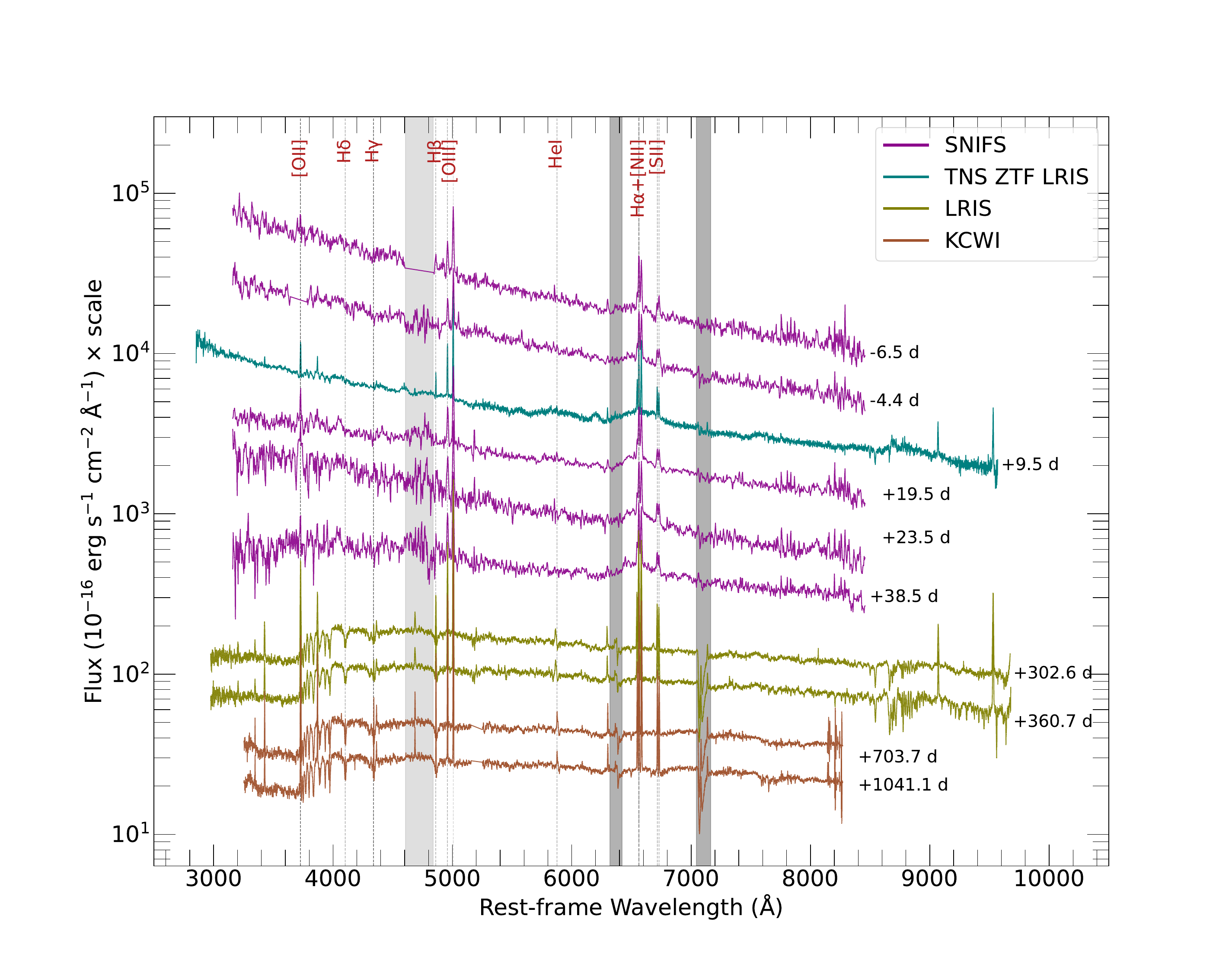}
\vspace{-3mm}
       \caption{SNIFS, LRIS and KCWI optical spectra of AT2021yky, where the time relative to the peak in the observed frame is given to the right of each spectrum. The spectra span from 7 days before peak (top) to $\sim$ 3 years after peak (bottom). Prominent emission lines are labeled. The dark gray-shaded regions show the strong telluric bands, and the light gray shows the dichroic region. The spectra are offset for visibility.} 
       \label{fig:kcwi}
    \end{center}
\end{figure*}

Recently, a growing number of nuclear transients have been observed whose physical origins remain unclear \citep[e.g.][]{kankare17, trakhtenbrot19b, neustadt20, Hinkle_2022, Hinkle_science_25}. These events are presently called ambiguous nuclear transients (ANTs) because they show some features of normal AGN variability and TDEs but differ in other ways \citep{Hinkle2024, Wiseman25}. Suggested origins include a TDE occurring within a pre-existing AGN accretion disk \citep{blanchard17-16dtm, Chan19}, or accretion disk instabilities \citep{Sniegowska23}. While the photometric evolution of ANTs closely resembles that of TDEs, their spectroscopic properties are more similar to those of AGN \citep{Wiseman25}. In particular, ANTs exhibit strong, high-equivalent-width emission lines, especially from the hydrogen Balmer series, whereas TDEs typically show broader but comparatively weaker emission features. Mid-infrared (MIR) flares are a common characteristic of ANTs \citep{Hinkle_2024}, whereas they are detected much less frequently in TDEs \citep{vanvelzen16}. However, there are MIR flares lacking optical counterparts that may include heavily obscured TDEs \citep{Masterson24}. The MIR emission observed is interpreted as dust reprocessing of the transient emission by circumnuclear material, often associated with the AGN "torus". Several ANTs exhibit no evidence for prior AGN activity and some have optical spectra that differ significantly from those of typical AGN \citep[e.g.,][]{Hinkle_2022, Oates24}. Notable examples include PS1-10adi \citep{kankare17}, ASASSN-18jd \citep{neustadt20}, ASASSN-18el \citep{trakhtenbrot19b, ricci20}, ASASSN-17jz \citep{Holoien_2022}, AT2018dyk \citep{frederick19}, ASASSN-20hx \citep{Hinkle_2022}, AT2021loi \citep{Makrygianni_2023}, AT2021lwx \citep{Wiseman2023, Subrayan_2023}, and AT2020adpi \citep{Wiseman25, Pandey_2025}. A distinct and especially energetic subclass, known as extreme nuclear transients (ENTs), are the most luminous events observed to date, and it is proposed that they arise from the tidal disruption of massive stars  \citep[$\sim 3$--$10~M_\odot$,][]{Hinkle_science_25, Graham_2025}.  

\begin{deluxetable}{cccc}
\tablewidth{240pt}
\tabletypesize{\footnotesize} 
\tablecaption{Photometry}
\tablehead{
\colhead{MJD} &
\colhead{Filter} &
\colhead{Magnitude (AB)} & 
\colhead{Uncertainty} }
\startdata
59503.49 &	Swift UV W2	& 19.277 & 0.137\\ 
59511.79 & Swift UV W2	& 19.493 & 0.102\\
59515.30 & Swift UV W2	& 19.665 & 0.148\\
59519.55 & Swift UV W2	& 19.545 & 0.118\\
\ldots  & \ldots & \ldots & \ldots  \\
61090.223 & ATLAS c	& 19.696 & 0.143 \\
61090.231 &	ATLAS c	& 19.922 & 0.388 \\
61090.941 &	ATLAS c & 19.841 & 0.079 
\enddata 
\tablecomments{Magnitudes and associated uncertainties for the photometry of AT2021yky. All magnitudes are reported in the AB magnitude system. Only a portion of the table is shown here to illustrate the format; the complete dataset is available online as an ancillary file.}
\label{tab:phot} 
\end{deluxetable}

Modern time domain surveys have also uncovered new classes of non-nuclear transients, such as fast blue optical transients \citep[FBOTs;][]{drout14, Inserra19, Liu23}. These are characterized by blue colors, rapid rise and decay times of days to weeks, and high peak luminosities. These transients are believed to originate from unusual kinds of stellar death, such as the direct collapse of a massive star, a magnetar-powered supernova, or a merger with a compact object companion \citep{Prentice18, Marguitti19, Metger22, Gottlieb22}. A luminous subclass, known as luminous fast blue optical transients \citep[LFBOTs;][]{Prentice18, Ho23, Sevilla26, Omand26, Lebaron26}, anchored by the prototype AT2018cow \citep{Prentice18, Marguitti19}, also have luminous X-ray and radio emission and a dense circumstellar medium. Possible origins of LFBOTs are massive star deaths \citep{Marguitti19}, stellar mergers \citep{Metger22}, and tidal disruption of a star by an intermediate-mass black hole \citep[e.g.,][] {perley19, Gutirrez24}.

In this paper, we present the multi-wavelength characterization of the nuclear transient AT2021yky, which exhibits a rapid rise in its UV and optical emission. AT2021yky (ZTF21abzciqh, RA = 01:07:50.224, Dec= +27:18:53.28) was discovered on September 9, 2021, by ZTF \citep{Munoz21} and classified a month later as a type II supernova at $z=0.076$ by \citet{ChuTNS2021}. Section \ref{sec:obs} presents the photometric and spectroscopic observations of the transient. In Section \ref{sec:results}, we analyse the host galaxy properties, construct the UV/optical SED, derive the key physical characteristics of the light curve, and finally discuss the unique spectroscopic evolution of the source. Section \ref{sec:discussion} discusses the nature and classification of AT2021yky. Throughout this work, we adopt a luminosity distance of $D_{L} = 346.4$~Mpc, for a flat cosmology with $h = 0.696$, $\Omega_{\rm M} = 0.286$, and $\Omega_{\Lambda} = 0.714$ \citep{wright06}. All photometric measurements have been corrected for Galactic extinction along the line of sight, using $A_{V} = 0.176$~mag \citep{Extinction2011}.

\section{Observations and Survey Data}\label{sec:obs}
This section describes the archival data available for the host galaxy, along with the photometric and spectroscopic observations of AT2021yky. All photometry has been corrected for Galactic extinction.

\vspace{-4mm}
\subsection{Optical Photometry}\label{sec:photometry}

We obtained the optical light curve of AT2021yky from the Zwicky Transient Facility \citep[ZTF;][]{Graham2019, bellm19}, spanning 17 March 2018 (MJD 58194) to 18 June 2026 (MJD 61209). The $g$, $r$, and $i$ band photometry was retrieved from the IPAC ZTF forced photometry server \citep{masci2023newforcedphotometryservice}, following the recommended procedures\footnote{\url{https:/irsa.ipac.caltech.edu/data/ZTF/docs/ztf_forced_photometry.pdf}}. The forced-photometry light curve does not include the reference image flux, which is the host galaxy flux plus any pre-existing AGN flux. To search for residual emission from the transient post-peak, we first computed a weighted mean baseline flux from pre-transient epochs in each ZTF band ($g,\,r,\,i$). Then, we divided the post-peak epochs into their respective seasonal windows and computed a weighted mean flux within each season after subtracting the pre-transient mean flux. We converted the resulting flux excess and its propagated uncertainty into a magnitude or a $3\sigma$ upper limit, depending on the detection significance.

We retrieved the ATLAS \citep{tonry18} host-subtracted forced photometry for AT2021yky from the ATLAS Forced Photometry server \citep{Shingles2021} spanning MJD 58328 to MJD 61209. We then binned the data in 10-day intervals using an inverse-variance weighted mean, retaining only bins with a signal-to-noise ratio (SNR) $>3$ as detections.

\subsection{Infrared Photometry}\label{sec:IR_photometry}

We also obtained mid-infrared (MIR) light curves in the $W1$ and $W2$ bands from the Wide-field Infrared Survey Explorer \citep[WISE;][]{Mainzer2011}, following the procedure outlined in \citet{WISE-pull-lc}. The data were binned into 24-hour intervals, and the flux and uncertainty in each bin were computed as the inverse-variance weighted mean and its associated uncertainty. Unfortunately, the data coverage between MJD 59350 and 60050, which coincides with the period of peak transient activity, is sparse, and we cannot characterize the MIR evolution near the peak.

\subsection{UV Photometry}

The Neil Gehrels Swift Observatory monitored the transient near peak between MJD 59502 and 59592, using all six filters of the Ultraviolet/Optical Telescope \citep[UVOT;][]{roming05}: $V$ (5468~\AA), $B$ (4392~\AA), $U$ (3465~\AA), $UVW1$ (2600~\AA), $UVM2$ (2246~\AA), and $UVW2$ (1928~\AA). At each epoch, multiple exposures were taken per filter and co-added into a single image using the \texttt{HEASoft} task \texttt{uvotimsum}. Source counts were extracted from a circular aperture of radius 5\farcs{0} centred on the transient position, while the background was estimated from a nearby source-free region with a radius of $\sim$50\farcs{0}. The count rates were then converted to fluxes and AB magnitudes using \texttt{HEASoft} \textit{version 6.34} with the calibration database release \texttt{2024-08-12\_V6.34} 
\citep{poole08, breeveld10}. 

Figure~\ref{fig:all_lc} shows the host-subtracted UV and optical light curves of AT2021yky, and Table~\ref{tab:phot} provides the photometric data. We show only the $W1, W2 \text{ and} ~M2$ bands of the Swift UVOT filters in Figure~\ref{fig:all_lc}. Binned detections and $3\sigma$ upper limits from the ZTF seasonal stacking are also shown. Residual emission is detected in the ZTF $g$ and $r$ bands, consistent with the late-time flux seen in the ATLAS $o$ band. We constrained the peak time to be MJD 59481.9 (Section~\ref{sec:LC}), and this date is used as the reference epoch for our light curves and spectra. 

\subsection{X-Ray Photometry}\label{sec:xray}

The \textit{Swift} X-Ray Telescope \citep{burrows05} monitored AT2021yky at the same time as the UVOT observations. No X-ray signal was detected in either the individual exposures or the combined dataset. We place $3\sigma$ upper limits on the X-ray flux by adopting an absorbed power-law spectral model of the form $dN/dE \propto E^{-\gamma}$ with $\gamma = 1.8$, a value commonly associated with AGN spectra \citep{ricci17}. We used a Galactic hydrogen column density of $N_{\rm H} = 6.15 \times 10^{20}~\mathrm{cm^{-2}}$ \citep{HI4PI16}. For the early-time observations taken around the peak (27.2~ks total across two different observation IDs), we obtain a $3\sigma$ flux upper limit of $f_X < 2.35 \times 10^{-14}$~erg~s$^{-1}$~cm$^{-2}$ in the 0.5--10.0~keV band, corresponding to $L_X \lesssim 3.4 \times 10^{41}$~erg~s$^{-1}$. For the 4.2~ks late-time observation taken $\sim 1000$ days post-flare when the transient had faded, we find $f_X < 7.0 \times 10^{-14}$~erg~s$^{-1}$~cm$^{-2}$ ($L_X \lesssim 1.0 \times 10^{42}$~erg~s$^{-1}$). Combining all of the observations, we obtain a limit of $f_X < 2.72 \times 10^{-14}$~erg~s$^{-1}$~cm$^{-2}$ ($L_X \lesssim 3.9 \times 10^{41}$~erg~s$^{-1}$).

\subsection{Radio Photometry}

We observed AT2021yky on MJD 59512.25 ($\sim$30 days post peak) at 850 $\mu m$ (353 GHz) with the Submillimetre Common-User Bolometer Array 2 \citep[SCUBA-2;][]{holland13} on the James Clerk Maxwell Telescope (JCMT). Since the source was expected to be unresolved at submillimeter wavelengths, we adopted a constant-velocity Daisy mapping strategy, which is optimized for compact point sources. The data were reduced and flux-calibrated using the standard SCUBA-2 pipeline hosted at the Canadian Astronomical Data Centre\footnote{\url{https://www.cadc-ccda.hia-iha.nrc-cnrc.gc.ca/en/jcmt/scubalc.html}}, with phase and flux calibrators observed on the same night as the target. The details of the analysis are presented in section~\ref{sec:radio}.

\subsection{Spectroscopic Observations}\label{spectra}

We obtained the spectroscopic data of the source from several facilities. Pre-peak and near-peak spectra (within 40 days of peak) were acquired from the Spectroscopic Classification of Astronomical Transients Survey (SCAT) \citep{Tucker_2022} using the SuperNova Integral Field Spectrograph \citep[SNIFS;][]{lantz04} on the 88-inch University of Hawai'i telescope (UH88). To get the host-galaxy subtracted spectra for further analysis, we used SCAT DR1 data \citep{Tucker2026}, which extracts spectra using point-spread-function (PSF) fitting to minimize host-galaxy contamination. Additionally, two spectra were obtained approximately 300 days after the optical peak using the Low-Resolution Imaging Spectrometer \citep[LRIS;][]{oke95} on the Keck~I telescope. We also include the LRIS spectrum from the Transient Name Server (TNS), obtained by ZTF $\approx 10$ days after peak \citep{ChuTNS2021}. Two additional spectra were taken with KCWI on Keck in August 2023 (MJD 60185.6) and July 2024 (MJD 60522.6), when the transient had faded. 

The SNIFS spectra were reduced and flux-calibrated using the SCAT pipeline, while the LRIS spectra were reduced with \texttt{PypeIt} \citep{pypeit:zenodo, pypeit:joss_arXiv, pypeit:joss_pub}.  We reduced the KCWI spectra using KCWI DRP\footnote{\url{https://kcwi-drp.readthedocs.io/en/latest/index.html}}. Figure~\ref{fig:kcwi} shows the optical spectroscopic evolution of AT2021yky spanning from approximately one week before peak to $\sim$1000 days post-peak in the observer frame.

\begin{figure}[t]
    \begin{center}
\includegraphics[width=9cm]{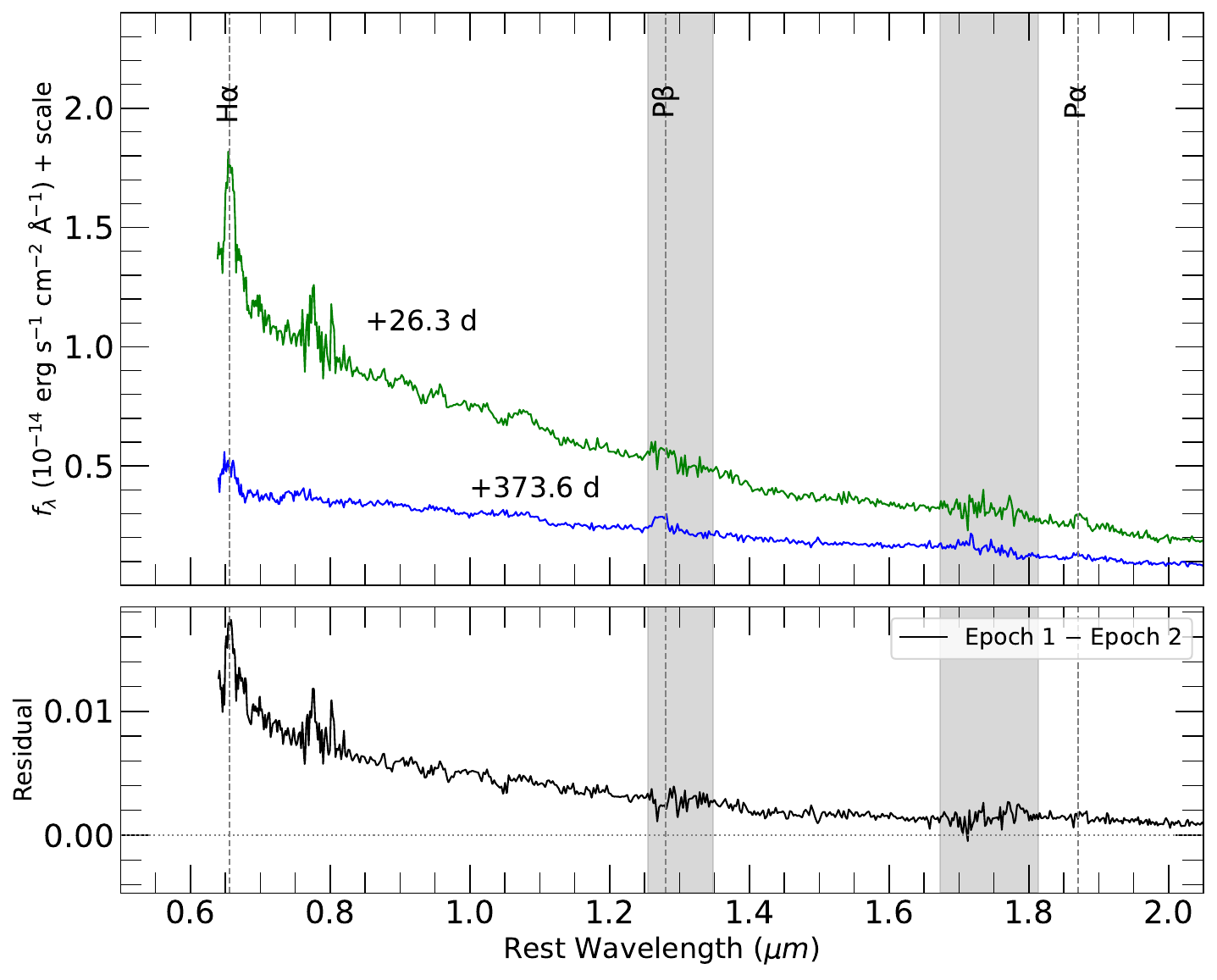}
       \caption{SpeX near-IR Spectra of AT2021yky, where the time after peak in the observed frame is given above the spectra. Prominent emission lines are labeled. The vertical gray bands mark strong atmospheric telluric absorption regions. The panel shows the difference spectrum found by subtracting epoch 2 from epoch 1. This should be the transient-only emission, since we expect the emission to come mostly from the host galaxy after 300 days. } 
       \label{fig:spex}
    \end{center}
\end{figure}

We additionally obtained two near-infrared (NIR) spectra of AT2021yky with the SpeX spectrograph \citep{rayner03} on the NASA Infrared Telescope Facility (IRTF), near peak and one year post-peak. These spectra were observed in prism mode at a spectral resolution of $R \sim 80$. We reduced the NIR spectra using the \texttt{SpeXtool} pipeline \citep{Cushing2004}. Figure~\ref{fig:spex} shows the near-IR spectroscopic evolution of AT2021yky.

\section{Results}\label{sec:results}
In this section, we present the results of our multiwavelength analysis of AT2021yky. We begin with the properties of the host galaxy (Section~\ref{sec:host}), followed by the optical and radio light curves (Section~\ref{sec:LC} and ~\ref{sec:radio}). We then characterize the evolution of the transient by modeling its spectral energy distribution (SED) (Section~\ref{sec:tsed}) and finally discuss the spectral properties derived from our optical and IR spectroscopic observations (Section~\ref{sec:spec}).

\begin{figure}[t]
\centering
\includegraphics[width=\columnwidth]{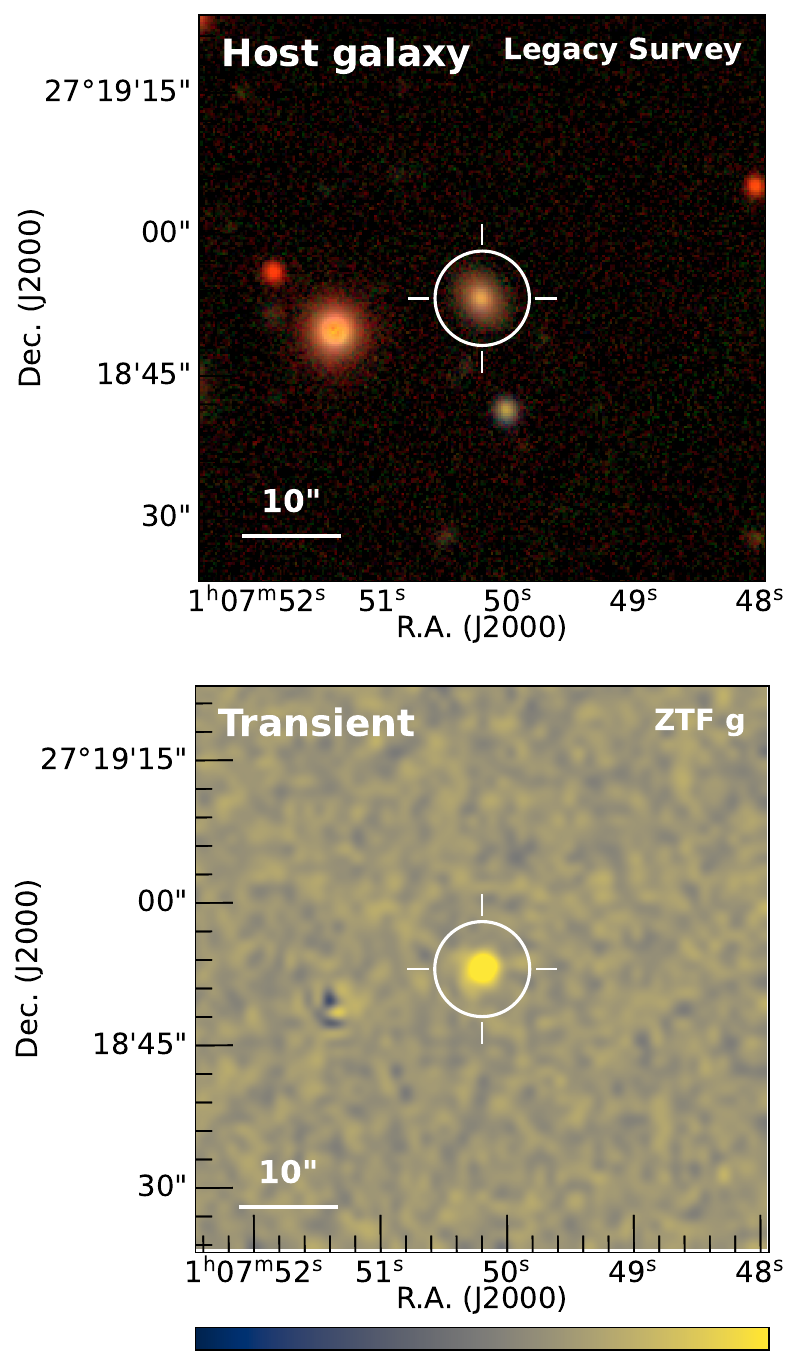}
  \centering
\caption{(\textit{Top:)} Pre-transient DESI Legacy Survey \citep{Dey19} image of the host galaxy of AT2021yky (circled) in the $g$, $r$, and $z$ bands. \textit{(Bottom:)} ZTF $g-$band difference image taken during the active phase of AT2021yky. The transient is clearly detected as a bright point source spatially coincident with the host nucleus. }
\label{fig:host_3color}
\end{figure}

\subsection{Host Galaxy}\label{sec:host}

The host galaxy of AT2021yky is identified as WISEA J010750.23+271853.2. We retrieved a three-color ($g,r,z$) image of the host from the DESI Legacy Survey \citep{Dey19}. Figure~\ref{fig:host_3color} presents the Legacy Survey image centered on the transient position, together with the ZTF $g$-band difference image of AT2021yky. The transient is coincident with the host nucleus within astrometric uncertainties. The circle in both panels has a fixed 5$''$ radius and is centered on the transient position. We also retrieved the host galaxy morphological parameters from the Legacy Survey DR9 Tractor catalog \citep{Dey19}. The catalog reports a Sérsic index of $n=6\pm0.1$, implying a bulge-like Sérsic morphology. The half-light radius is $R_e=0.62''\pm0.002''$, corresponding to $\approx 0.9$ kpc.

To quantify the spatial coincidence between the transient and the host nucleus, we downloaded a ZTF g-band reference image built with 15 frames obtained prior to the transient (over MJD 58280 to 58352) and a science image obtained 6 days after peak (MJD 59488.3). We used the {\tt ISIS} image subtraction package (\citealt{Alard1998}, \citealt{Alard2000}) to analyze a $400^2$ pixel region centered on the transient. The interpolation to match the science frame to the reference image used 33 stars with RMS residuals of $0\farcs16$ (the pixel scale is $1\farcs00$).  The position of the transient in the subtracted image is offset from the position of the host in the reference image by $0\farcs17\pm 0\farcs16 $ using the RMS residuals from the interpolation as the error estimate. The position of the source in the science image and that in the subtracted image match to $0\farcs04$.  We used this approach because the transient does not completely dominate the g band emission at peak. However, we also simply found the best linear transform between the pixel positions of 16 non-transient sources within 2~arcmin of the transient on the reference and science images. Combining the residuals of the transformation with the offset between the host galaxy in the reference image and the transient in the science image, we find that the transient is offset from the host center by $0\farcs08\pm0\farcs13$. If we adopt the larger values from the image subtraction results, the transient offset is $0.27\pm0.26$ of the host half-light radius, or $247\pm 232$~pc given our adopted distance.

\begin{figure}[t]
    \begin{center} 
    \includegraphics[width=\columnwidth]{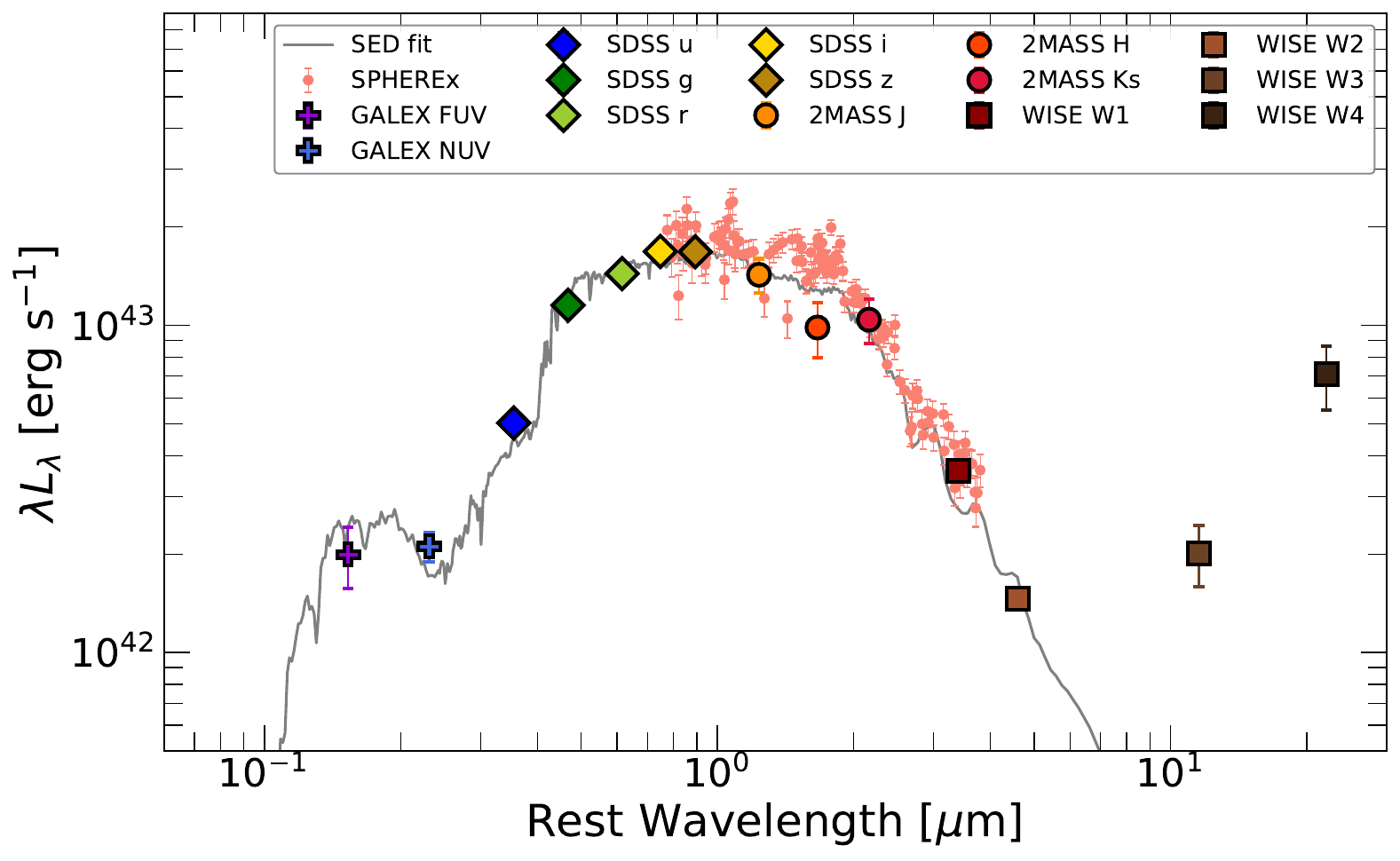}
        \caption{Spectral Energy Distribution (SED) of the host galaxy. The FAST++ SED fit is shown in gray. The SPHEREx spectrum (truncated beyond $3.8\mu m$ due to low SNR) is shown as the salmon points. The far-IR emission is not included in the model, since FAST does not model dust emission.}  
       \label{fig:HOST_SED}
    \end{center}  
\end{figure}

We characterize the host galaxy by fitting the pre-transient photometric SED with the Fitting and Assessment of Synthetic Templates code \citep[FAST++;][]{kriek09}. We fit photometry from GALEX \citep{chris_2005_galex}, SDSS \citep{ahumada20}, 2MASS \citep{2MASS_2006}, and WISE ($W1$ and $W2$) \citep{wright10}, as listed in Table~\ref{tab:host_phot}. We assumed a \citet{cardelli89} extinction law with $R_V = 3.1$, a \citet{salpeter55} initial mass function, and \citet{bruzual03} stellar population synthesis models with an exponentially declining star formation history at Solar metallicity. The best-fit model yields a stellar mass of $\log(M_*/M_\odot) = 9.69^{+0.04}_{-0.22}$, a stellar population age of $\log(\rm age/yr) = 9.0^{+0.15}_{-1.0}$ and a star-formation rate (SFR) of $\approx 0.5~M_\odot~\rm yr^{-1}$ which is consistent with a modestly star-forming galaxy undergoing a gradual decline in its SFR. The best-fit SED is shown in Figure~\ref{fig:HOST_SED}, and the model parameters are given in Table~\ref{tab:fast}. While FAST does not model dust emission, the relatively high $A_V \simeq 0.58$ of the FAST model suggests that there should be dust emission comparable to the optical-UV emission, and this likely explains the rise towards the far-IR seen in the WISE $W3$ and $W4$ bands in Figure~\ref{fig:HOST_SED}.

We also retrieved spectrophotometric observations of the host galaxy from the SPHEREx telescope \citep[SpectroPhotometer for the History of the Universe, Epoch of Reionization, and Ices eXplorer;][]{Bock_2026} using their Spectrophotometry Tool\footnote{\url{https://irsa.ipac.caltech.edu/applications/spherex/tool-spectrophotometry}}. The SPHEREx spectra cover 0.75--5.0~$\mu$m with spectral resolutions of $R \sim 41$ over 0.75--4.2~$\mu$m and $R \sim 135$ over 4.2--5.0~$\mu$m. The SPHEREx data span MJD 60870--61035, approximately 3.8 years after the transient peak, and we therefore expect the spectra to be dominated by host galaxy emission. We retrieved 200 individual flux measurements at the host position, of which 170 remained after applying the clean source quality flag. We restricted the analysis to wavelengths $\lambda \leq 3.8\,\mu$m, as fluxes at longer wavelengths are very noisy. The SPHEREx data are overplotted in salmon in Figure~\ref{fig:HOST_SED}. While the host SED was constructed from pre-transient photometry, the SPHEREx observations were obtained approximately four years after peak. The close agreement between the two indicates that the near-IR emission at this epoch is dominated by the host galaxy, with no significant residual contribution from the transient.

\begin{table}
\centering
\caption{Host Photometry}
\label{tab:host_phot}
\begin{tabular}{lcccc}
\hline\hline
Instrument & Band & $\lambda_{\rm eff}$ ($\mu$m) & Magnitude & Uncertainty \\
\hline
GALEX      & FUV        & 0.153  & 21.83 & 0.23 \\
GALEX      & NUV        & 0.231  & 21.32 & 0.11 \\
SDSS & $u$ & 0.355 & 19.91 & 0.06 \\
SDSS & $g$ & 0.468 & 18.71 & 0.01 \\
SDSS & $r$ & 0.616 & 18.17 & 0.01 \\
SDSS & $i$ & 0.748 & 17.79 & 0.01 \\
SDSS & $z$ & 0.893 & 17.60 & 0.03 \\
ZTF  & $g$ & 0.478  & 18.67 & 0.04 \\
ZTF  & $r$ & 0.642  &  18.17 & 0.03 \\
ZTF  & $i$ & 0.787  & 17.79 & 0.04 \\
2MASS & $J$ & 1.24  & 16.53 & 0.13 \\
2MASS & $H$ & 1.66  & 16.13 & 0.21 \\
2MASS & $K_s$ & 2.16  & 15.32 & 0.17 \\
WISE & $W1$ & 3.35  & 15.14 & 0.03 \\
WISE & $W2$ & 4.60  & 15.15 & 0.08 \\
WISE & $W3$ & 11.7 & 11.88 & 0.23 \\
WISE & $W4$ & 22.1 & 8.45  & 0.24 \\
\hline
\end{tabular}
\tablecomments{Archival photometric measurements of the host galaxy WISEA J010750.23+271853.2. SDSS data are Model magnitudes corrected for Galactic extinction. All magnitudes are in the AB system except 2MASS (Vega) and WISE (Vega).}
\end{table}

\begin{table}[t]
\centering
\caption{Stellar Population Parameters from FAST++}
\label{tab:fast}
\begin{tabular}{lc}
\hline\hline
Parameter & Value \\
\hline
Star formation timescale ($\log(\tau/\mathrm{yr})$)       & $8.40^{+0.07}_{-1.81}$ \\[4pt]
Stellar population age ($\log(\mathrm{age/yr})$)           & $9.00^{+0.15}_{-1.00}$ \\[4pt]
Dust attenuation $A_V$ (mag)                               & $0.58^{+0.73}_{-0.40}$ \\[4pt]
Stellar mass ($\log\,M_{\odot}$)                           & $9.69^{+0.04}_{-0.22}$ \\[4pt]
Star formation rate ($\log(M_{\odot}\,\mathrm{yr}^{-1})$) & $-0.31^{+0.29}_{-7.32}$ \\[4pt]
Specific star formation rate ($\log\,\mathrm{yr}^{-1}$)   & $-10.00^{+0.33}_{-7.12}$ \\[4pt]
\hline
\end{tabular}

\tablecomments{Stellar population parameters derived by fitting the host galaxy photometry with FAST++ \citep[v1.3;][]{kriek09} using \citet{bruzual03} stellar population templates, a Salpeter IMF, an exponentially declining star formation history ($\mathrm{SFR} \propto e^{-t/\tau}$), and a Milky Way dust attenuation law with $R_V = 3.1$, fixed at the spectroscopic redshift $z = 0.076$. Metallicity was fixed at $Z = 0.02$. Uncertainties represent 68\% confidence intervals.}
\end{table}

To constrain the black hole mass, we use the FAST++ stellar mass and the empirical $M_{\rm BH}$--$M_*$ scaling relation of \citet{reines15}, calibrated for AGN host galaxies. This gives us a central black hole mass of $\log(M_{\rm BH}/M_\odot) =6.07_{-0.28}^{+0.17}$. This is broadly consistent with estimates from the $M_{\rm BH}$--$M_{\rm bulge}$ relations of \citet{mcconnell13}, which yield $\log(M_{\rm BH}/M_\odot) = 6.90_{-0.45}^{+0.33}$, assuming a bulge-to-total mass ratio of $B/T = 0.7$ appropriate for an early-type galaxy (Sérsic index $n=6\pm0.1$). The offset between the two estimates is within the intrinsic scatter of both relations ($\sim 0.5$~dex for \citealt{mcconnell13} and $\sim 0.55$~dex for \citealt{reines15}).

\begin{figure}[t]
    \begin{center} 
    \includegraphics[width=\columnwidth]{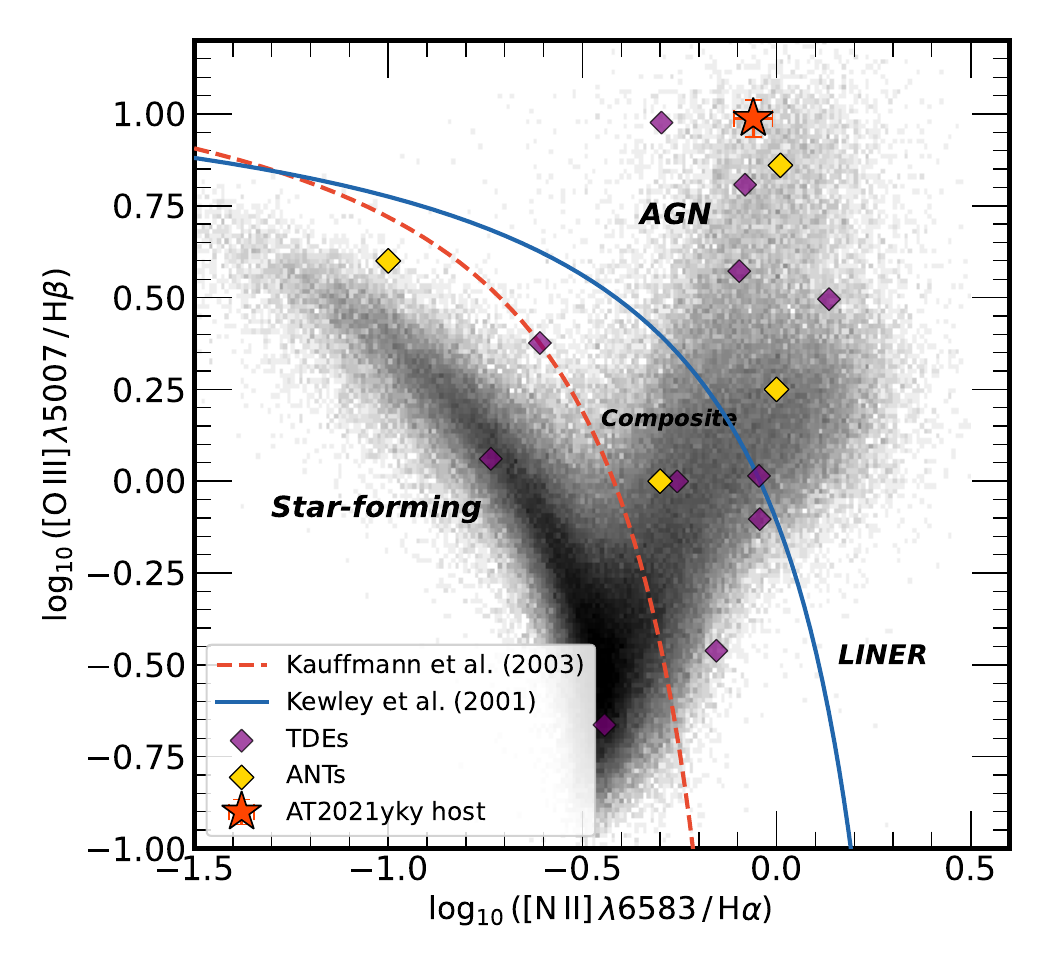}
        \caption{BPT Diagram for the host galaxy, placing it in the AGN regime. The solid curve marks the theoretical upper limit for pure star formation \citep{kewley01}, and the dashed line is the empirical AGN classification boundary of \citet{kauffmann03}. The gray points show the BPT diagram of SDSS DR8 emission-line galaxies from \citet{brinchmann04}. The purple diamonds are TDE host galaxies from \citet{french20}, and the yellow diamonds are ANT host galaxies showing PS1-10adi \citep{kankare17}, ASASSN-18jd \citep{neustadt20}, AT2018dyk \citep{frederick19}, and AT2021fez \citep{Wiseman25}.}  
       \label{fig:bpt}
    \end{center}  
\end{figure}

To investigate the ionization state of the host galaxy and the possible presence of pre-existing nuclear activity, we measured the narrow emission line fluxes from the late-time KCWI spectrum obtained $\approx$3 years after the transient peak, by which point the transient contribution to the spectrum had significantly faded. We measured fluxes for H$\beta$, [O\,{\sc iii}]~$\lambda5007~\text{\AA}$, [N\,{\sc ii}]~$\lambda6583~\text{\AA}$, and H$\alpha$ by fitting Gaussian profiles to each line using {\tt scipy.optimize.curve$\_$fit}. Figure~\ref{fig:bpt} shows the resulting position of the host galaxy on the Baldwin-Phillips-Terlevich \citep[BPT;][]{baldwin81} diagnostic diagram, with $\log(\rm [O\,III]/H\beta) = 0.988 \pm 0.014$ and $\log(\rm [N\,II]/H\alpha) =-0.060 \pm 0.005$. The host lies above both the \citet{kauffmann03} and \citet{kewley01} demarcation lines, placing it in the AGN region of the diagram. We also show host galaxies associated with TDEs \citep{french20} and ANTs \citep{kankare17, neustadt20, frederick19, Wiseman25} for comparison.

\begin{figure}[t]
    \begin{center} 
    \includegraphics[width=9.3cm]{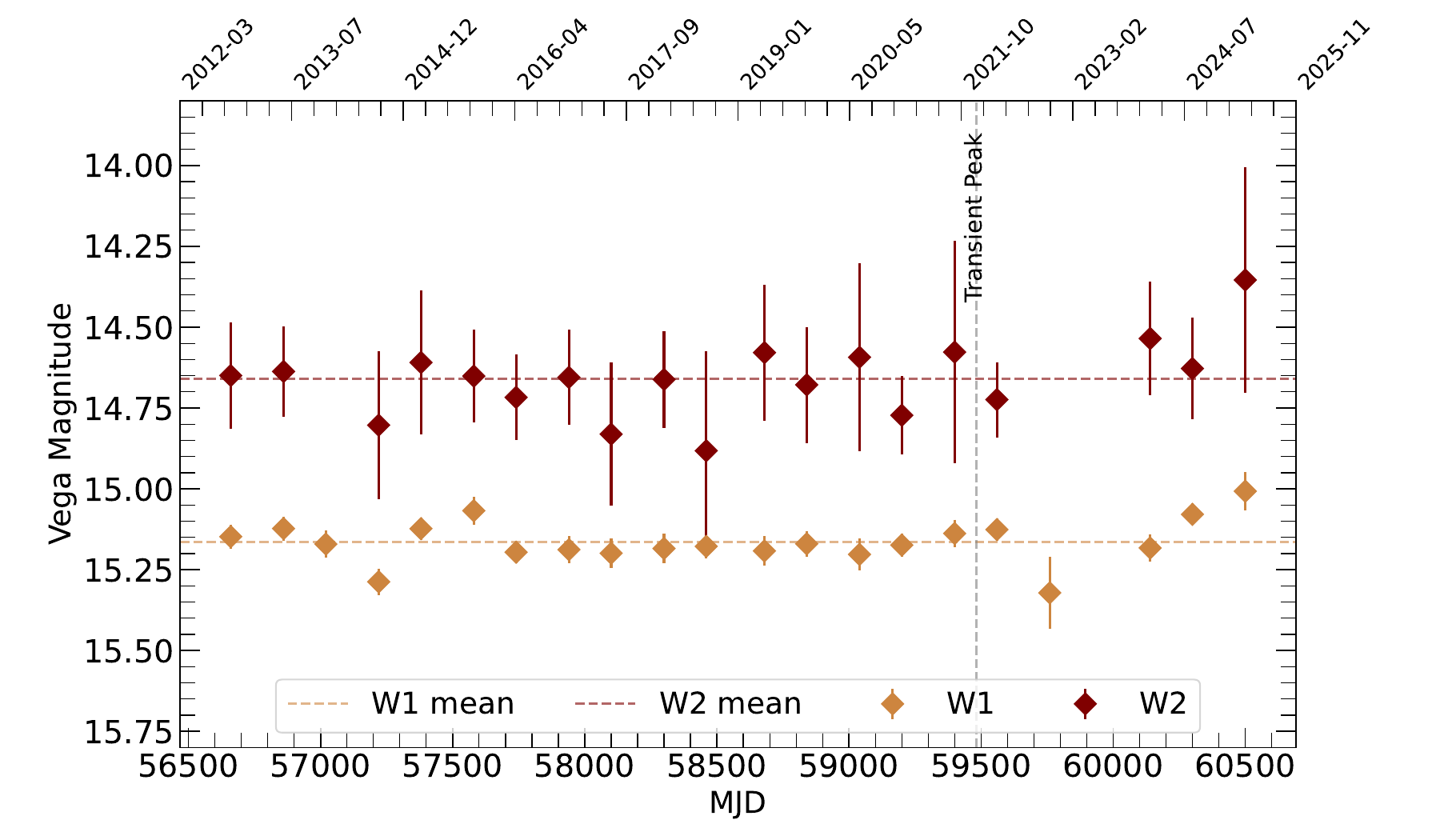}
        \caption{ WISE $W1$ and $W2$ light curves spanning 11 years of data from July 2013 -- July 2024, binned in 100-day bin sizes. No significant variability can be seen during this time, even near the transient peak.}  
       \label{fig:host_wiseLC}
    \end{center}  
\end{figure}

We searched for evidence of prior AGN activity in archival WISE data. Figure~\ref{fig:host_wiseLC} shows the WISE $W1$ and $W2$ light curves binned in 100-day intervals over 11 years (2013--2024), showing no significant variability. The $W1-W2 = -0.05 \pm 0.12$ color is well below the AGN threshold of $W1-W2 > 0.8$ \citep{stern12} throughout this period. Therefore, the host of AT2021yky harbors a low-luminosity AGN which is active enough to drive the $\rm [O\,III]/H\beta$ and $\rm [N\,II]/H\alpha$ ratios into the Seyfert region, but far less luminous than the host in the mid-IR and optical.

\subsection{Light curve}\label{sec:LC}

In this section, we present the optical photometry of AT2021yky and characterize its temporal evolution.
We estimate the time of peak using the ZTF $g$-band light curve by fitting a weighted quadratic function to the data within a $\pm15$-day window around the brightest observed point. The maximum of the resulting parabola gives a peak time of MJD $59481.98 \pm 1.01$ and a peak $g$-band luminosity of $L_{\rm peak} = (2.34 \pm 0.15) \times 10^{43}$~erg~s$^{-1}$. Next, we fit the rise of the ZTF $g$-band light curve as a power law with
\begin{equation}
L(t) =
\begin{cases}
k, & t < t_{1} \\[6pt]
k + h\left(t - t_{1}\right)^\alpha, & t \geq t_{1},
\end{cases}
\end{equation}
\noindent where $k$ is the baseline luminosity, $t_1$ is the time of first light, $\alpha$ is the power-law index, and $h$ is a normalization constant. We adopt MJD~$59481.9$ as the reference epoch corresponding to the observed peak, and we fix the baseline luminosity to the host galaxy ZTF $g$-band value (18.67 $\pm$ 0.04 mag), $k = (1.17 \pm 0.20)\times10^{43}$~erg~s$^{-1}$. We fit the model to data in the range $-80$ to $+10$~days relative to peak using \texttt{scipy.optimize.curve\_fit}, obtaining $t_1 = \mathrm{MJD}~59462.94 \pm 0.31$ and $\alpha = 1.33 \pm 0.07$. This is similar to the power-law slopes of the faint and fast TDEs, such as ASASSN-23bd \citep[$\approx$1.24;][]{Hoogendam_2024}. The rise from first light to peak is $18.2 \pm 0.7$ rest-frame days. This is significantly shorter than the rise times of typical optically selected TDEs, which generally rise over 30--50 days \citep[e.g.,][]{hammerstein23, Yao2023}, and is instead comparable to the rapid rise timescales seen in LFBOTs \citep{Sevilla26}. We discuss the implications of this fast rise time and compare it to the broader transient population in Section~\ref{sec:discussion}.

\begin{figure}[t]
\centering
\includegraphics[width=\columnwidth]{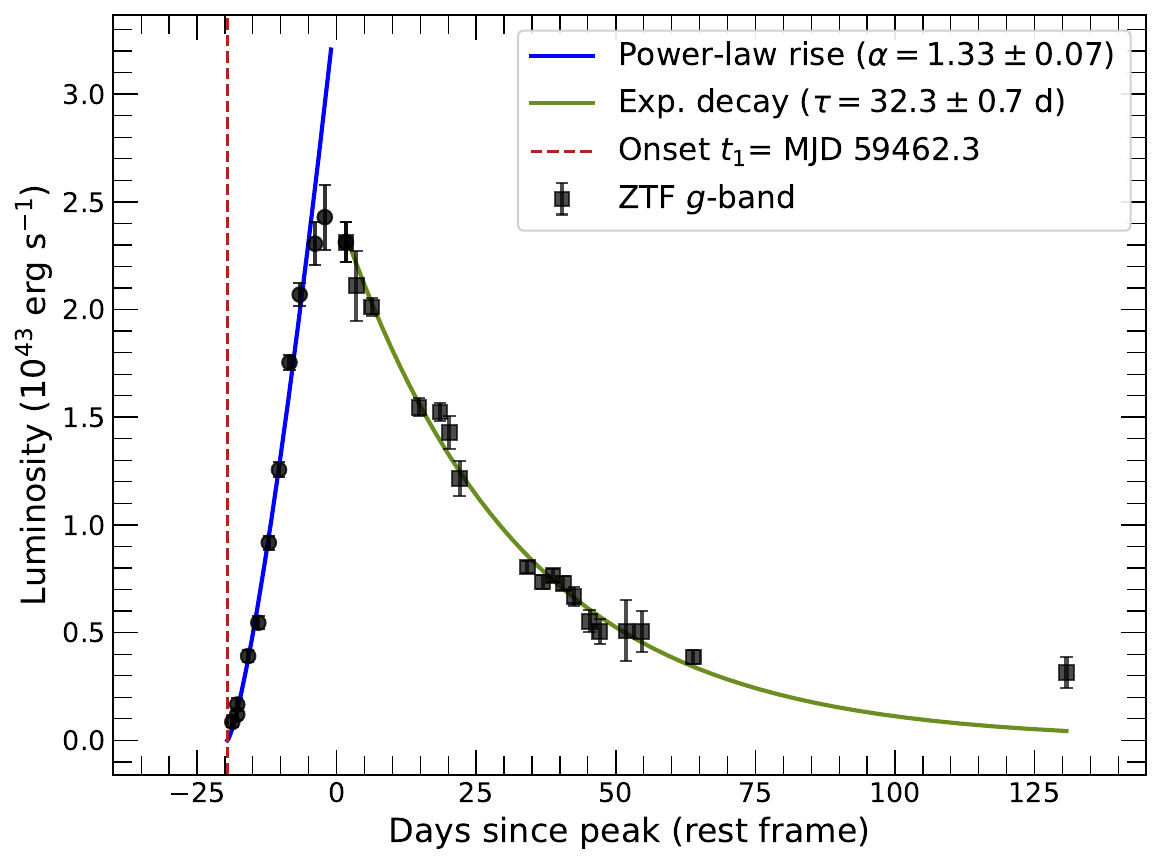}
\caption{Power-law (exponential) models for the light curve rise (fall). The best-fit power-law index is $\alpha = 1.33 \pm 0.07$ with the rise starting at $t_1=$ MJD $59462.9 \pm 0.3$ (vertical dashed line). The best-fit decay rate is $\tau \approx 32$~days.}
\label{fig:lightcurve_fits}
\end{figure}

We model the post-peak decline of AT2021yky using an exponential decay
\begin{equation}
L(t) = L_0 \, e^{-t/\tau},
\end{equation}

\noindent where $L_0$ is the luminosity at peak and $\tau$ is the decay rate. We fit this model to the ZTF $g$-band light curve over the interval 1--150 days post-peak in rest-frame and find $\tau = 32.32 \pm 0.72 $~rest-frame days and $L_0 = (2.46 \pm 0.04)\times10^{43}$~erg~s$^{-1}$. Typical TDE decline rates span $20$--$60$~days \citep{Yao2023, vanvelzen21, hinkle20a}, and our decay rate falls within this range. We discuss this further in Section~\ref{sec:discussion}. We show the power-law and exponential model fits to the lightcurve in Figure~\ref{fig:lightcurve_fits}.

\subsection{Submillimeter Constraints}\label{sec:radio}

In this section, we derive radio luminosity constraints from our JCMT/SCUBA-2 observations and place AT2021yky in the context of the TDE radio luminosity distribution. AT2021yky was not detected at 850 $\mu$m (353 GHz) with JCMT/SCUBA-2 at $\sim 30$ days post-peak, yielding a 3$\sigma$ upper limit of $\sim$ 5 mJy. To place this limit in the context of TDE radio observations, which are typically obtained at gigahertz frequencies, we converted our 353 GHz flux upper limit to an equivalent 5 GHz flux following the procedure of \cite{Hinkle_2022}. We used the radio light curve of AT2019dsg \citep{cendes21} as a template for the radio spectral energy distribution of a normal non-relativistic TDE, adopting $\nu_m = 0.1$ and $ p= 2.7$. Because our JCMT epoch at $\Delta t \approx 30$ days precedes the first well-constrained radio epoch of AT2019dsg (60 days), we extrapolated the self-absorption frequency and peak flux to 30 days using the power-law trends measured at 60–80 days ($\nu_a \propto t^{-0.96}$, $F_p \propto t^{0.62}$), yielding $\nu_a \approx 35$ GHz and $F_p \approx 0.44$ mJy at this epoch. Using the measured spectral shape to compute the flux ratio $f_{353\,\mathrm{GHz}}/f_{5\,\mathrm{GHz}} \approx 18.9$, our 353 GHz upper limit converts to a $3\sigma$ upper limit of $\lesssim 0.3$ mJy at 5 GHz. This corresponds to a 5 GHz radio luminosity of $L_{5\,\mathrm{GHz}} \lesssim 4 \times 10^{28}$ erg s$^{-1}$ Hz$^{-1}$, or $\nu L_\nu \lesssim 2 \times 10^{38}$ erg s$^{-1}$. This limit is broadly consistent with the range of radio luminosities observed in non-relativistic TDE outflows \citep[$L_R \sim 10^{27}-10^{29}$ erg s$^{-1}$  Hz$^{-1}$; e.g., ][]{cendes21, alexander20}, and does not rule out such an outflow at early times. However, it excludes any relativistic outflow or luminous radio flare comparable to the jetted TDE Sw J1644+57 with $L_R \sim 10^{31}-10^{33}$ erg s$^{-1}$ Hz$^{-1}$ \citep{zauderer11}.

\subsection{Transient SED Evolution}\label{sec:tsed}

\begin{figure}[t]
    \begin{center} 
    \includegraphics[width=9.3cm]{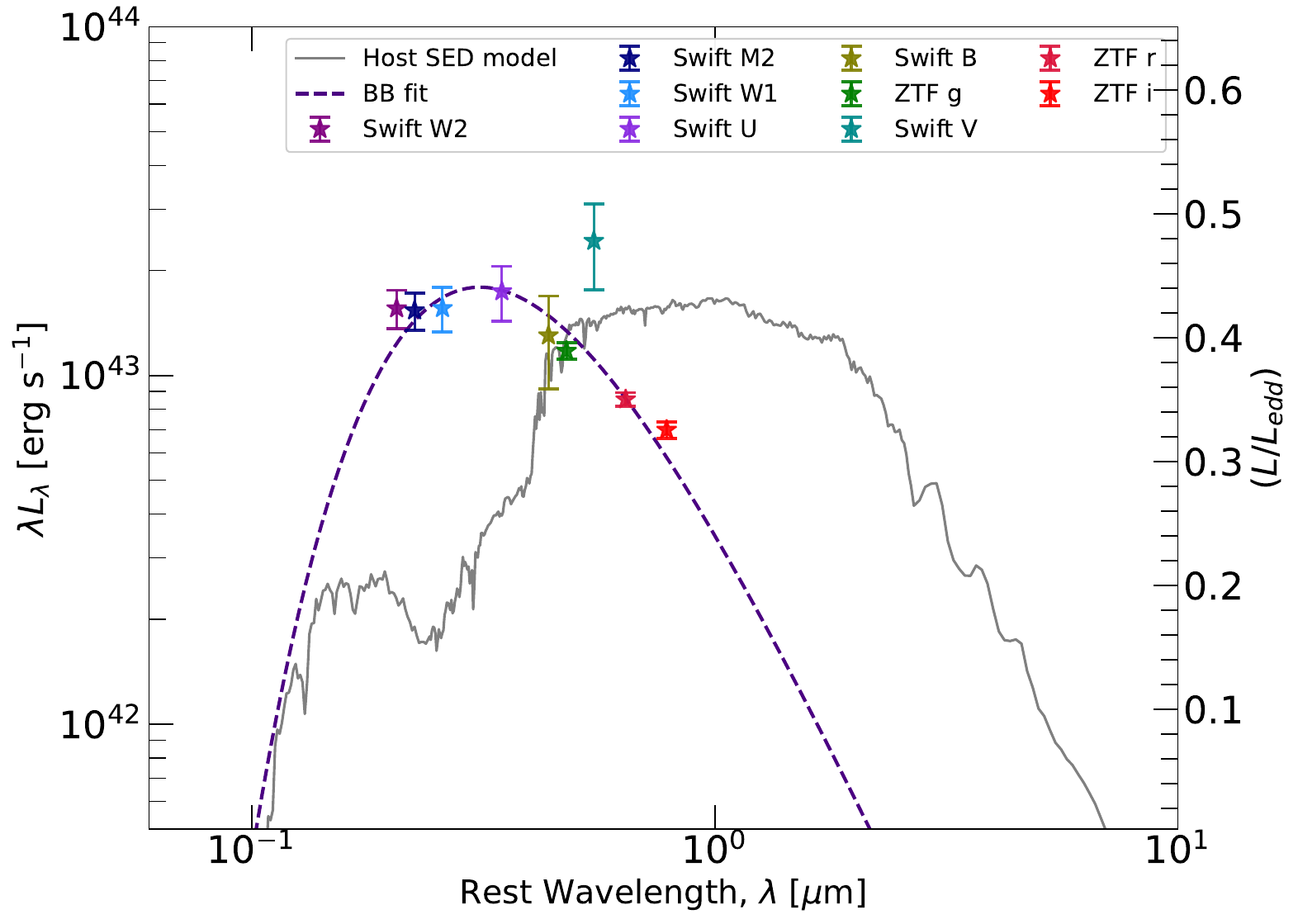}
        \caption{Spectral Energy Distribution (SED) of the host-subtracted transient $\sim20$ days post peak (MJD 59503). A blackbody fit to the transient is shown by the dashed indigo line and the host galaxy SED model from Fig~\ref{fig:HOST_SED} is shown by a solid gray line. The SED is well described by a single-temperature blackbody model with best-fit parameters of $T \approx 14000$~K and $R \approx 1.3 \times 10^{15}$~cm.}  
       \label{fig:transient_SED}
    \end{center}  
\end{figure}

Figure~\ref{fig:transient_SED} shows the host-subtracted and Galactic extinction-corrected UV/optical SED of AT2021yky constructed from photometric data obtained $\approx$20 days after peak (MJD~59503, the first Swift epoch). We fit a single-temperature blackbody to the corrected photometric data at each epoch following the method of \citet{Hinkle_2022}. We use Markov Chain Monte Carlo (MCMC) methods to fit each of the epochs, constraining the bolometric luminosity, temperature, and effective radius of the transient. The UV/optical SED of AT2021yky is well described by a slowly evolving blackbody over the $\sim88$ days of Swift coverage. For reference, the SED constructed with data $\approx$20 days after peak (Figure~\ref{fig:transient_SED}) is well described by a single-temperature blackbody model with best-fit parameters of $T = 13{,}800\pm 1360$~K, and $R = (1.33 \pm 0.05) \times 10^{15}$~cm.

A UV-dominated SED well fit by a blackbody is typical of TDEs \citep[e.g.,][]{holoien14a, holoien16-15oi, vanvelzen21}. For a black hole mass of $\log(M_{\rm BH}/M_\odot) \approx 6$, the Eddington luminosity is $L_{\rm Edd} \approx 1.5 \times 10^{44}$~erg~s$^{-1}$. The peak bolometric luminosity of $L_{\rm bol}^{\rm peak} = (4.1\pm 1.1)\times10^{43}$~erg~s$^{-1}$ corresponds to an Eddington ratio of $\lambda_{\rm Edd} \equiv L_{\rm bol}^{\rm peak}/L_{\rm Edd} \approx 0.26$, consistent with the near-Eddington accretion rates \citep[$\lambda_{\rm Edd} \gtrsim 0.2$,][]{wevers19} inferred for optical TDEs. AT2021yky is among the least luminous well-characterised nuclear transients, and its peak luminosity is comparable to those of the TDEs AT2019qiz \citep{nicholl20}, iPTF16fnl \citep{blagorodnova17, brown18}, and the ANT ASASSN-20hx \citep{Hinkle_2022}. Integrating the bolometric luminosity over the first 100 days of observations in the rest frame, we find a total radiated energy of $E_{\rm rad} = (1.05 \pm 0.04) \times 10^{50}$~erg, although this is a lower limit as it does not account for emission outside the UV/optical bands.

The X-ray non-detection places a $3\sigma$ upper limit of $L_X \lesssim 3.9 \times 10^{41}$~erg~s$^{-1}$ for the combined data, corresponding to less than $1\%$ of $L_{\rm bol}^{\rm peak}$ and $\lesssim 0.3\%$ of $L_{\rm Edd}$. X-ray faintness relative to the optical/UV emission is a typical property of optically selected TDEs and ANTs \citep{vanvelzen21, Hinkle_2022, auchettl17, Guolo24}. When detected, $L_{\rm BB}/L_X$ spans a wide range (0.5, 3000) at early times among optically selected TDEs \citep{Guolo24}, but the many X-ray non-detections imply even larger ratios for the bulk of the optically discovered population. The most X-ray faint events are consistent with reprocessing models in which soft X-rays from the inner accretion disk are absorbed and re-emitted at UV/optical wavelengths \citep{auchettl17, dai18, Thomsen22, Guolo24}. 

 \begin{figure}[t]
    \begin{center} 
    \includegraphics[width=\columnwidth]{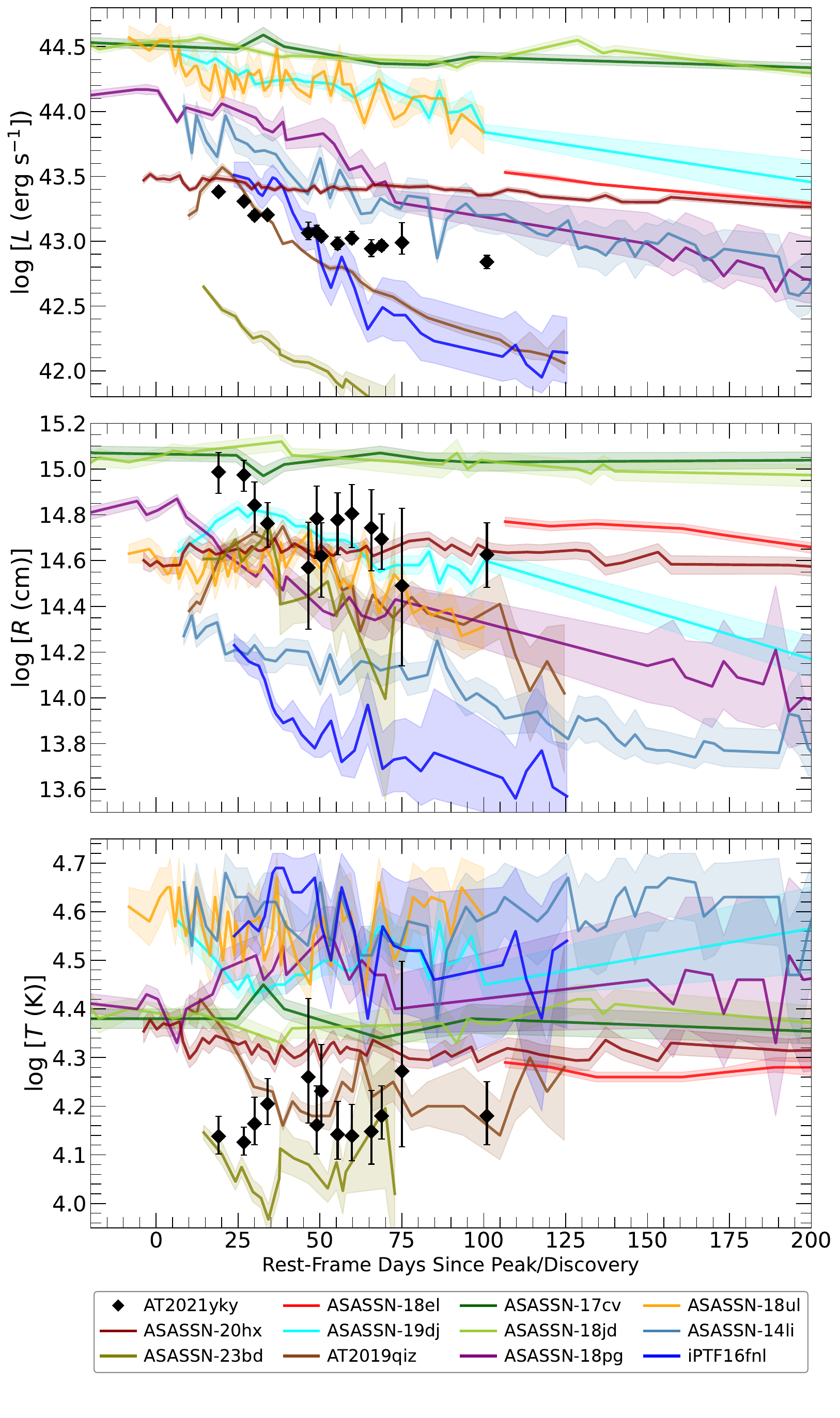}
        \caption{
Evolution of the UV/optical blackbody luminosity (top panel), radius (middle panel), and temperature (bottom panel) for AT2021yky (black squares) compared to a sample of well-studied TDEs and ANTs. The comparison sample includes the TDEs ASASSN-19dj \citep[cyan;][]{hinkle21b}, AT2019qiz \citep[brown;][]{nicholl20}, AT2018fyk \citep[orange;][]{wevers19a}, ASASSN-18pg \citep[purple;][]{holoien20}, ASASSN-14li \citep[steel blue;][]{brown17-14li}, ASASSN-23bd \citep[olive;][]{Hoogendam_2024}, and iPTF16fnl \citep[dark blue;][]{blagorodnova17}, as well as the nuclear transients ASASSN-18el \citep[red;][]{trakhtenbrot19a}, ASASSN-17cv \citep[dark green;][]{trakhtenbrot19b}, ASASSN-18jd \citep[light green;][]{neustadt20}, and ASASSN-20hx \citep[dark red;][]{Hinkle_2022}. All curves are smoothed for visual clarity. Time is in rest-frame days relative to the peak for objects with pre-peak coverage, and relative to discovery otherwise.}  
       \label{fig:BB_SED_evolution}
    \end{center}  
\end{figure}

Figure~\ref{fig:BB_SED_evolution} shows the evolution of the blackbody parameters of AT2021yky. The light curves are linearly interpolated to a uniform time grid for clarity. We compare AT2021yky to several well-studied TDEs: ASASSN-19dj/AT2019azh \citep[cyan;][]{hinkle21b},  AT2019qiz \citep[ZTF19abzrhgq; brown;][]{nicholl20}, ASASSN-18ul \citep[AT2018fyk; orange;][]{wevers19a}, ASASSN-18pg \citep[purple;][]{holoien20}, ASASSN-14li \citep[steel blue;][]{brown17-14li}, ASASSN-23bd \citep[olive;][]{Hoogendam_2024} and iPTF16fnl \citep[dark blue;][]{blagorodnova17}. We also include comparisons to the nuclear transients ASASSN-18el/AT2018zf \citep[red;][]{trakhtenbrot19a, ricci20}, ASASSN-17cv \citep[AT2017bgt, dark green;][]{trakhtenbrot19b}, ASASSN-18jd \citep[light green;][]{neustadt20}, and ASASSN-20hx \citep[dark red;][]{Hinkle_2022}. Time is in rest-frame days relative to peak for objects with well-constrained peaks, and relative to discovery otherwise. The blackbody photospheric radius of AT2021yky, $\log[R/\mathrm{cm}] \approx 14.6-15.0$, is large compared to the majority of the optically selected TDEs in our comparison sample, which occupy $\log[R/\mathrm{cm}] \approx 13.6-14.8$ over the same phase range, and is instead comparable to the AGN-associated nuclear transients ASASSN-17cv \citep{trakhtenbrot19b} and ASASSN-18jd \citep{neustadt20}. The effective temperature of $T \sim 14{,}000$~K near peak is cool relative to the typical $\sim 20{,}000$--$30{,}000$~K temperatures of optically selected TDEs and ANTs, but is consistent with the faint and fast (FaF) TDE subclass, ASASSN-23bd \citep{Hoogendam_2024} had $\log[T/\mathrm{K}] \approx 4.0-4.15$ over a comparable phase range, while AT2019qiz \citep{nicholl20} had $\log[T/\mathrm{K}] \approx 4.2-4.3$. The bolometric luminosity likewise places AT2021yky toward the faint end of the sample, declining from $\log[L/\mathrm{erg~s^{-1}}] \approx 43.4$ near peak to $\approx 42.9$ by $\sim 100$~days, comparable to AT2019qiz and ASASSN-20hx \citep{Hinkle_2022} and roughly an order of magnitude below the luminous ANTs ASASSN-17cv and ASASSN-18jd, though brighter than ASASSN-23bd. Together, the cool temperature and low luminosity are most consistent with the FaF TDE population. We discuss this further in section~\ref{sec:tde}. At peak, the photospheric radius of $R_\mathrm{BB} \sim 10^{15}$~cm is $\approx 10^{4}$ times the gravitational radius $R_g = GM_\bullet/c^2 \approx 1.5 \times 10^{11}$~cm for a $10^6~M_\odot$ black hole, indicating that the emitting region is far from the innermost accretion flow and is instead consistent with a reprocessing layer or debris envelope. The absence of significant photospheric cooling disfavors a supernova interpretation, and we discuss this further in Section~\ref{sec:discussion}.

\subsection{Spectral Properties} \label{sec:spec}

Most core-collapse (CC) supernovae exhibit strong blue continua in the first few days after explosion, arising from the shock cooling of the massive-star envelope \citep[e.g., ][]{Waxman17, Morag23}. This is the likely origin of the early-time classification of AT2021yky as a Type~II SN, as the initial spectrum was obtained shortly after discovery when the transient was still rising. However, as the source approached peak, the optical spectra evolved into a blue, largely featureless continuum inconsistent with a normal CC~SN at comparable phases. This spectral appearance is instead reminiscent of LFBOTs \citep{Sevilla26} and featureless nuclear transients such as ASASSN-20hx \citep{Hinkle_2022} and the superluminous supernova ASASSN-15lh \citep{dong16}, with narrow emission lines from the host galaxy superimposed on the continuum.

\begin{figure}
    \begin{center}    
    \includegraphics[width=\columnwidth]{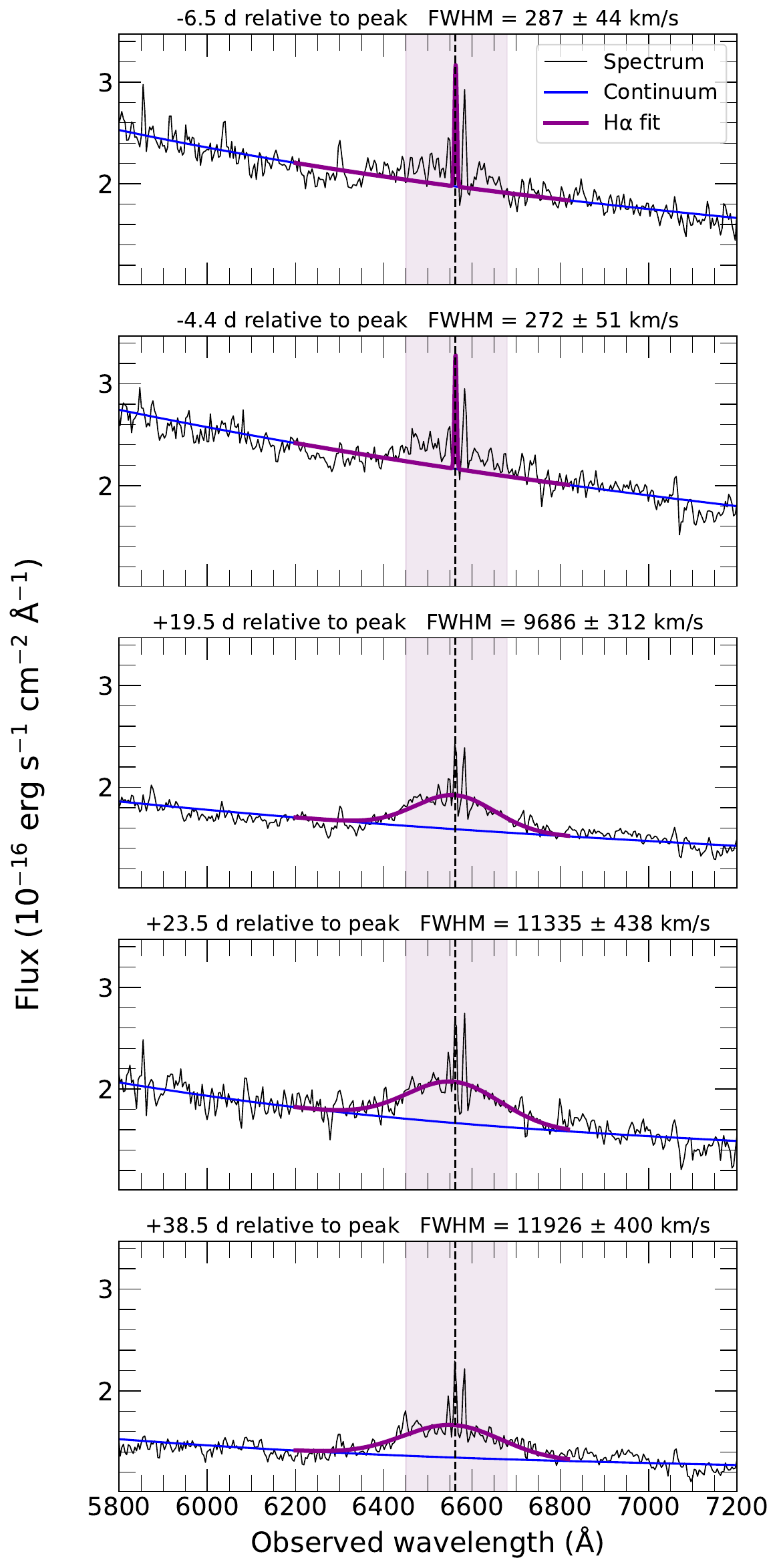}
        \caption{Host-subtracted SNIFS spectra of AT2021yky spanning from a week before the transient peak to 40 days post-peak. The H$\alpha$ emission line broadens post-peak with the FWHM reaching $\sim11,000~\rm km~s^{-1}$. A Gaussian fit is shown in purple, and the corresponding broad component FWHM and time relative to peak for each spectrum are given.}  
       \label{fig:spectral_properties}
    \end{center}  
\end{figure}

\begin{figure}
    \begin{center} 
    \includegraphics[width=9.8cm, height=6.2cm]{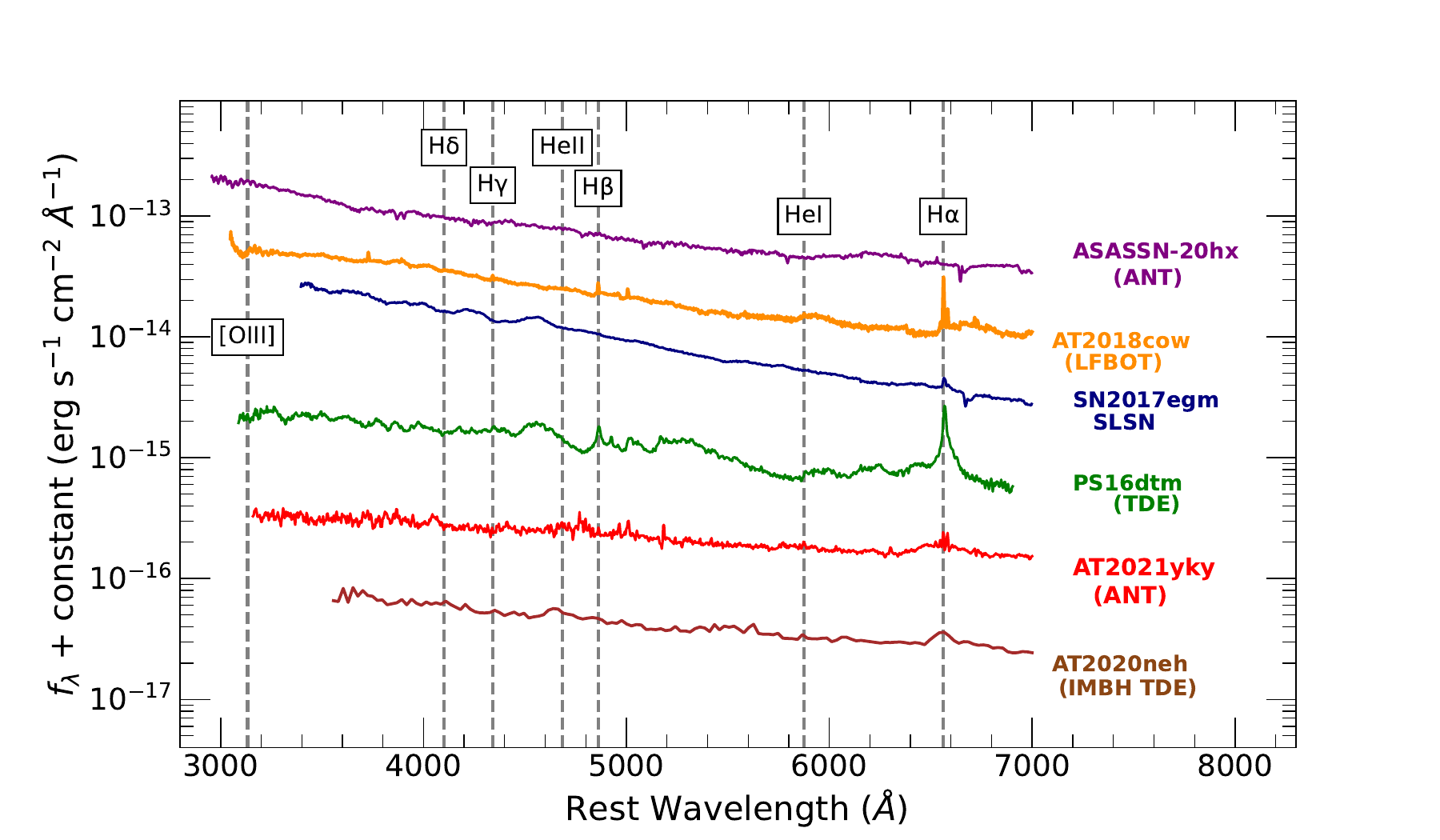}
        \caption{The near-peak host-subtracted SNIFS spectrum of AT2021yky compared with well studied transients: the ANT ASASSN-20hx \citep{Hinkle_2022}, the LFBOT AT2018cow \citep{Prentice18}, the SLSN-I SN2017egm \citep{bose18c},  the TDE PS16dtm \citep{blanchard17-16dtm}, and the IMBH TDE candidate AT2020neh \citep{Angus22}. All transient spectra are chosen to correspond to epochs near peak emission.}  
       \label{fig:compare_spectra}
    \end{center}  
\end{figure}

What distinguishes AT2021yky from these events is the subsequent emergence of a broad H$\alpha$ component. We use \texttt{scipy.optimize.curve\_fit} to fit a Gaussian to the H$\alpha$ line after masking the narrow component of the line to determine the Full Width at Half Maximum (FWHM). The Gaussian fits and corresponding FWHM of $\approx$ 11,000 km~s$^{-1}$ for the 5 SNIFS spectra are shown in Figure~\ref{fig:spectral_properties}, and we report the FWHM, central wavelength, and line luminosity corresponding to the transient phase in Table~\ref{tab:halpha_properties}. Other typical emission features, such as H$\beta$, He II, [O \,III], or Bowen fluorescence lines, are not detected at these epochs. This single-line spectroscopic evolution is unusual among both TDEs, which typically exhibit multiple broad Balmer and/or helium features \citep{Charalampopoulos_2022, vanvelzen21}, and ANTs, which tend to show multiple narrow and broad Balmer lines with [O \,III] \citep{Wiseman25}. The most analogous behavior is seen in AT2020neh, which is a rapidly rising TDE from a candidate intermediate-mass black hole (IMBH). The spectra of this transient are largely featureless after the peak until a broad H$\alpha$ component emerges at $\sim 40$ days \citep{Angus22}, though the line FWHM in AT2021yky is significantly broader. In Figure~\ref{fig:compare_spectra}, we compare the host-subtracted SNIFS spectrum of AT2021yky near peak to optical spectra of well-studied luminous transients spanning a range of physical classes: the ANT ASASSN-20hx \citep{Hinkle_2022}, the LFBOT AT2018cow \citep{Prentice18}, the SLSN-I SN2017egm \citep{bose18c}, the TDE PS16dtm \citep{blanchard17-16dtm}, and AT2020neh \citep{Angus22}. We also see H$\alpha$, P$\alpha$, and P$\beta$ lines in the near-IR spectrum (see Fig~\ref{fig:spex}). The H$\alpha$ FWHM in the host-subtracted SpeX spectrum at $\sim 26$ days is $ 9537 \pm 506$~km~s$^{-1}$, which is broadly consistent with what we observe in the optical SNIFS spectra around the same epoch.

Interpreted as a virial velocity, the broad H$\alpha$ width of $\sim 11{,}000~\rm km~s^{-1}$ implies a radius of $R \sim GM_\bullet / v^2 \sim 10^{14}$ cm from the black hole, which is smaller than the photospheric radius inferred from the blackbody fits (Section~\ref{sec:tsed}). We also find weak evidence for line broadening (the FWHM increases from $\approx9600~\rm km~s^{-1}$ to $\approx12000~\rm km~s^{-1}$) as the transient fades (from 20 days to 40 days post-peak, shown in Figure~\ref{fig:spectral_properties}), which is typically seen in AGN \citep{peterson04, denney09, Shen2012}. TDEs show opposite behaviour, where line-width scales with continuum flux \citep{holoien16-14li}.

\begin{table}
    \centering
    \caption{H$\alpha$ Line Evolution}
    \label{tab:halpha_properties}
    \begin{tabular}{cccc}
        \hline\hline
        Phase & FWHM (km s$^{-1}$) & $\lambda_{\rm cen}$ (\AA) & $L(\mathrm{H}\alpha)$ (erg s$^{-1}$) \\
        \hline
        
        $-6.5$ d  & $287 \pm 44$     & 6562.90 & $(1.21 \pm 0.15) \times 10^{40}$ \\
        $-4.4$ d  & $272 \pm 51$     & 6562.62 & $(1.08 \pm 0.14) \times 10^{40}$ \\
        $+19.5$ d & $9686 \pm 312$   & 6560.76 & $(1.14 \pm 0.033) \times 10^{41}$ \\
        $+23.5$ d & $11335 \pm 438$  & 6557.43 & $(1.62 \pm 0.054) \times 10^{41}$ \\
        $+38.5$ d & $11926 \pm 400$  & 6554.10 & $(1.35 \pm 0.039) \times 10^{41}$ \\
        \hline
    \end{tabular}
    \tablecomments{Evolution of the H$\alpha$ line properties with respect to peak time MJD 59481.9. We show the respective FWHM (km s$^{-1}$), central wavelength $\lambda_{\rm cen}$ (\AA), and line luminosity $L(\mathrm{H}\alpha)$ (erg s$^{-1}$) for each phase. }
\end{table}

We also consider whether the late-time narrow-line emission seen (LRIS and KCWI spectra in Figure~\ref{fig:kcwi}) can be interpreted as host excitation driven by the transient itself. If the narrow-line emission originates from a sufficiently extended region, the short-lived transient cannot have significantly illuminated the bulk of the emitting gas, and the observed narrow lines would instead reflect pre-existing conditions in the host galaxy. In spherical polar coordinates with radius $r$ and $x = \cos\theta$, the region that has been exposed to the transient after an elapsed time $\Delta t$ is bounded by $ r < (c\,\Delta t)/(1 - x)$. For a spherical narrow-line region (NLR) of radius $r_m$, we define the critical angle as $x_m = 1 - (c\Delta t/{r_m})$. For $x > x_m$, the transient has illuminated the NLR gas all the way out to $r_m$; for $x < x_m$, the illuminated volume is limited by light travel time. The total volume of NLR gas that has been irradiated is therefore
\begin{equation}
    V_{\rm irr} = \frac{2\pi}{3} \left[ 
        \int_{-1}^{x_m} \frac{(c\,\Delta t)^3}{(1-x)^3}\, dx 
        + \int_{x_m}^{1} r_m^3\, dx 
    \right].
\end{equation}
Dividing by the total NLR volume $4\pi r_m^3/3$, the illuminated fraction is
\begin{equation}
    f_{\rm irr} = \frac{3}{4}\frac{c\,\Delta t}{r_m}
    \left(1 - \frac{c^2\Delta t^2}{12\,r_m^2}\right),
    \label{eq:firr}
\end{equation}
which approaches unity when $c\,\Delta t = 2r_m$ (twice the light travel time across the NLR),  and the entire sphere has been illuminated. For $c\,\Delta t \ll r_m$, the second term in Equation~\ref{eq:firr} is negligible, and the illuminated fraction simplifies to
\begin{equation}
    f_{\rm irr} \simeq \frac{3}{4}\frac{c\,\Delta t}{r_m}.
\end{equation}
For a transient lifetime of $\Delta t \sim 1~\mathrm{yr}$, $c\,\Delta t \approx 0.3~\mathrm{pc}$. Typical NLR radii in Seyfert galaxies are $r_m \sim 10$--$1000~\mathrm{pc}$, implying $f_{\rm irr} \lesssim 0.02$ even for a very compact NLR.  The late-time spectra (see Figure~\ref{fig:kcwi}) are also a good model of the narrow emission lines at all epochs, while we would expect evolution in the narrow lines if the emission were being driven by the transient. Based on these two considerations, the late-time narrow-line spectrum should closely resemble the pre-transient spectrum, and thus is evidence for prior AGN activity.

\section{Discussion and Conclusions}\label{sec:discussion}

In this section, we discuss the possible physical origins of AT2021yky. To summarize the key observational properties established above, AT2021yky reached a peak bolometric luminosity of $L_{\rm peak} = (4.1 \pm 1.1) \times 10^{43}$~erg~s$^{-1}$, placing it at the fainter end of the optically selected TDE population, following a rapid rise of $\sim18$~days. Its early spectra are characterized by a largely featureless blue continuum, with a single broad ($\mathrm{FWHM} \sim 11000~\rm km~s^{-1}$) H$\alpha$ emission feature emerging at later times. The host galaxy has weak prior AGN activity, and we infer a black hole mass of
$\log(M_{\rm BH}/M_{\odot}) \approx 6.07$. This combination places AT2021yky in an ambiguous region of transient parameter space. We systematically consider five physical scenarios: a core-collapse supernova, an LFBOT, an AGN flare, a TDE, and an ANT. For each scenario, we assess the degree to which the multiwavelength observations presented in Section~\ref{sec:results} support or disfavor that interpretation.

\subsection{AT2021yky as a core-collapse supernova}

AT2021yky was initially classified as a Type II supernova \citep{ChuTNS2021}, but several lines of evidence argue against a Type II supernova classification. First, the light curve morphology is inconsistent with a Type IIP SN, which exhibits a characteristic plateau phase lasting $\sim100$ days powered by hydrogen recombination, followed by a steep drop onto the radioactive tail \citep[e.g.,][]{anderson14, Valenti16}. AT2021yky shows no such plateau. Instead, it rises rapidly over $\sim 18$ days as compared to the expected steeper rise for IIPs and decays smoothly (see Figure~\ref{fig:lightcurve_fits}).  Second, the decline timescale places AT2021yky firmly in the TDE regime rather than among the SN population. As shown in Figure~\ref{fig:fbots_comparison}, the characteristic rise and decay times of AT2021yky are broadly consistent with the TDE region rather than typical core-collapse SNe, which fade on timescales of weeks to a few months. The slow, sustained decline is more naturally explained by accretion-powered emission than by radioactive decay of $\rm Ni^{56}$.

Third, the spectroscopic evolution is atypical for any known SN subclass. As shown in Figure~\ref{fig:spectral_properties} and Table~\ref{tab:halpha_properties}, the broad H$\alpha$ (FWHM$\sim 11,000~\rm km~s^{-1}$) appears only post-peak and strengthens with time, which is the opposite behavior from Type II Supernovae, where Balmer lines are present from the earliest epochs and narrow as the ejecta cool and recombine. The emergence of a broad component on a timescale of weeks to months after peak, without accompanying absorption troughs or P Cygni profiles, is instead characteristic of TDE-like reprocessing or AGN-like broad-line region emission, not SN ejecta. 

Fourth, the blackbody temperature evolution is inconsistent with a supernova interpretation. Typical Type II supernovae cool rapidly from $\sim 10^4$ K at peak to $\sim 5000-6000$ K within a few weeks as the photosphere recedes into the ejecta \citep{filippenko97, Dessart11}. AT2021yky maintains a nearly constant temperature of $T_\mathrm{BB} \sim 14,000 $ K over at least $\sim100$ days (Figure~\ref{fig:BB_SED_evolution}), showing no evidence for cooling. While this temperature is somewhat cooler than the hottest TDEs, the evolution is inconsistent with the SN scenario. AT2021yky is also unlikely to be a superluminous supernova (SLSN), as its peak absolute magnitude ($-19.64\pm0.12$~mag, $g$ band) falls below the $M<-21$ mag threshold typically used to define the SLSN class \citep{Quimby11}.

\subsection{AT2021yky as a Luminous Fast Blue Optical Transient}\label{sec:comp_fbots}

\begin{figure}[t]
\includegraphics[width=0.45\textwidth]{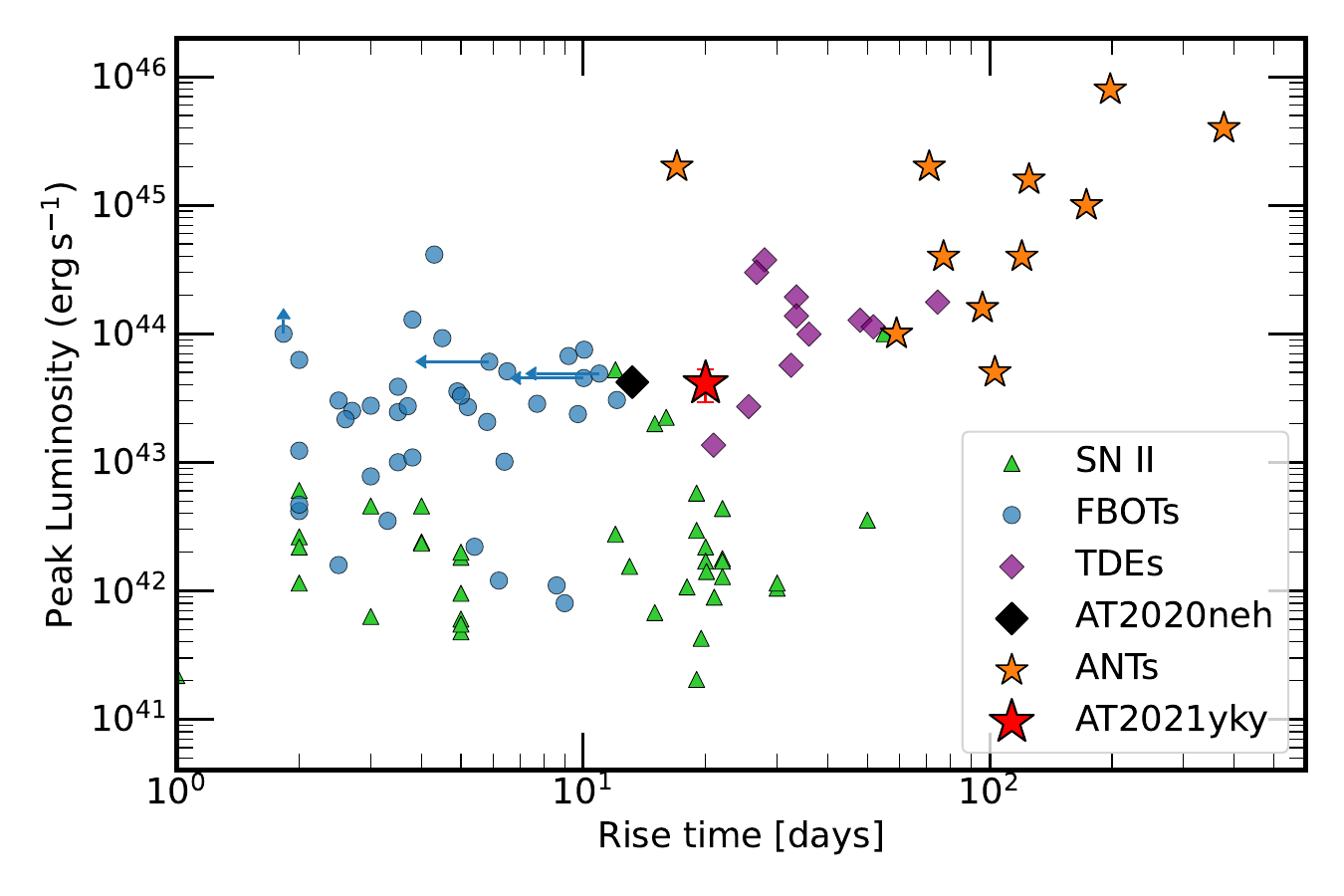} \includegraphics[width=0.45\textwidth]{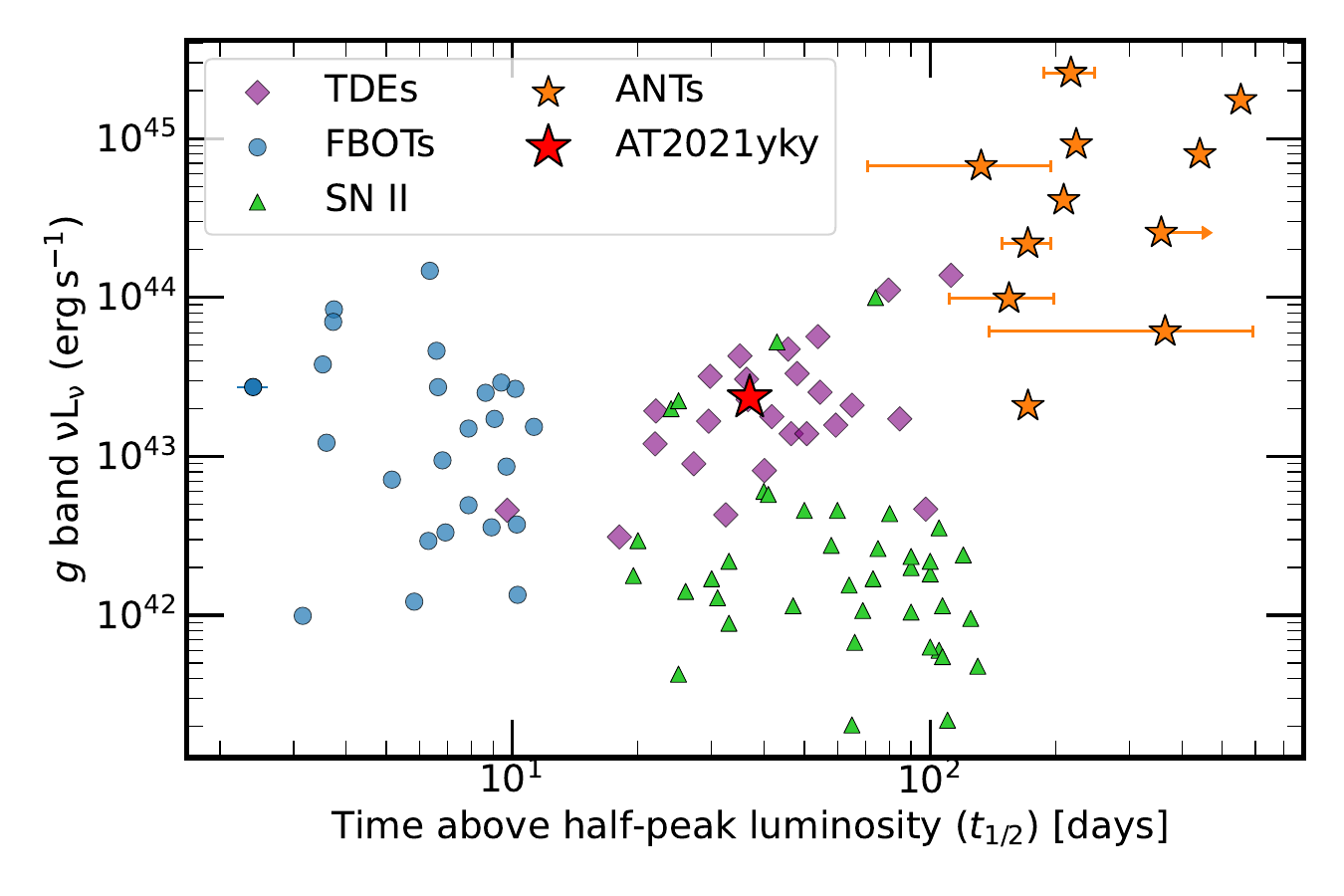}
\caption{
(Top) Peak luminosity versus rise time for various classes of optical transients. AT2021yky (red star, this work) is shown alongside AT2020neh (black diamond) for comparison. Other transients include FBOTs and LFBOTs (blue circles; \citealt{drout14, Arcavi16, Pursiainen18}), SNe~II (green triangles), TDEs (purple diamonds), and the sample of ambiguous nuclear transients (ANTs; orange stars) of \citet{Wiseman25}.  AT2021yky lies in an intermediate parameter space between the TDE and FBOT populations, at rise times shorter than any ANT in the comparison sample. Adapted from \citet{Lebaron26} and \citet{Angus22}.
(Bottom) Peak $g$-band luminosity versus time above half-peak luminosity for a similar sample, including FBOTs and LFBOTs (blue circles), SNe~II (green triangles), TDEs (purple diamonds), and ANTs (orange stars). ANT horizontal error bars reflect the width of the observing gap bracket in ZTF data. AT2021yky is shown as a red star and lies in the TDE regime of parameter space, more than a factor of three shorter in $t_{1/2}$ than any ANT. Adapted from \citet{Somalwar25}.}
\label{fig:fbots_comparison}
\end{figure}

AT2021yky shares some similarities with LFBOTs \citep{Ho23, Sevilla26}. These include the peak absolute magnitude $M_g \approx -20$, total radiated energy ($\sim 1 \times 10^{50}$ erg), the fast rise time ($\approx$ 18 days), and featureless spectra with broad H$\alpha$ \citep{Lebaron26}, which are broadly consistent with the LFBOT population (see Figure~\ref{fig:fbots_comparison} for rise time comparison). However, it is inconsistent with this class on several key criteria. The light-curve decay is far too slow: LFBOTs are defined by a time above half-peak luminosity of $t_{1/2} \leq 12$ days \citep{drout14}, whereas AT2021yky has $t_{1/2} \sim 40$ days (see Figure~\ref{fig:fbots_comparison}). Next, most of the LFBOTs show a blueshifted H$\alpha$ component 
\citep[e.g.,][]{Gutirrez24}, which we do not see in AT2021yky.  Furthermore, nearly all confirmed LFBOTs are detected at X-ray and/or radio wavelengths, with X-ray luminosities of $10^{42}-10^{44}$ erg s$^{-1}$ and radio luminosities of $10^{38}-10^{40}$ erg s$^{-1}$ at 10 GHz \citep{Sevilla26}, which exceed the X-ray and radio limits for AT2021yky from Sec~\ref{sec:xray} and \ref{sec:radio}. Together, these properties rule out an LFBOT classification.

\subsection{AT2021yky as an AGN flare}

Standard AGN variability also fails to account for the observed properties of AT2021yky. Typical AGN spectra contain prominent broad Balmer lines and narrow forbidden lines such as [O\,\textsc{iii}] and [N\,\textsc{ii}] \citep{vandenberk01, frederick20, sheng2020}, of which only broad H$\alpha$ is detected here. The UV/optical SED of AT2021yky is better described by a blackbody than the power-law continuum characteristic of AGN emission \citep{vandenberk01, temple2023}. Furthermore, the 11-year NEOWISE $W1$ and $W2$ light curves show no significant variability prior to the transient 
(Figure~\ref{fig:host_wiseLC}). This argues against the host harboring a strongly accreting AGN before the event. Together, these properties are inconsistent with AT2021yky arising from typical AGN variability.

\begin{figure}[t]
  \begin{center} 
  \includegraphics[width=\columnwidth]{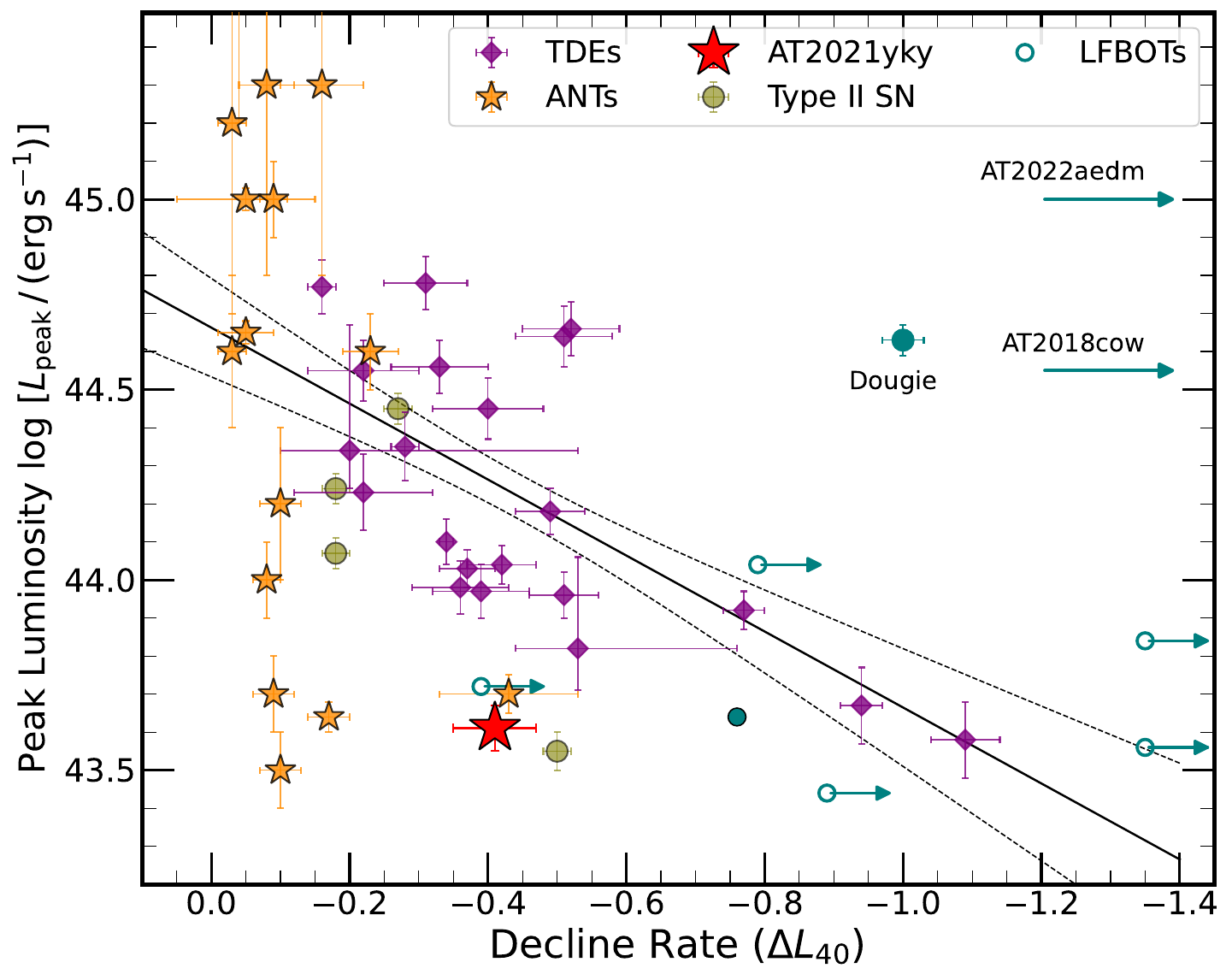}
       \caption{Peak luminosity versus decline rate $\Delta L_{40}$, which is the logarithmic change in bolometric luminosity between peak and 40 days post-peak, for a sample of TDEs, ANTs and other luminous transients. Purple diamonds are TDEs, orange stars are ANTs \citep{Wiseman25}, green circles are Type II SNe \citep{hinkle20a, hinkle21a}, and teal circles are LFBOTs \citep{Sevilla26}, including the luminous transient Dougie \citep{vinko15}. Open teal circles with arrows denote LFBOTs that are undetected 40 days post-peak, representing upper limits on $\Delta L_{40}$. AT2018cow \citep{Marguitti19} and AT2022aedm \citep{Nicholl23} are also shown with arrows as their rapid decline rates place $\Delta L_{40}$ beyond the plot range. The solid and dashed black lines show the best-fit relation for TDEs and its uncertainty. The red star marks AT2021yky. ANT peak luminosities are from \citet{Wiseman25}, Table 4; the large uncertainties for some sources reflect blackbody fits constrained by optical photometry alone.}  
       \label{fig:L40}
    \end{center}  
\end{figure}

\subsection{AT2021yky as a TDE}\label{sec:tde}

A combination of strong UV emission, sustained blue colors, and a blackbody SED makes AT2021yky a compelling TDE candidate during its early evolution. Despite being less luminous than the majority of known TDEs, AT2021yky exhibits substantial UV emission, reaching a peak absolute UV magnitude of $M_{\mathrm{UV}} \approx -18.8$~mag. Its SED is well characterized by a blackbody model, consistent with the thermal emission commonly observed in TDEs \citep[e.g.,][]{holoien16-14li, holoien19b, Gezari2021}. The derived blackbody temperature remains approximately constant at $14{,}000$~K throughout the observed evolution, placing AT2021yky among the coolest of TDEs relative to the typically hotter temperatures of $\sim 30,000$~K \citep[e.g.,][]{holoien14-14ae, holoien16-14li, holoien19c, holoien20, Gezari2021, hinkle21a, hammerstein23, Yao2023}.

AT2021yky's peak bolometric luminosity of $(4.1 \pm 1.1) \times 10^{43}$~erg~s$^{-1}$ is low compared with the broader TDE population, and its $\sim18$~day rise is correspondingly rapid. Both properties place it in the parameter space occupied by the growing class of faint and fast (FaF) TDEs
\citep{blagorodnova17, nicholl20, Hoogendam_2024}. In Table~\ref{tab:faf_comparison} we compare AT2021yky to five well-studied members of this class. The inferred black hole mass of $\log(M_{\rm BH}/M_\odot) = 6.07^{+0.17}_{-0.28}$ is consistent with the range spanned by these objects, which cluster around $\log(M_{\rm BH}/M_\odot) \sim 5$--$6.5$, and the X-ray non-detection is likewise consistent within this sample as FaF TDEs are predominantly X-ray faint.

Where AT2021yky differs from the FaF population is spectroscopic. Each of the comparison objects in Table~\ref{tab:faf_comparison} shows hydrogen together with helium, and in several cases N\,{\sc iii} Bowen fluorescence features. AT2021yky instead shows a largely featureless blue continuum with a single broad H$\alpha$ component emerging after peak, and it lacks the broad H$\beta$, He\,{\sc ii}, [O\,{\sc iii}], and Bowen fluorescence lines commonly associated with TDEs \citep{Charalampopoulos_2022, vanvelzen21}. The host environment is also distinct: the FaF comparison objects are mostly hosted by quiescent Balmer-strong, spiral, and dwarf galaxies, whereas the host of AT2021yky shows signatures of pre-existing low-luminosity AGN activity.

\begin{table*}
\centering
\caption{Comparison of different properties of AT2021yky with well-studied faint and fast TDEs (iPTF16fnl, AT 2019qiz, ASASSN-23bd, AT2020wey and AT2020neh).}
\label{tab:faf_comparison}
\begin{tabular}{lcccccc}
\hline \hline \\
& AT2021yky & ASASSN-23bd & AT 2020wey & iPTF16fnl & AT 2019qiz & AT 2020neh \\
\hline
$^{(a)}\log(M_{\rm BH}/M_\odot)$ & $^{(b)}6.07_{-0.28}^{+0.17}$ &$6.4^{+0.1}_{-0.1}$  & -- & $5.5^{+0.42}_{-0.42}$ & $^{(c)}(5.75-6.52)^{+0.45}_{-0.45}$ & $^{(d)}4.8^{+0.5}_{-0.9}$ \\[4pt]
$L_{\rm X}$ ($10^{40}$\,erg\,s$^{-1}$) & $<34$ & $<1.75$ & $<6.1$ & $<0.24$ & $5.1$ & $<45$ \\[4pt]
Spectral type & Featureless with broad H$\alpha$ &  H+He+N\,{\sc iii} & H+He & H+He+N\,{\sc iii} & H+He+N\,{\sc iii} & H+He+N\,{\sc iii} \\[4pt]
Host galaxy & weak AGN & LINER & quiescent Balmer-
strong (QBS) & QBS & Spiral & Dwarf \\
\hline
\end{tabular}
\begin{flushleft}
\footnotesize
$^{(a)}$Based on the $M-\sigma$ relation. (b) We calculated the mass using $M_{\rm BH}$--$M_*$ scaling relation. (c) There is a range in the BH masses because three different relations are used \citep{nicholl20}. (d) Calculated from light-curve models. Table adapted from Table~4 of \citet{Charalampopoulos23}.
\end{flushleft}
\end{table*}

The closest spectroscopic analog is AT2020neh \citep{Angus22}, the only optical TDE candidate associated with an IMBH. Both events have similar peak luminosities and rise times (see Figure~\ref{fig:fbots_comparison}), and both exhibit a blue, featureless continuum at early times that subsequently develops a broad H$\alpha$ emission line with no evidence for elements heavier than hydrogen at late epochs, a spectral sequence that \citet{Angus22} identify as a hallmark of TDE classification. AT2020neh shows a blueshifted H$\alpha$ component that we do not detect in AT2021yky. The late-time optical light curves of both events show persistent low-level residual emission above the
pre-transient host baseline, suggestive of ongoing accretion. The two differ in central mass and environment: the black hole mass we infer for AT2021yky places it above the IMBH range \citep[$2 < \log (M_\mathrm{BH}/M_\odot) < 6$;][]{Ebisuzaki01} and above the mass inferred for AT2020neh from light curve models. Unlike the quiescent star-forming dwarf host of AT2020neh, the host of AT2021yky is weakly active. AT2021yky may therefore represent an analog of AT2020neh occurring around a more massive black hole in a more massive, weakly active host.

\subsection{AT2021yky as an ANT}

In the end, the available data for AT2021yky do not converge on a single physical interpretation. The observations point to a nuclear transient powered by some form of accretion onto the central SMBH. We cannot firmly establish whether that accretion was triggered by the disruption of a star or by a change in the accretion state of the pre-existing low-luminosity AGN. The photometric properties of AT2021yky are broadly consistent with the population of faint and fast TDEs. However, the spectroscopic evolution is less TDE-like. Unlike the majority of optically selected TDEs, AT2021yky lacks the characteristic emission-line properties of TDEs, displaying only a single broad H$\alpha$ component without accompanying He~II or higher-order Balmer emission. We further find that this component broadens as the transient fades, a behaviour more typical of changes in the broad-line region of an AGN than of the spectral evolution of a TDE. The X-ray non-detection does not distinguish between these possibilities, as both X-ray-faint TDEs and low-luminosity AGN flares are consistent with the observed upper limits. These are precisely the ambiguities that define the ANT classification.

AT2021yky is not the first object to present this combination. ASASSN-20hx shows a comparable fast rise (20$-$30 days) and TDE-like blackbody decline at a similarly low luminosity and temperature, in a host with a low-luminosity AGN. The two events differ in some ways, as ASASSN-20hx was detected in X-rays near peak, and its spectra showed no emission lines at all, in contrast to the broad H$\alpha$ seen in AT2021yky. The different spectroscopic behavior of these two ANTs suggests that a TDE/AGN-flare in weakly active hosts may manifest differently from object to object, rather than resolving to a single mechanism common to the class.

The physical mechanism driving ANTs remains an open question, with proposed explanations including TDEs in AGN environments \citep{Chan19}, extreme accretion-rate changes in low-luminosity AGN \citep{trakhtenbrot19a}, and partial tidal disruptions \citep[e.g., PS1-11af;][]{chornock14}. Continued discovery and systematic spectroscopic monitoring of a larger population of these nuclear flares will be necessary to determine whether they share a common physical origin or represent several distinct channels converging on similar observational signatures. Ongoing surveys like ASAS-SN, ZTF, and ATLAS, as well as the new surveys, such as the Vera C. Rubin Observatory's Legacy Survey of Space and Time \citep[LSST;][]{Ivezic19} and the Nancy Grace Roman Space Telescope \citep{Spergel15}, will steadily build larger samples of these exotic nuclear transients. This will enable a more complete census of the accretion-driven phenomena occupying the AGN/TDE boundary.

\section*{Acknowledgements}
PP is thankful to Rick Pogge for his valuable feedback and the participants of IAU Symposium 406 for insightful discussions. PP is thankful to Natalie LeBaron, Raffaella Margutti, and Yuhan Yao for their feedback on the manuscript. PP is additionally grateful to Bobo for his unwavering support and encouragement throughout this project. J.T.H. acknowledges support from NASA through the NASA Hubble Fellowship grant HST-HF2-51577.001-A, awarded by STScI. STScI is operated by the Association of Universities for Research in Astronomy, Incorporated, under NASA contract NAS5-26555. CSK is supported by NSF grants AST-2307385 and AST-2407206. Parts of this research were supported by the Australian Research Council Discovery Early Career Researcher Award (DECRA) through project number DE230101069. W.B.H. acknowledges support from the National Science Foundation Graduate Research Fellowship Program under Grant No. 2236415. This work is based in part on observations made by ATLAS, Keck, Pan-STARRS, and UH88. The authors wish to recognize and acknowledge the very significant cultural role and reverence that the summits of Maunakea, Haleakal\=a, and Mauna Loa have always had within the indigenous Hawaiian community.  We are most fortunate to have the opportunity to conduct observations from these mountains. This work has made use of data from the Asteroid Terrestrial-impact Last Alert System (ATLAS) project. ATLAS project is primarily funded to search for near-earth asteroids through NASA grants NN12AR55G, 80NSSC18K0284, and 80NSSC18K1575; byproducts of the NEO search include images and catalogs from the survey area. This work was partially funded by Kepler/K2 grant J1944/80NSSC19K0112 and HST GO-15889, and STFC grants ST/T000198/1 and ST/S006109/1. The ATLAS science products have been made possible through the contributions of the University of Hawaii Institute for Astronomy, Queen’s University Belfast, the Space Telescope Science Institute, the South African Astronomical Observatory, and the Millennium Institute of Astrophysics (MAS), Chile.

\newpage

\bibliographystyle{mnras}
\bibliography{biblio} 

@ARTICLE{Bock_2026,
       author = {{Bock}, James J. and {Aboobaker}, Asad M. and {Adamo}, Joseph and {Akeson}, Rachel and {Alred}, John M. and {Alibay}, Farah and {Ashby}, Matthew L.~N. and {Bach}, Yoonsoo P. and {Bleem}, Lindsey E. and {Bolton}, Douglas and {Braun}, David F. and {Bruton}, Sean and {Bryan}, Sean A. and {Chang}, Tzu-Ching and {Chen}, Shuang-Shuang and {Cheng}, Yun-Ting and {Cheshire}, IV, James R. and {Chiang}, Yi-Kuan and {de Janvry}, Jean Choppin and {Condon}, Samuel and {Cook}, Walter R. and {Cooray}, Asantha and {Crill}, Brendan P. and {Cukierman}, Ari J. and {Dor{\'e}}, Olivier and {Dowell}, C. Darren and {Dubois-Felsmann}, Gregory P. and {Eifler}, Tim and {Everett}, Spencer and {Fabinsky}, Beth E. and {Faisst}, Andreas L. and {Fanson}, James L. and {Farrington}, Allen H. and {Fatahi}, Tamim and {Fazar}, Candice M. and {Feder}, Richard M. and {Frater}, Eric H. and {Grasshorn Gebhardt}, Henry S. and {Giri}, Utkarsh and {Goldina}, Tatiana and {Gorjian}, Varoujan and {Habib}, Salman and {Hart}, William G. and {Heinrich}, Chen and {Hora}, Joseph L. and {Huai}, Zhaoyu and {Hui}, Howard and {Jo}, Young-Soo and {Jeong}, Woong-Seob and {Kang}, Jae Hwan and {Kang}, Miju and {Kecman}, Branislav and {Kim}, Chul-Hwan and {Kim}, Jaeyeong and {Kim}, Minjin and {Kim}, Young-Jun and {Kim}, Yongjung and {Kirkpatrick}, J. Davy and {Kobayashi}, Yosuke and {Korngut}, Phil M. and {Krause}, Elisabeth and {Lee}, Bomee and {Lee}, Ho-Gyu and {Lee}, Jae-Joon and {Lee}, Jeong-Eun and {Lisse}, Carey M. and {Mariani}, Giacomo and {Masters}, Daniel C. and {Mauskopf}, Philip D. and {Melnick}, Gary J. and {Minasyan}, Mary H. and {Mirocha}, Jordan and {Miyasaka}, Hiromasa and {Moore}, Anne and {Moore}, Bradley D. and {Murgia}, Giulia and {Naylor}, Bret J. and {Nelson}, Christina and {Nguyen}, Chi H. and {Nguyen}, Hien T. and {Noh}, Jinyoung K. and {Padin}, Stephen and {Paladini}, Roberta and {Park}, Sung-Joon and {Penanen}, Konstantin I. and {Putnam}, Dustin S. and {Pyo}, Jeonghyun and {Ramachandra}, Nesar and {Ramanathan}, Keshav and {Rustamkulov}, Zafar and {Reiley}, Daniel J. and {Rice}, Eric B. and {Rocca}, Jennifer M. and {Seok}, Ji Yeon and {Smith}, Roger and {Stober}, Jeremy and {Susca}, Sara and {Teplitz}, Harry I. and {Thelen}, Michael P. and {Tolls}, Volker and {Torrini}, Gabriela and {Trangsrud}, Amy R. and {Unwin}, Stephen and {Velicheti}, Phani and {Wang}, Pao-Yu and {Wen}, Robin Y. and {Werner}, Michael W. and {Williams}, Abby E. and {Williamson}, Ross and {Wincentsen}, James and {Windhorst}, Rogier A. and {Yang}, Soung-Chul and {Yang}, Yujin and {Zemcov}, Michael},
        title = "{The SPHEREx Satellite Mission}",
      journal = {\apj},
         year = 2026,
        month = mar,
       volume = {999},
       number = {1},
          eid = {139},
        pages = {139},
          doi = {10.3847/1538-4357/ae2be2},
archivePrefix = {arXiv},
       eprint = {2511.02985},
 primaryClass = {astro-ph.IM},
       adsurl = {https://ui.adsabs.harvard.edu/abs/2026ApJ...999..139B}
}

@ARTICLE{Pandey_2025,
       author = {{Pandey}, Paarmita and {Hinkle}, Jason and {Kochanek}, Christopher and {Tucker}, Michael and {Reynolds}, Mark and {Neustadt}, Jack and {Thompson}, Todd and {Auchettl}, Katie and {Shappee}, Benjamin and {Do}, Aaron and {Desai}, Dhvanil and {Hoogendam}, W. and {Ashall}, C. and {Lowe}, Thomas and {Shahbandeh}, Melissa and {Payne}, Anna},
        title = "{Unraveling the Nature of the Nuclear Transient AT2020adpi}",
      journal = {The Open Journal of Astrophysics},
         year = 2025,
        month = dec,
       volume = {8},
        pages = {51453},
          doi = {10.33232/001c.151453},
archivePrefix = {arXiv},
       eprint = {2509.03593},
 primaryClass = {astro-ph.HE},
       adsurl = {https://ui.adsabs.harvard.edu/abs/2025OJAp....851453P}
}

@article{Graham_2025,
   title={An extremely luminous flare recorded from a supermassive black hole},
   ISSN={2397-3366},
   url={http://dx.doi.org/10.1038/s41550-025-02699-0},
   DOI={10.1038/s41550-025-02699-0},
   journal={Nature Astronomy},
   publisher={Springer Science and Business Media LLC},
   author={Graham, Matthew J. and McKernan, Barry and Ford, K. E. Saavik and Stern, Daniel and Cantiello, Matteo and Drake, Andrew J. and Ding, Yuanze and Kasliwal, Mansi and Koss, Mike and Margutti, Raffaella and Rose, Sam and Somalwar, Jean and Wiseman, Phil and Djorgovski, S. G. and Veres, Patrik M. and Bellm, Eric C. and Chen, Tracy X. and Groom, Steven L. and Kulkarni, Shrinivas R. and Mahabal, Ashish},
   year={2025},
   month=nov }

@article{Subrayan_2023,
   title={Scary Barbie: An Extremely Energetic, Long-duration Tidal Disruption Event Candidate without a Detected Host Galaxy at z = 0.995},
   volume={948},
   ISSN={2041-8213},
   url={http://dx.doi.org/10.3847/2041-8213/accf1a},
   DOI={10.3847/2041-8213/accf1a},
   number={2},
   journal={The Astrophysical Journal Letters},
   publisher={American Astronomical Society},
   author={Subrayan, Bhagya M. and Milisavljevic, Dan and Chornock, Ryan and Margutti, Raffaella and Alexander, Kate D. and Ramakrishnan, Vandana and Duffell, Paul C. and Dickinson, Danielle A. and Lee, Kyoung-Soo and Giannios, Dimitrios and Lentner, Geoffery and Linvill, Mark and Garretson, Braden and Graham, Matthew J. and Stern, Daniel and Brethauer, Daniel and Duong, Tien and Jacobson-Galán, Wynn and LeBaron, Natalie and Matthews, David and Sears, Huei and Venkatraman, Padma},
   year={2023},
   month=may, pages={L19} }

@article{Makrygianni_2023,
   title={AT 2021loi: A Bowen Fluorescence Flare with a Rebrightening Episode Occurring in a Previously Known AGN},
   volume={953},
   ISSN={1538-4357},
   url={http://dx.doi.org/10.3847/1538-4357/ace1ee},
   DOI={10.3847/1538-4357/ace1ee},
   number={1},
   journal={The Astrophysical Journal},
   publisher={American Astronomical Society},
   author={Makrygianni, Lydia and Trakhtenbrot, Benny and Arcavi, Iair and Ricci, Claudio and Lam, Marco C. and Horesh, Assaf and Sfaradi, Itai and Bostroem, K. Azalee and Hosseinzadeh, Griffin and Howell, D. Andrew and Pellegrino, Craig and Fender, Rob and Green, David A. and Williams, David R. A. and Bright, Joe},
   year={2023},
   month=aug, pages={32} }

@ARTICLE{Wiseman2023,
       author = {{Wiseman}, P. and {Wang}, Y. and {H{\"o}nig}, S. and {Castro-Segura}, N. and {Clark}, P. and {Frohmaier}, C. and {Fulton}, M.~D. and {Leloudas}, G. and {Middleton}, M. and {M{\"u}ller-Bravo}, T.~E. and {Mummery}, A. and {Pursiainen}, M. and {Smartt}, S.~J. and {Smith}, K. and {Sullivan}, M. and {Anderson}, J.~P. and {Acosta Pulido}, J.~A. and {Charalampopoulos}, P. and {Banerji}, M. and {Dennefeld}, M. and {Galbany}, L. and {Gromadzki}, M. and {Guti{\'e}rrez}, C.~P. and {Ihanec}, N. and {Kankare}, E. and {Lawrence}, A. and {Mockler}, B. and {Moore}, T. and {Nicholl}, M. and {Onori}, F. and {Petrushevska}, T. and {Ragosta}, F. and {Rest}, S. and {Smith}, M. and {Wevers}, T. and {Carini}, R. and {Chen}, T. -W. and {Chambers}, K. and {Gao}, H. and {Huber}, M. and {Inserra}, C. and {Magnier}, E. and {Makrygianni}, L. and {Toy}, M. and {Vincentelli}, F. and {Young}, D.~R.},
        title = "{Multiwavelength observations of the extraordinary accretion event AT2021lwx}",
      journal = {\mnras},
         year = 2023,
        month = jul,
       volume = {522},
       number = {3},
        pages = {3992-4002},
          doi = {10.1093/mnras/stad1000},
archivePrefix = {arXiv},
       eprint = {2303.04412},
 primaryClass = {astro-ph.HE},
       adsurl = {https://ui.adsabs.harvard.edu/abs/2023MNRAS.522.3992W}
}

@article{Holoien_2022,
   title={Investigating the Nature of the Luminous Ambiguous Nuclear Transient ASASSN-17jz},
   volume={933},
   ISSN={1538-4357},
   url={http://dx.doi.org/10.3847/1538-4357/ac74b9},
   DOI={10.3847/1538-4357/ac74b9},
   number={2},
   journal={The Astrophysical Journal},
   publisher={American Astronomical Society},
   author={Holoien, Thomas W.-S. and Neustadt, Jack M. M. and Vallely, Patrick J. and Auchettl, Katie and Hinkle, Jason T. and Romero-Cañizales, Cristina and Shappee, Benjamin. J. and Kochanek, Christopher S. and Stanek, K. Z. and Chen, Ping and Dong, Subo and Prieto, Jose L. and Thompson, Todd A. and Brink, Thomas G. and Filippenko, Alexei V. and Zheng, WeiKang and Bersier, David and Bose, Subhash and Burgasser, Adam J. and Channa, Sanyum and de Jaeger, Thomas and Hestenes, Julia and Im, Myungshin and Jeffers, Benjamin and Jun, Hyunsung D. and Lansbury, George and Post, Richard S. and Ross, Timothy W. and Stern, Daniel and Tang, Kevin and Tucker, Michael A. and Valenti, Stefano and Yunus, Sameen and Zhang, Keto D.},
   year={2022},
   month=jul, pages={196} }

@ARTICLE{Extinction2011,
       author = {{Schlafly}, Edward F. and {Finkbeiner}, Douglas P.},
        title = "{Measuring Reddening with Sloan Digital Sky Survey Stellar Spectra and Recalibrating SFD}",
      journal = {\apj},
         year = 2011,
        month = aug,
       volume = {737},
       number = {2},
          eid = {103},
        pages = {103},
          doi = {10.1088/0004-637X/737/2/103},
archivePrefix = {arXiv},
       eprint = {1012.4804},
 primaryClass = {astro-ph.GA},
       adsurl = {https://ui.adsabs.harvard.edu/abs/2011ApJ...737..103S}
}

@article{Hoogendam_2024,
   title={Discovery and follow-up of ASASSN-23bd (AT 2023clx): the lowest redshift and luminosity optically selected tidal disruption event},
   volume={530},
   ISSN={1365-2966},
   url={http://dx.doi.org/10.1093/mnras/stae1121},
   DOI={10.1093/mnras/stae1121},
   number={4},
   journal={Monthly Notices of the Royal Astronomical Society},
   publisher={Oxford University Press (OUP)},
   author={Hoogendam, W B and Hinkle, J T and Shappee, B J and Auchettl, K and Kochanek, C S and Stanek, K Z and Maksym, W P and Tucker, M A and Huber, M E and Morrell, N and Burns, C R and Hey, D and Holoien, T W -S and Prieto, J L and Stritzinger, M and Do, A and Polin, A and Ashall, C and Brown, P J and DerKacy, J M and Ferrari, L and Galbany, L and Hsiao, E Y and Kumar, S and Lu, J and Stevens, C P},
   year={2024},
   month=apr, pages={4501–4518} }

@ARTICLE{rayner03,
       author = {{Rayner}, J.~T. and {Toomey}, D.~W. and {Onaka}, P.~M. and {Denault}, A.~J. and {Stahlberger}, W.~E. and {Vacca}, W.~D. and {Cushing}, M.~C. and {Wang}, S.},
        title = "{SpeX: A Medium-Resolution 0.8-5.5 Micron Spectrograph and Imager for the NASA Infrared Telescope Facility}",
      journal = {\pasp},
         year = 2003,
        month = mar,
       volume = {115},
       number = {805},
        pages = {362-382},
          doi = {10.1086/367745},
       adsurl = {https://ui.adsabs.harvard.edu/abs/2003PASP..115..362R}
}

@ARTICLE{bose18c,
       author = {{Bose}, Subhash and {Dong}, Subo and {Pastorello}, A. and {Filippenko}, Alexei V. and {Kochanek}, C.~S. and {Mauerhan}, Jon and {Romero-Ca{\~n}izales}, C. and {Brink}, Thomas G. and {Chen}, Ping and {Prieto}, J.~L. and {Post}, R. and {Ashall}, Christopher and {Grupe}, Dirk and {Tomasella}, L. and {Benetti}, Stefano and {Shappee}, B.~J. and {Stanek}, K.~Z. and {Cai}, Zheng and {Falco}, E. and {Lundqvist}, Peter and {Mattila}, Seppo and {Mutel}, Robert and {Ochner}, Paolo and {Pooley}, David and {Stritzinger}, M.~D. and {Villanueva}, S., Jr. and {Zheng}, WeiKang and {Beswick}, R.~J. and {Brown}, Peter J. and {Cappellaro}, E. and {Davis}, Scott and {Fraser}, Morgan and {de Jaeger}, Thomas and {Elias-Rosa}, N. and {Gall}, C. and {Gaudi}, B. Scott and {Herczeg}, Gregory J. and {Hestenes}, Julia and {Holoien}, T.~W. -S. and {Hosseinzadeh}, Griffin and {Hsiao}, E.~Y. and {Hu}, Shaoming and {Jaejin}, Shin and {Jeffers}, Ben and {Koff}, R.~A. and {Kumar}, Sahana and {Kurtenkov}, Alexander and {Lau}, Marie Wingyee and {Prentice}, Simon and {Reynolds}, T. and {Rudy}, Richard J. and {Shahbandeh}, Melissa and {Somero}, Auni and {Stassun}, Keivan G. and {Thompson}, Todd A. and {Valenti}, Stefano and {Woo}, Jong-Hak and {Yunus}, Sameen},
        title = "{Gaia17biu/SN 2017egm in NGC 3191: The Closest Hydrogen-poor Superluminous Supernova to Date Is in a {\textquotedblleft}Normal,{\textquotedblright} Massive, Metal-rich Spiral Galaxy}",
      journal = {\apj},
         year = 2018,
        month = jan,
       volume = {853},
       number = {1},
          eid = {57},
        pages = {57},
          doi = {10.3847/1538-4357/aaa298},
archivePrefix = {arXiv},
       eprint = {1708.00864},
 primaryClass = {astro-ph.HE},
       adsurl = {https://ui.adsabs.harvard.edu/abs/2018ApJ...853...57B}
}

@ARTICLE{perley19,
       author = {{Perley}, Daniel A.},
        title = "{Fully Automated Reduction of Longslit Spectroscopy with the Low Resolution Imaging Spectrometer at the Keck Observatory}",
      journal = {\pasp},
         year = 2019,
        month = aug,
       volume = {131},
       number = {1002},
        pages = {084503},
          doi = {10.1088/1538-3873/ab215d},
archivePrefix = {arXiv},
       eprint = {1903.07629},
 primaryClass = {astro-ph.IM},
       adsurl = {https://ui.adsabs.harvard.edu/abs/2019PASP..131h4503P}
}

@ARTICLE{cendes21,
       author = {{Cendes}, Y. and {Alexander}, K.~D. and {Berger}, E. and {Eftekhari}, T. and {Williams}, P.~K.~G. and {Chornock}, R.},
        title = "{Radio Observations of an Ordinary Outflow from the Tidal Disruption Event AT2019dsg}",
      journal = {arXiv e-prints},
         year = 2021,
        month = mar,
          eid = {arXiv:2103.06299},
        pages = {arXiv:2103.06299},
archivePrefix = {arXiv},
       eprint = {2103.06299},
 primaryClass = {astro-ph.HE},
       adsurl = {https://ui.adsabs.harvard.edu/abs/2021arXiv210306299C}
}

@ARTICLE{holland13,
       author = {{Holland}, W.~S. and {Bintley}, D. and {Chapin}, E.~L. and {Chrysostomou}, A. and {Davis}, G.~R. and {Dempsey}, J.~T. and {Duncan}, W.~D. and {Fich}, M. and {Friberg}, P. and {Halpern}, M. and {Irwin}, K.~D. and {Jenness}, T. and {Kelly}, B.~D. and {MacIntosh}, M.~J. and {Robson}, E.~I. and {Scott}, D. and {Ade}, P.~A.~R. and {Atad-Ettedgui}, E. and {Berry}, D.~S. and {Craig}, S.~C. and {Gao}, X. and {Gibb}, A.~G. and {Hilton}, G.~C. and {Hollister}, M.~I. and {Kycia}, J.~B. and {Lunney}, D.~W. and {McGregor}, H. and {Montgomery}, D. and {Parkes}, W. and {Tilanus}, R.~P.~J. and {Ullom}, J.~N. and {Walther}, C.~A. and {Walton}, A.~J. and {Woodcraft}, A.~L. and {Amiri}, M. and {Atkinson}, D. and {Burger}, B. and {Chuter}, T. and {Coulson}, I.~M. and {Doriese}, W.~B. and {Dunare}, C. and {Economou}, F. and {Niemack}, M.~D. and {Parsons}, H.~A.~L. and {Reintsema}, C.~D. and {Sibthorpe}, B. and {Smail}, I. and {Sudiwala}, R. and {Thomas}, H.~S.},
        title = "{SCUBA-2: the 10 000 pixel bolometer camera on the James Clerk Maxwell Telescope}",
      journal = {\mnras},
         year = 2013,
        month = apr,
       volume = {430},
       number = {4},
        pages = {2513-2533},
          doi = {10.1093/mnras/sts612},
archivePrefix = {arXiv},
       eprint = {1301.3650},
 primaryClass = {astro-ph.IM},
       adsurl = {https://ui.adsabs.harvard.edu/abs/2013MNRAS.430.2513H}
}

@ARTICLE{vandenberk01,
       author = {{Vanden Berk}, Daniel E. and {Richards}, Gordon T. and {Bauer}, Amanda and {Strauss}, Michael A. and {Schneider}, Donald P. and {Heckman}, Timothy M. and {York}, Donald G. and {Hall}, Patrick B. and {Fan}, Xiaohui and {Knapp}, G.~R. and {Anderson}, Scott F. and {Annis}, James and {Bahcall}, Neta A. and {Bernardi}, Mariangela and {Briggs}, John W. and {Brinkmann}, J. and {Brunner}, Robert and {Burles}, Scott and {Carey}, Larry and {Castander}, Francisco J. and {Connolly}, A.~J. and {Crocker}, J.~H. and {Csabai}, Istv{\'a}n and {Doi}, Mamoru and {Finkbeiner}, Douglas and {Friedman}, Scott and {Frieman}, Joshua A. and {Fukugita}, Masataka and {Gunn}, James E. and {Hennessy}, G.~S. and {Ivezi{\'c}}, {\v{Z}}eljko and {Kent}, Stephen and {Kunszt}, Peter Z. and {Lamb}, D.~Q. and {Leger}, R. French and {Long}, Daniel C. and {Loveday}, Jon and {Lupton}, Robert H. and {Meiksin}, Avery and {Merelli}, Aronne and {Munn}, Jeffrey A. and {Newberg}, Heidi Jo and {Newcomb}, Matt and {Nichol}, R.~C. and {Owen}, Russell and {Pier}, Jeffrey R. and {Pope}, Adrian and {Rockosi}, Constance M. and {Schlegel}, David J. and {Siegmund}, Walter A. and {Smee}, Stephen and {Snir}, Yehuda and {Stoughton}, Chris and {Stubbs}, Christopher and {SubbaRao}, Mark and {Szalay}, Alexander S. and {Szokoly}, Gyula P. and {Tremonti}, Christy and {Uomoto}, Alan and {Waddell}, Patrick and {Yanny}, Brian and {Zheng}, Wei},
        title = "{Composite Quasar Spectra from the Sloan Digital Sky Survey}",
      journal = {\aj},
         year = 2001,
        month = aug,
       volume = {122},
       number = {2},
        pages = {549-564},
          doi = {10.1086/321167},
archivePrefix = {arXiv},
       eprint = {astro-ph/0105231},
 primaryClass = {astro-ph},
       adsurl = {https://ui.adsabs.harvard.edu/abs/2001AJ....122..549V}
}

@ARTICLE{alexander20,
       author = {{Alexander}, Kate D. and {van Velzen}, Sjoert and {Horesh}, Assaf and {Zauderer}, B. Ashley},
        title = "{Radio Properties of Tidal Disruption Events}",
      journal = {\ssr},
         year = 2020,
        month = jun,
       volume = {216},
       number = {5},
          eid = {81},
        pages = {81},
          doi = {10.1007/s11214-020-00702-w},
archivePrefix = {arXiv},
       eprint = {2006.01159},
 primaryClass = {astro-ph.HE},
       adsurl = {https://ui.adsabs.harvard.edu/abs/2020SSRv..216...81A}
}

@ARTICLE{frederick19,
       author = {{Frederick}, Sara and {Gezari}, Suvi and {Graham}, Matthew J. and {Cenko}, S. Bradley and {van Velzen}, Sjoert and {Stern}, Daniel and {Blagorodnova}, Nadejda and {Kulkarni}, Shrinivas R. and {Yan}, Lin and {De}, Kishalay and {Fremling}, U. Christoffer and {Hung}, Tiara and {Kara}, Erin and {Shupe}, David L. and {Ward}, Charlotte and {Bellm}, Eric C. and {Dekany}, Richard and {Duev}, Dmitry A. and {Feindt}, Ulrich and {Giomi}, Matteo and {Kupfer}, Thomas and {Laher}, Russ R. and {Masci}, Frank J. and {Miller}, Adam A. and {Neill}, James D. and {Ngeow}, Chow-Choong and {Patterson}, Maria T. and {Porter}, Michael and {Rusholme}, Ben and {Sollerman}, Jesper and {Walters}, Richard},
        title = "{A New Class of Changing-look LINERs}",
      journal = {\apj},
         year = 2019,
        month = sep,
       volume = {883},
       number = {1},
          eid = {31},
        pages = {31},
          doi = {10.3847/1538-4357/ab3a38},
archivePrefix = {arXiv},
       eprint = {1904.10973},
 primaryClass = {astro-ph.HE},
       adsurl = {https://ui.adsabs.harvard.edu/abs/2019ApJ...883...31F}
}

@ARTICLE{frederick20,
       author = {{Frederick}, Sara and {Gezari}, Suvi and {Graham}, Matthew J. and {Sollerman}, Jesper and {van Velzen}, Sjoert and {Perley}, Daniel A. and {Stern}, Daniel and {Ward}, Charlotte and {Hammerstein}, Erica and {Hung}, Tiara and {Yan}, Lin and {Andreoni}, Igor and {Bellm}, Eric C. and {Duev}, Dmitry A. and {Kowalski}, Marek and {Mahabal}, Ashish A. and {Masci}, Frank J. and {Medford}, Michael and {Rusholme}, Ben and {Walters}, Richard},
        title = "{A Family Tree of Optical Transients from Narrow-Line Seyfert 1 Galaxies}",
      journal = {arXiv e-prints},
         year = 2020,
        month = oct,
          eid = {arXiv:2010.08554},
        pages = {arXiv:2010.08554},
archivePrefix = {arXiv},
       eprint = {2010.08554},
 primaryClass = {astro-ph.HE},
       adsurl = {https://ui.adsabs.harvard.edu/abs/2020arXiv201008554F}
}

@ARTICLE{macleod16,
       author = {{MacLeod}, Chelsea L. and {Ross}, Nicholas P. and {Lawrence}, Andy and {Goad}, Mike and {Horne}, Keith and {Burgett}, William and {Chambers}, Ken C. and {Flewelling}, Heather and {Hodapp}, Klaus and {Kaiser}, Nick and {Magnier}, Eugene and {Wainscoat}, Richard and {Waters}, Christopher},
        title = "{A systematic search for changing-look quasars in SDSS}",
      journal = {\mnras},
         year = 2016,
        month = mar,
       volume = {457},
       number = {1},
        pages = {389-404},
          doi = {10.1093/mnras/stv2997},
archivePrefix = {arXiv},
       eprint = {1509.08393},
 primaryClass = {astro-ph.GA},
       adsurl = {https://ui.adsabs.harvard.edu/abs/2016MNRAS.457..389M}
}

@ARTICLE{bianchi05,
       author = {{Bianchi}, S. and {Guainazzi}, M. and {Matt}, G. and {Chiaberge}, M. and {Iwasawa}, K. and {Fiore}, F. and {Maiolino}, R.},
        title = "{A search for changing-look AGN in the Grossan catalog}",
      journal = {\aap},
         year = 2005,
        month = oct,
       volume = {442},
       number = {1},
        pages = {185-194},
          doi = {10.1051/0004-6361:20053389},
archivePrefix = {arXiv},
       eprint = {astro-ph/0507323},
 primaryClass = {astro-ph},
       adsurl = {https://ui.adsabs.harvard.edu/abs/2005A&A...442..185B}
}

@ARTICLE{hinkle21b,
       author = {{Hinkle}, Jason T. and {Holoien}, Thomas W. -S. and {Shappee}, Benjamin. J. and {Auchettl}, Katie},
        title = "{A Swift Fix for Nuclear Outbursts}",
      journal = {\apj},
         year = 2021,
        month = apr,
       volume = {910},
       number = {2},
          eid = {83},
        pages = {83},
          doi = {10.3847/1538-4357/abe4d8},
archivePrefix = {arXiv},
       eprint = {2012.08521},
 primaryClass = {astro-ph.HE},
       adsurl = {https://ui.adsabs.harvard.edu/abs/2021ApJ...910...83H}
}

@ARTICLE{trakhtenbrot19b,
       author = {{Trakhtenbrot}, Benny and {Arcavi}, Iair and {Ricci}, Claudio and {Tacchella}, Sandro and {Stern}, Daniel and {Netzer}, Hagai and {Jonker}, Peter G. and {Horesh}, Assaf and {Mej{\'\i}a-Restrepo}, Juli{\'a}n Esteban and {Hosseinzadeh}, Griffin and {Hallefors}, Valentina and {Howell}, D. Andrew and {McCully}, Curtis and {Balokovi{\'c}}, Mislav and {Heida}, Marianne and {Kamraj}, Nikita and {Lansbury}, George Benjamin and {Wyrzykowski}, {\L}ukasz and {Gromadzki}, Mariusz and {Hamanowicz}, Aleksandra and {Cenko}, S. Bradley and {Sand}, David J. and {Hsiao}, Eric Y. and {Phillips}, Mark M. and {Diamond}, Tiara R. and {Kara}, Erin and {Gendreau}, Keith C. and {Arzoumanian}, Zaven and {Remillard}, Ron},
        title = "{A new class of flares from accreting supermassive black holes}",
      journal = {Nature Astronomy},
         year = 2019,
        month = jan,
       volume = {3},
        pages = {242-250},
          doi = {10.1038/s41550-018-0661-3},
archivePrefix = {arXiv},
       eprint = {1901.03731},
 primaryClass = {astro-ph.GA},
       adsurl = {https://ui.adsabs.harvard.edu/abs/2019NatAs...3..242T}
}

@ARTICLE{Gottlieb22,
       author = {{Gottlieb}, Ore and {Tchekhovskoy}, Alexander and {Margutti}, Raffaella},
        title = "{Shocked jets in CCSNe can power the zoo of fast blue optical transients}",
      journal = {\mnras},
         year = 2022,
        month = jul,
       volume = {513},
       number = {3},
        pages = {3810-3817},
          doi = {10.1093/mnras/stac910},
archivePrefix = {arXiv},
       eprint = {2201.04636},
 primaryClass = {astro-ph.HE},
       adsurl = {https://ui.adsabs.harvard.edu/abs/2022MNRAS.513.3810G}
}

@ARTICLE{Liu23,
       author = {{Liu}, Jian-Feng and {Liu}, Liang-Duan and {Yu}, Yun-Wei and {Zhu}, Jin-Ping},
        title = "{A Population Study of the Radio Emission of Fast Blue Optical Transients}",
      journal = {\apj},
         year = 2023,
        month = mar,
       volume = {946},
       number = {1},
          eid = {35},
        pages = {35},
          doi = {10.3847/1538-4357/acbb04},
archivePrefix = {arXiv},
       eprint = {2301.06403},
 primaryClass = {astro-ph.HE},
       adsurl = {https://ui.adsabs.harvard.edu/abs/2023ApJ...946...35L}
}

@ARTICLE{Metger22,
       author = {{Metzger}, Brian D.},
        title = "{Luminous Fast Blue Optical Transients and Type Ibn/Icn SNe from Wolf-Rayet/Black Hole Mergers}",
      journal = {\apj},
         year = 2022,
        month = jun,
       volume = {932},
       number = {2},
          eid = {84},
        pages = {84},
          doi = {10.3847/1538-4357/ac6d59},
archivePrefix = {arXiv},
       eprint = {2203.04331},
 primaryClass = {astro-ph.HE},
       adsurl = {https://ui.adsabs.harvard.edu/abs/2022ApJ...932...84M}
}

@ARTICLE{Marguitti19,
       author = {{Margutti}, R. and {Metzger}, B.~D. and {Chornock}, R. and {Vurm}, I. and {Roth}, N. and {Grefenstette}, B.~W. and {Savchenko}, V. and {Cartier}, R. and {Steiner}, J.~F. and {Terreran}, G. and {Margalit}, B. and {Migliori}, G. and {Milisavljevic}, D. and {Alexander}, K.~D. and {Bietenholz}, M. and {Blanchard}, P.~K. and {Bozzo}, E. and {Brethauer}, D. and {Chilingarian}, I.~V. and {Coppejans}, D.~L. and {Ducci}, L. and {Ferrigno}, C. and {Fong}, W. and {G{\"o}tz}, D. and {Guidorzi}, C. and {Hajela}, A. and {Hurley}, K. and {Kuulkers}, E. and {Laurent}, P. and {Mereghetti}, S. and {Nicholl}, M. and {Patnaude}, D. and {Ubertini}, P. and {Banovetz}, J. and {Bartel}, N. and {Berger}, E. and {Coughlin}, E.~R. and {Eftekhari}, T. and {Frederiks}, D.~D. and {Kozlova}, A.~V. and {Laskar}, T. and {Svinkin}, D.~S. and {Drout}, M.~R. and {MacFadyen}, A. and {Paterson}, K.},
        title = "{An Embedded X-Ray Source Shines through the Aspherical AT 2018cow: Revealing the Inner Workings of the Most Luminous Fast-evolving Optical Transients}",
      journal = {\apj},
         year = 2019,
        month = feb,
       volume = {872},
       number = {1},
          eid = {18},
        pages = {18},
          doi = {10.3847/1538-4357/aafa01},
archivePrefix = {arXiv},
       eprint = {1810.10720},
 primaryClass = {astro-ph.HE},
       adsurl = {https://ui.adsabs.harvard.edu/abs/2019ApJ...872...18M}
}

@ARTICLE{Inserra19,
       author = {{Inserra}, C.},
        title = "{Observational properties of extreme supernovae}",
      journal = {Nature Astronomy},
         year = 2019,
        month = aug,
       volume = {3},
        pages = {697-705},
          doi = {10.1038/s41550-019-0854-4},
archivePrefix = {arXiv},
       eprint = {1908.02314},
 primaryClass = {astro-ph.HE},
       adsurl = {https://ui.adsabs.harvard.edu/abs/2019NatAs...3..697I}
}

@ARTICLE{Angus22,
       author = {{Angus}, C.~R. and {Baldassare}, V.~F. and {Mockler}, B. and {Foley}, R.~J. and {Ramirez-Ruiz}, E. and {Raimundo}, S.~I. and {French}, K.~D. and {Auchettl}, K. and {Pfister}, H. and {Gall}, C. and {Hjorth}, J. and {Drout}, M.~R. and {Alexander}, K.~D. and {Dimitriadis}, G. and {Hung}, T. and {Jones}, D.~O. and {Rest}, A. and {Siebert}, M.~R. and {Taggart}, K. and {Terreran}, G. and {Tinyanont}, S. and {Carroll}, C.~M. and {DeMarchi}, L. and {Earl}, N. and {Gagliano}, A. and {Izzo}, L. and {Villar}, V.~A. and {Zenati}, Y. and {Arendse}, N. and {Cold}, C. and {de Boer}, T.~J.~L. and {Chambers}, K.~C. and {Coulter}, D.~A. and {Khetan}, N. and {Lin}, C.~C. and {Magnier}, E.~A. and {Rojas-Bravo}, C. and {Wainscoat}, R.~J. and {Wojtak}, R.},
        title = "{A fast-rising tidal disruption event from a candidate intermediate-mass black hole}",
      journal = {Nature Astronomy},
         year = 2022,
        month = dec,
       volume = {6},
        pages = {1452-1463},
          doi = {10.1038/s41550-022-01811-y},
archivePrefix = {arXiv},
       eprint = {2209.00018},
 primaryClass = {astro-ph.HE},
       adsurl = {https://ui.adsabs.harvard.edu/abs/2022NatAs...6.1452A}
}

@ARTICLE{trakhtenbrot19a,
       author = {{Trakhtenbrot}, Benny and {Arcavi}, Iair and {MacLeod}, Chelsea L. and {Ricci}, Claudio and {Kara}, Erin and {Graham}, Melissa L. and {Stern}, Daniel and {Harrison}, Fiona A. and {Burke}, Jamison and {Hiramatsu}, Daichi and {Hosseinzadeh}, Griffin and {Howell}, D. Andrew and {Smartt}, Stephen J. and {Rest}, Armin and {Prieto}, Jose L. and {Shappee}, Benjamin J. and {Holoien}, Thomas W. -S. and {Bersier}, David and {Filippenko}, Alexei V. and {Brink}, Thomas G. and {Zheng}, WeiKang and {Li}, Ruancun and {Remillard}, Ronald A. and {Loewenstein}, Michael},
        title = "{1ES 1927+654: An AGN Caught Changing Look on a Timescale of Months}",
      journal = {\apj},
         year = 2019,
        month = sep,
       volume = {883},
       number = {1},
          eid = {94},
        pages = {94},
          doi = {10.3847/1538-4357/ab39e4},
archivePrefix = {arXiv},
       eprint = {1903.11084},
 primaryClass = {astro-ph.GA},
       adsurl = {https://ui.adsabs.harvard.edu/abs/2019ApJ...883...94T}
}

@ARTICLE{edelson15,
       author = {{Edelson}, R. and {Gelbord}, J.~M. and {Horne}, K. and {McHardy}, I.~M. and {Peterson}, B.~M. and {Ar{\'e}valo}, P. and {Breeveld}, A.~A. and {De Rosa}, G. and {Evans}, P.~A. and {Goad}, M.~R. and {Kriss}, G.~A. and {Brandt}, W.~N. and {Gehrels}, N. and {Grupe}, D. and {Kennea}, J.~A. and {Kochanek}, C.~S. and {Nousek}, J.~A. and {Papadakis}, I. and {Siegel}, M. and {Starkey}, D. and {Uttley}, P. and {Vaughan}, S. and {Young}, S. and {Barth}, A.~J. and {Bentz}, M.~C. and {Brewer}, B.~J. and {Crenshaw}, D.~M. and {Dalla Bont{\`a}}, E. and {De Lorenzo-C{\'a}ceres}, A. and {Denney}, K.~D. and {Dietrich}, M. and {Ely}, J. and {Fausnaugh}, M.~M. and {Grier}, C.~J. and {Hall}, P.~B. and {Kaastra}, J. and {Kelly}, B.~C. and {Korista}, K.~T. and {Lira}, P. and {Mathur}, S. and {Netzer}, H. and {Pancoast}, A. and {Pei}, L. and {Pogge}, R.~W. and {Schimoia}, J.~S. and {Treu}, T. and {Vestergaard}, M. and {Villforth}, C. and {Yan}, H. and {Zu}, Y.},
        title = "{Space Telescope and Optical Reverberation Mapping Project. II. Swift and HST Reverberation Mapping of the Accretion Disk of NGC 5548}",
      journal = {\apj},
         year = 2015,
        month = jun,
       volume = {806},
       number = {1},
          eid = {129},
        pages = {129},
          doi = {10.1088/0004-637X/806/1/129},
archivePrefix = {arXiv},
       eprint = {1501.05951},
 primaryClass = {astro-ph.GA},
       adsurl = {https://ui.adsabs.harvard.edu/abs/2015ApJ...806..129E}
}

@ARTICLE{ahumada20,
       author = {{Ahumada}, Romina and {Allende Prieto}, Carlos and {Almeida}, Andr{\'e}s and {Anders}, Friedrich and {Anderson}, Scott F. and {Andrews}, Brett H. and {Anguiano}, Borja and {Arcodia}, Riccardo and {Armengaud}, Eric and {Aubert}, Marie and {Avila}, Santiago and {Avila-Reese}, Vladimir and {Badenes}, Carles and {Balland}, Christophe and {Barger}, Kat and {Barrera-Ballesteros}, Jorge K. and {Basu}, Sarbani and {Bautista}, Julian and {Beaton}, Rachael L. and {Beers}, Timothy C. and {Benavides}, B. Izamar T. and {Bender}, Chad F. and {Bernardi}, Mariangela and {Bershady}, Matthew and {Beutler}, Florian and {Bidin}, Christian Moni and {Bird}, Jonathan and {Bizyaev}, Dmitry and {Blanc}, Guillermo A. and {Blanton}, Michael R. and {Boquien}, M{\'e}d{\'e}ric and {Borissova}, Jura and {Bovy}, Jo and {Brandt}, W.~N. and {Brinkmann}, Jonathan and {Brownstein}, Joel R. and {Bundy}, Kevin and {Bureau}, Martin and {Burgasser}, Adam and {Burtin}, Etienne and {Cano-D{\'\i}az}, Mariana and {Capasso}, Raffaella and {Cappellari}, Michele and {Carrera}, Ricardo and {Chabanier}, Sol{\`e}ne and {Chaplin}, William and {Chapman}, Michael and {Cherinka}, Brian and {Chiappini}, Cristina and {Doohyun Choi}, Peter and {Chojnowski}, S. Drew and {Chung}, Haeun and {Clerc}, Nicolas and {Coffey}, Damien and {Comerford}, Julia M. and {Comparat}, Johan and {da Costa}, Luiz and {Cousinou}, Marie-Claude and {Covey}, Kevin and {Crane}, Jeffrey D. and {Cunha}, Katia and {da Silva Ilha}, Gabriele and {Dai}, Yu Sophia and {Damsted}, Sanna B. and {Darling}, Jeremy and {Davidson}, James W., Jr. and {Davies}, Roger and {Dawson}, Kyle and {De}, Nikhil and {de la Macorra}, Axel and {De Lee}, Nathan and {de Andrade Queiroz}, Anna B{\'a}rbara and {Deconto Machado}, Alice and {de la Torre}, Sylvain and {Dell'Agli}, Flavia and {du Mas des Bourboux}, H{\'e}lion and {Diamond-Stanic}, Aleksandar M. and {Dillon}, Sean and {Donor}, John and {Drory}, Niv and {Duckworth}, Chris and {Dwelly}, Tom and {Ebelke}, Garrett and {Eftekharzadeh}, Sarah and {Eigenbrot}, Arthur Davis and {Elsworth}, Yvonne P. and {Eracleous}, Mike and {Erfanianfar}, Ghazaleh and {Escoffier}, Stephanie and {Fan}, Xiaohui and {Farr}, Emily and {Fern{\'a}ndez-Trincado}, Jos{\'e} G. and {Feuillet}, Diane and {Finoguenov}, Alexis and {Fofie}, Patricia and {Fraser-McKelvie}, Amelia and {Frinchaboy}, Peter M. and {Fromenteau}, Sebastien and {Fu}, Hai and {Galbany}, Llu{\'\i}s and {Garcia}, Rafael A. and {Garc{\'\i}a-Hern{\'a}ndez}, D.~A. and {Garma Oehmichen}, Luis Alberto and {Ge}, Junqiang and {Geimba Maia}, Marcio Antonio and {Geisler}, Doug and {Gelfand}, Joseph and {Goddy}, Julian and {Gonzalez-Perez}, Violeta and {Grabowski}, Kathleen and {Green}, Paul and {Grier}, Catherine J. and {Guo}, Hong and {Guy}, Julien and {Harding}, Paul and {Hasselquist}, Sten and {Hawken}, Adam James and {Hayes}, Christian R. and {Hearty}, Fred and {Hekker}, S. and {Hogg}, David W. and {Holtzman}, Jon A. and {Horta}, Danny and {Hou}, Jiamin and {Hsieh}, Bau-Ching and {Huber}, Daniel and {Hunt}, Jason A.~S. and {Ider Chitham}, J. and {Imig}, Julie and {Jaber}, Mariana and {Jimenez Angel}, Camilo Eduardo and {Johnson}, Jennifer A. and {Jones}, Amy M. and {J{\"o}nsson}, Henrik and {Jullo}, Eric and {Kim}, Yerim and {Kinemuchi}, Karen and {Kirkpatrick}, Charles C., IV and {Kite}, George W. and {Klaene}, Mark and {Kneib}, Jean-Paul and {Kollmeier}, Juna A. and {Kong}, Hui and {Kounkel}, Marina and {Krishnarao}, Dhanesh and {Lacerna}, Ivan and {Lan}, Ting-Wen and {Lane}, Richard R. and {Law}, David R. and {Le Goff}, Jean-Marc and {Leung}, Henry W. and {Lewis}, Hannah and {Li}, Cheng and {Lian}, Jianhui and {Lin}, Lihwai and {Long}, Dan and {Longa-Pe{\~n}a}, Pen{\'e}lope and {Lundgren}, Britt and {Lyke}, Brad W. and {Ted Mackereth}, J. and {MacLeod}, Chelsea L. and {Majewski}, Steven R. and {Manchado}, Arturo and {Maraston}, Claudia and {Martini}, Paul and {Masseron}, Thomas and {Masters}, Karen L. and {Mathur}, Savita and {McDermid}, Richard M. and {Merloni}, Andrea and {Merrifield}, Michael and {M{\'e}sz{\'a}ros}, Szabolcs and {Miglio}, Andrea and {Minniti}, Dante and {Minsley}, Rebecca and {Miyaji}, Takamitsu and {Mohammad}, Faizan Gohar and {Mosser}, Benoit and {Mueller}, Eva-Maria and {Muna}, Demitri and {Mu{\~n}oz-Guti{\'e}rrez}, Andrea and {Myers}, Adam D. and {Nadathur}, Seshadri and {Nair}, Preethi and {Nandra}, Kirpal and {do Nascimento}, Janaina Correa and {Nevin}, Rebecca Jean and {Newman}, Jeffrey A. and {Nidever}, David L. and {Nitschelm}, Christian and {Noterdaeme}, Pasquier and {O'Connell}, Julia E. and {Olmstead}, Matthew D. and {Oravetz}, Daniel and {Oravetz}, Audrey and {Osorio}, Yeisson and {Pace}, Zachary J. and {Padilla}, Nelson and {Palanque-Delabrouille}, Nathalie and {Palicio}, Pedro A. and {Pan}, Hsi-An and {Pan}, Kaike and {Parker}, James and {Paviot}, Romain and {Peirani}, Sebastien and {Pe{\~n}a Ram{\'r}ez}, Karla and {Penny}, Samantha and {Percival}, Will J. and {Perez-Fournon}, Ismael and {P{\'e}rez-R{\`a}fols}, Ignasi and {Petitjean}, Patrick and {Pieri}, Matthew M. and {Pinsonneault}, Marc and {Poovelil}, Vijith Jacob and {Povick}, Joshua Tyler and {Prakash}, Abhishek and {Price-Whelan}, Adrian M. and {Raddick}, M. Jordan and {Raichoor}, Anand and {Ray}, Amy and {Rembold}, Sandro Barboza and {Rezaie}, Mehdi and {Riffel}, Rogemar A. and {Riffel}, Rog{\'e}rio and {Rix}, Hans-Walter and {Robin}, Annie C. and {Roman-Lopes}, A. and {Rom{\'a}n-Z{\'u}{\~n}iga}, Carlos and {Rose}, Benjamin and {Ross}, Ashley J. and {Rossi}, Graziano and {Rowlands}, Kate and {Rubin}, Kate H.~R. and {Salvato}, Mara and {S{\'a}nchez}, Ariel G. and {S{\'a}nchez-Menguiano}, Laura and {S{\'a}nchez-Gallego}, Jos{\'e} R. and {Sayres}, Conor and {Schaefer}, Adam and {Schiavon}, Ricardo P. and {Schimoia}, Jaderson S. and {Schlafly}, Edward and {Schlegel}, David and {Schneider}, Donald P. and {Schultheis}, Mathias and {Schwope}, Axel and {Seo}, Hee-Jong and {Serenelli}, Aldo and {Shafieloo}, Arman and {Shamsi}, Shoaib Jamal and {Shao}, Zhengyi and {Shen}, Shiyin and {Shetrone}, Matthew and {Shirley}, Raphael and {Silva Aguirre}, V{\'\i}ctor and {Simon}, Joshua D. and {Skrutskie}, M.~F. and {Slosar}, An{\v{z}}e and {Smethurst}, Rebecca and {Sobeck}, Jennifer and {Sodi}, Bernardo Cervantes and {Souto}, Diogo and {Stark}, David V. and {Stassun}, Keivan G. and {Steinmetz}, Matthias and {Stello}, Dennis and {Stermer}, Julianna and {Storchi-Bergmann}, Thaisa and {Streblyanska}, Alina and {Stringfellow}, Guy S. and {Stutz}, Amelia and {Su{\'a}rez}, Genaro and {Sun}, Jing and {Taghizadeh-Popp}, Manuchehr and {Talbot}, Michael S. and {Tayar}, Jamie and {Thakar}, Aniruddha R. and {Theriault}, Riley and {Thomas}, Daniel and {Thomas}, Zak C. and {Tinker}, Jeremy and {Tojeiro}, Rita and {Toledo}, Hector Hernandez and {Tremonti}, Christy A. and {Troup}, Nicholas W. and {Tuttle}, Sarah and {Unda-Sanzana}, Eduardo and {Valentini}, Marica and {Vargas-Gonz{\'a}lez}, Jaime and {Vargas-Maga{\~n}a}, Mariana and {V{\'a}zquez-Mata}, Jose Antonio and {Vivek}, M. and {Wake}, David and {Wang}, Yuting and {Weaver}, Benjamin Alan and {Weijmans}, Anne-Marie and {Wild}, Vivienne and {Wilson}, John C. and {Wilson}, Robert F. and {Wolthuis}, Nathan and {Wood-Vasey}, W.~M. and {Yan}, Renbin and {Yang}, Meng and {Y{\`e}che}, Christophe and {Zamora}, Olga and {Zarrouk}, Pauline and {Zasowski}, Gail and {Zhang}, Kai and {Zhao}, Cheng and {Zhao}, Gongbo and {Zheng}, Zheng and {Zheng}, Zheng and {Zhu}, Guangtun and {Zou}, Hu},
        title = "{The 16th Data Release of the Sloan Digital Sky Surveys: First Release from the APOGEE-2 Southern Survey and Full Release of eBOSS Spectra}",
      journal = {\apjs},
         year = 2020,
        month = jul,
       volume = {249},
       number = {1},
          eid = {3},
        pages = {3},
          doi = {10.3847/1538-4365/ab929e},
archivePrefix = {arXiv},
       eprint = {1912.02905},
 primaryClass = {astro-ph.GA},
       adsurl = {https://ui.adsabs.harvard.edu/abs/2020ApJS..249....3A}
}

@ARTICLE{vanvelzen20b,
       author = {{van Velzen}, Sjoert and {Holoien}, Thomas W. -S. and {Onori}, Francesca and {Hung}, Tiara and {Arcavi}, Iair},
        title = "{Optical-Ultraviolet Tidal Disruption Events}",
      journal = {\ssr},
         year = 2020,
        month = oct,
       volume = {216},
       number = {8},
          eid = {124},
        pages = {124},
          doi = {10.1007/s11214-020-00753-z},
archivePrefix = {arXiv},
       eprint = {2008.05461},
 primaryClass = {astro-ph.HE},
       adsurl = {https://ui.adsabs.harvard.edu/abs/2020SSRv..216..124V}
}

@ARTICLE{ricci20,
       author = {{Ricci}, C. and {Kara}, E. and {Loewenstein}, M. and {Trakhtenbrot}, B. and {Arcavi}, I. and {Remillard}, R. and {Fabian}, A.~C. and {Gendreau}, K.~C. and {Arzoumanian}, Z. and {Li}, R. and {Ho}, L.~C. and {MacLeod}, C.~L. and {Cackett}, E. and {Altamirano}, D. and {Gandhi}, P. and {Kosec}, P. and {Pasham}, D. and {Steiner}, J. and {Chan}, C. -H.},
        title = "{The Destruction and Recreation of the X-Ray Corona in a Changing-look Active Galactic Nucleus}",
      journal = {\apjl},
         year = 2020,
        month = jul,
       volume = {898},
       number = {1},
          eid = {L1},
        pages = {L1},
          doi = {10.3847/2041-8213/ab91a1},
archivePrefix = {arXiv},
       eprint = {2007.07275},
 primaryClass = {astro-ph.HE},
       adsurl = {https://ui.adsabs.harvard.edu/abs/2020ApJ...898L...1R}
}

@ARTICLE{hinkle21a,
       author = {{Hinkle}, Jason T. and {Holoien}, T.~W. -S. and {Auchettl}, K. and {Shappee}, B.~J. and {Neustadt}, J.~M.~M. and {Payne}, A.~V. and {Brown}, J.~S. and {Kochanek}, C.~S. and {Stanek}, K.~Z. and {Graham}, M.~J. and {Tucker}, M.~A. and {Do}, A. and {Anderson}, J.~P. and {Bose}, S. and {Chen}, P. and {Coulter}, D.~A. and {Dimitriadis}, G. and {Dong}, Subo and {Foley}, R.~J. and {Huber}, M.~E. and {Hung}, T. and {Kilpatrick}, C.~D. and {Pignata}, G. and {Piro}, A.~L. and {Rojas-Bravo}, C. and {Siebert}, M.~R. and {Stalder}, B. and {Thompson}, Todd A. and {Tonry}, J.~L. and {Vallely}, P.~J. and {Wisniewski}, J.~P.},
        title = "{Discovery and follow-up of ASASSN-19dj: an X-ray and UV luminous TDE in an extreme post-starburst galaxy}",
      journal = {\mnras},
         year = 2021,
        month = jan,
       volume = {500},
       number = {2},
        pages = {1673-1696},
          doi = {10.1093/mnras/staa3170},
archivePrefix = {arXiv},
       eprint = {2006.06690},
 primaryClass = {astro-ph.HE},
       adsurl = {https://ui.adsabs.harvard.edu/abs/2021MNRAS.500.1673H}
}

@ARTICLE{nicholl20,
       author = {{Nicholl}, M. and {Wevers}, T. and {Oates}, S.~R. and {Alexand
        er}, K.~D. and {Leloudas}, G. and {Onori}, F. and {Jerkstrand}, A. and
         {Gomez}, S. and {Campana}, S. and {Arcavi}, I. and
         {Charalampopoulos}, P. and {Gromadzki}, M. and {Ihanec}, N. and
         {Jonker}, P.~G. and {Lawrence}, A. and {Mandel}, I. and {Short}, P. and
         {Burke}, J. and {Hiramatsu}, D. and {Howell}, D.~A. and
         {Pellegrino}, C. and {Abbot}, H. and {Anderson}, J.~P. and
         {Berger}, E. and {Blanchard}, P.~K. and {Cannizzaro}, G. and
         {Chen}, T. -W. and {Dennefeld}, M. and {Galbany}, L. and
         {Gonzalez-Gaitan}, S. and {Hosseinzadeh}, G. and {Inserra}, C. and
         {Irani}, I. and {Kuin}, P. and {Muller-Bravo}, T. and {Pineda}, J. and
         {Ross}, N.~P. and {Roy}, R. and {Tucker}, B. and {Wyrzykowski}, L. and
         {Young}, D.~R.},
        title = "{An outflow powers the optical rise of the nearby, fast-evolving tidal disruption event AT2019qiz}",
      journal = {arXiv e-prints},
         year = 2020,
        month = jun,
          eid = {arXiv:2006.02454},
        pages = {arXiv:2006.02454},
archivePrefix = {arXiv},
       eprint = {2006.02454},
 primaryClass = {astro-ph.HE},
       adsurl = {https://ui.adsabs.harvard.edu/abs/2020arXiv200602454N}
}

@ARTICLE{french20,
       author = {{French}, K. Decker and {Arcavi}, Iair and {Zabludoff}, Ann I. and
         {Stone}, Nicholas and {Hiramatsu}, Daichi and {van Velzen}, Sjoert and
         {McCully}, Curtis and {Jiang}, Ning},
        title = "{The Structure of Tidal Disruption Event Host Galaxies on Scales of Tens to Thousands of Parsecs}",
      journal = {\apj},
         year = 2020,
        month = mar,
       volume = {891},
       number = {1},
          eid = {93},
        pages = {93},
          doi = {10.3847/1538-4357/ab7450},
archivePrefix = {arXiv},
       eprint = {2002.02498},
 primaryClass = {astro-ph.HE},
       adsurl = {https://ui.adsabs.harvard.edu/abs/2020ApJ...891...93F}
}

@ARTICLE{holoien20,
       author = {{Holoien}, Thomas W. -S. and {Auchettl}, Katie and {Tucker}, Michael A. and
         {Shappee}, Benjamin J. and {Patel}, Shannon G. and
         {Miller-Jones}, James C.~A. and {Mockler}, Brenna and
         {Groenewald}, Dani{\`e}l N. and {Brown}, Jonathan S. and
         {Kochanek}, Christopher S. and {Stanek}, K.~Z. and {Chen}, Ping and
         {Dong}, Subo and {Prieto}, Jose L. and {Thompson}, Todd A. and
         {Beaton}, Rachael L. and {Connor}, Thomas and
         {Cowperthwaite}, Philip S. and {Dahmen}, Linnea and
         {French}, K. Decker and {Morrell}, Nidia and {Buckley}, David A.~H. and
         {Gromadzki}, Mariusz and {Roy}, Rupak and {Coulter}, David A. and
         {Dimitriadis}, Georgios and {Foley}, Ryan J. and
         {Kilpatrick}, Charles D. and {Piro}, Anthony L. and
         {Rojas-Bravo}, C{\'e}sar and {Siebert}, Matthew R. and
         {van Velzen}, Sjoert},
        title = "{The Rise and Fall of ASASSN-18pg: Following a TDE from Early To Late Times}",
      journal = {arXiv e-prints},
         year = 2020,
        month = mar,
          eid = {arXiv:2003.13693},
        pages = {arXiv:2003.13693},
archivePrefix = {arXiv},
       eprint = {2003.13693},
 primaryClass = {astro-ph.HE},
       adsurl = {https://ui.adsabs.harvard.edu/abs/2020arXiv200313693H}
}

@ARTICLE{vanvelzen21,
       author = {{van Velzen}, Sjoert and {Gezari}, Suvi and {Hammerstein}, Erica and {Roth}, Nathaniel and {Frederick}, Sara and {Ward}, Charlotte and {Hung}, Tiara and {Cenko}, S. Bradley and {Stein}, Robert and {Perley}, Daniel A. and {Taggart}, Kirsty and {Foley}, Ryan J. and {Sollerman}, Jesper and {Blagorodnova}, Nadejda and {Andreoni}, Igor and {Bellm}, Eric C. and {Brinnel}, Valery and {De}, Kishalay and {Dekany}, Richard and {Feeney}, Michael and {Fremling}, Christoffer and {Giomi}, Matteo and {Golkhou}, V. Zach and {Graham}, Matthew J. and {Ho}, Anna. Y.~Q. and {Kasliwal}, Mansi M. and {Kilpatrick}, Charles D. and {Kulkarni}, Shrinivas R. and {Kupfer}, Thomas and {Laher}, Russ R. and {Mahabal}, Ashish and {Masci}, Frank J. and {Miller}, Adam A. and {Nordin}, Jakob and {Riddle}, Reed and {Rusholme}, Ben and {van Santen}, Jakob and {Sharma}, Yashvi and {Shupe}, David L. and {Soumagnac}, Maayane T.},
        title = "{Seventeen Tidal Disruption Events from the First Half of ZTF Survey Observations: Entering a New Era of Population Studies}",
      journal = {\apj},
         year = 2021,
        month = feb,
       volume = {908},
       number = {1},
          eid = {4},
        pages = {4},
          doi = {10.3847/1538-4357/abc258},
archivePrefix = {arXiv},
       eprint = {2001.01409},
 primaryClass = {astro-ph.HE},
       adsurl = {https://ui.adsabs.harvard.edu/abs/2021ApJ...908....4V}
}

@ARTICLE{hinkle20a,
       author = {{Hinkle}, Jason T. and {Holoien}, Thomas W. -S. and
         {Shappee}, Benjamin. J. and {Auchettl}, Katie and
         {Kochanek}, Christopher S. and {Stanek}, K.~Z. and {Payne}, Anna V. and
         {Thompson}, Todd A.},
        title = "{Peak-Luminosity/Decline-Rate Relationship for Tidal Disruption Events}",
      journal = {arXiv e-prints},
         year = 2020,
        month = jan,
          eid = {arXiv:2001.08215},
        pages = {arXiv:2001.08215},
archivePrefix = {arXiv},
       eprint = {2001.08215},
 primaryClass = {astro-ph.HE},
       adsurl = {https://ui.adsabs.harvard.edu/abs/2020arXiv200108215H}
}

@ARTICLE{wevers19a,
       author = {{Wevers}, Thomas and {Stone}, Nicholas C. and {van Velzen}, Sjoert and
         {Jonker}, Peter G. and {Hung}, Tiara and {Auchettl}, Katie and
         {Gezari}, Suvi and {Onori}, Francesca and {Mata S{\'a}nchez}, Daniel and
         {Kostrzewa-Rutkowska}, Zuzanna and {Casares}, Jorge},
        title = "{Black hole masses of tidal disruption event host galaxies II}",
      journal = {\mnras},
         year = 2019,
        month = aug,
       volume = {487},
       number = {3},
        pages = {4136-4152},
          doi = {10.1093/mnras/stz1602},
archivePrefix = {arXiv},
       eprint = {1902.04077},
 primaryClass = {astro-ph.HE},
       adsurl = {https://ui.adsabs.harvard.edu/abs/2019MNRAS.487.4136W}
}

@ARTICLE{HI4PI16,
       author = {{HI4PI Collaboration} and {Ben Bekhti}, N. and {Fl{\"o}er}, L. and
         {Keller}, R. and {Kerp}, J. and {Lenz}, D. and {Winkel}, B. and
         {Bailin}, J. and {Calabretta}, M.~R. and {Dedes}, L. and {Ford}, H.~A. and
         {Gibson}, B.~K. and {Haud}, U. and {Janowiecki}, S. and
         {Kalberla}, P.~M.~W. and {Lockman}, F.~J. and
         {McClure-Griffiths}, N.~M. and {Murphy}, T. and {Nakanishi}, H. and
         {Pisano}, D.~J. and {Staveley-Smith}, L.},
        title = "{HI4PI: A full-sky H I survey based on EBHIS and GASS}",
      journal = {\aap},
         year = "2016",
        month = "Oct",
       volume = {594},
          eid = {A116},
        pages = {A116},
          doi = {10.1051/0004-6361/201629178},
archivePrefix = {arXiv},
       eprint = {1610.06175},
 primaryClass = {astro-ph.GA},
       adsurl = {https://ui.adsabs.harvard.edu/abs/2016A&A...594A.116H}
}

@ARTICLE{brinchmann04,
       author = {{Brinchmann}, J. and {Charlot}, S. and {White}, S.~D.~M. and
         {Tremonti}, C. and {Kauffmann}, G. and {Heckman}, T. and
         {Brinkmann}, J.},
        title = "{The physical properties of star-forming galaxies in the low-redshift Universe}",
      journal = {\mnras},
         year = "2004",
        month = "Jul",
       volume = {351},
       number = {4},
        pages = {1151-1179},
          doi = {10.1111/j.1365-2966.2004.07881.x},
archivePrefix = {arXiv},
       eprint = {astro-ph/0311060},
 primaryClass = {astro-ph},
       adsurl = {https://ui.adsabs.harvard.edu/abs/2004MNRAS.351.1151B}
}

@ARTICLE{vanvelzen16,
       author = {{van Velzen}, S. and {Anderson}, G.~E. and {Stone}, N.~C. and
         {Fraser}, M. and {Wevers}, T. and {Metzger}, B.~D. and {Jonker}, P.~G. and
         {van der Horst}, A.~J. and {Staley}, T.~D. and {Mendez}, A.~J. and
         {Miller-Jones}, J.~C.~A. and {Hodgkin}, S.~T. and {Campbell}, H.~C. and
         {Fender}, R.~P.},
        title = "{A radio jet from the optical and x-ray bright stellar tidal disruption flare ASASSN-14li}",
      journal = {Science},
         year = "2016",
        month = "Jan",
       volume = {351},
       number = {6268},
        pages = {62-65},
          doi = {10.1126/science.aad1182},
archivePrefix = {arXiv},
       eprint = {1511.08803},
 primaryClass = {astro-ph.HE},
       adsurl = {https://ui.adsabs.harvard.edu/abs/2016Sci...351...62V}
}

@ARTICLE{Alard2000,
       author = {{Alard}, C.},
        title = "{Image subtraction using a space-varying kernel}",
      journal = {\aaps},
         year = "2000",
        month = "Jun",
       volume = {144},
        pages = {363-370},
          doi = {10.1051/aas:2000214},
       adsurl = {https://ui.adsabs.harvard.edu/abs/2000A&AS..144..363A}
}

@ARTICLE{salpeter55,
       author = {{Salpeter}, Edwin E.},
        title = "{The Luminosity Function and Stellar Evolution.}",
      journal = {\apj},
         year = "1955",
        month = "Jan",
       volume = {121},
        pages = {161},
          doi = {10.1086/145971},
       adsurl = {https://ui.adsabs.harvard.edu/abs/1955ApJ...121..161S}
}

@ARTICLE{holoien19c,
       author = {{Holoien}, Thomas W. -S. and {Vallely}, Patrick J. and
         {Auchettl}, Katie and {Stanek}, K.~Z. and {Kochanek}, Christopher S. and
         {French}, K. Decker and {Prieto}, Jose L. and {Shappee}, Benjamin J. and
         {Brown}, Jonathan S. and {Fausnaugh}, Michael M. and {Dong}, Subo and
         {Thompson}, Todd A. and {Bose}, Subhash and {Neustadt}, Jack M.~M. and
         {Cacella}, P. and {Brimacombe}, J. and {Kendurkar}, Malhar R. and
         {Beaton}, Rachael L. and {Boutsia}, Konstantina and {Chomiuk}, Laura and
         {Connor}, Thomas and {Morrell}, Nidia and {Newman}, Andrew B. and
         {Rudie}, Gwen C. and {Shishkovksy}, Laura and {Strader}, Jay},
        title = "{Discovery and Early Evolution of ASASSN-19bt, the First TDE Detected by TESS}",
      journal = {\apj},
         year = "2019",
        month = "Oct",
       volume = {883},
       number = {2},
          eid = {111},
        pages = {111},
          doi = {10.3847/1538-4357/ab3c66},
archivePrefix = {arXiv},
       eprint = {1904.09293},
 primaryClass = {astro-ph.HE},
       adsurl = {https://ui.adsabs.harvard.edu/abs/2019ApJ...883..111H}
}

@ARTICLE{holoien19b,
       author = {{Holoien}, T.~W. -S. and {Huber}, M.~E. and {Shappee}, B.~J. and
         {Eracleous}, M. and {Auchettl}, K. and {Brown}, J.~S. and
         {Tucker}, M.~A. and {Chambers}, K.~C. and {Kochanek}, C.~S. and
         {Stanek}, K.~Z. and {Rest}, A. and {Bersier}, D. and {Post}, R.~S. and
         {Aldering}, G. and {Ponder}, K.~A. and {Simon}, J.~D. and
         {Kankare}, E. and {Dong}, D. and {Hallinan}, G. and {Reddy}, N.~A. and
         {Sanders}, R.~L. and {Topping}, M.~W. and {Pan-STARRS} and
         {Bulger}, J. and {Lowe}, T.~B. and {Magnier}, E.~A. and
         {Schultz}, A.~S.~B. and {Waters}, C.~Z. and {Willman}, M. and
         {Wright}, D. and {Young}, D.~R. and {ASAS-SN} and {Dong}, Subo and
         {Prieto}, J.~L. and {Thompson}, Todd A. and {ATLAS} and {Denneau}, L. and
         {Flewelling}, H. and {Heinze}, A.~N. and {Smartt}, S.~J. and
         {Smith}, K.~W. and {Stalder}, B. and {Tonry}, J.~L. and {Weiland}, H.},
        title = "{PS18kh: A New Tidal Disruption Event with a Non-axisymmetric Accretion Disk}",
      journal = {\apj},
         year = "2019",
        month = "Aug",
       volume = {880},
       number = {2},
          eid = {120},
        pages = {120},
          doi = {10.3847/1538-4357/ab2ae1},
archivePrefix = {arXiv},
       eprint = {1808.02890},
 primaryClass = {astro-ph.HE},
       adsurl = {https://ui.adsabs.harvard.edu/abs/2019ApJ...880..120H}
}

@ARTICLE{holoien14a,
   author = {{Holoien}, T.~W.-S. and {Prieto}, J.~L. and {Stanek}, K.~Z. and 
    {Kochanek}, C.~S. and {Shappee}, B.~J. and {Zhu}, Z. and {Sicilia-Aguilar}, A. and 
    {Grupe}, D. and {Croxall}, K. and {Adams}, J.~J. and {Simon}, J.~D. and 
    {Morrell}, N. and {McGraw}, S.~M. and {Wagner}, R.~M. and {Basu}, U. and 
    {Beacom}, J.~F. and {Bersier}, D. and {Brimacombe}, J. and {Jencson}, J. and 
    {Pojmanski}, G. and {Starrfield}, S.~G. and {Szczygie{\l}}, D.~M. and 
    {Woodward}, C.~E.},
    title = "{Discovery and Observations of ASASSN-13db, an EX Lupi-type Accretion Event on a Low-mass T Tauri Star}",
  journal = {\apjl},
archivePrefix = "arXiv",
   eprint = {1401.3335},
 primaryClass = "astro-ph.SR",
     year = 2014,
    month = apr,
   volume = 785,
      eid = {L35},
    pages = {L35},
      doi = {10.1088/2041-8205/785/2/L35},
   adsurl = {http://adsabs.harvard.edu/abs/2014ApJ...785L..35H}
}

@ARTICLE{brown18,
   author = {{Brown}, J.~S. and {Kochanek}, C.~S. and {Holoien}, T.~W.-S. and 
	{Stanek}, K.~Z. and {Auchettl}, K. and {Shappee}, B.~J. and 
	{Prieto}, J.~L. and {Morrell}, N. and {Falco}, E. and {Strader}, J. and 
	{Chomiuk}, L. and {Post}, R. and {Villanueva}, Jr., S. and {Mathur}, S. and 
	{Dong}, S. and {Chen}, P. and {Bose}, S.},
    title = "{The ultraviolet spectroscopic evolution of the low-luminosity tidal disruption event iPTF16fnl}",
  journal = {\mnras},
archivePrefix = "arXiv",
   eprint = {1704.02321},
 primaryClass = "astro-ph.HE",
     year = 2018,
    month = jan,
   volume = 473,
    pages = {1130-1144},
      doi = {10.1093/mnras/stx2372},
   adsurl = {http://adsabs.harvard.edu/abs/2018MNRAS.473.1130B}
}

@ARTICLE{dong16,
   author = {{Dong}, S. and {Shappee}, B.~J. and {Prieto}, J.~L. and {Jha}, S.~W. and 
    {Stanek}, K.~Z. and {Holoien}, T.~W.-S. and {Kochanek}, C.~S. and 
    {Thompson}, T.~A. and {Morrell}, N. and {Thompson}, I.~B. and 
    {Basu}, U. and {Beacom}, J.~F. and {Bersier}, D. and {Brimacombe}, J. and 
    {Brown}, J.~S. and {Chen}, P. and {Conseil}, E. and {Danilet}, A.~B. and 
    {Falco}, E. and {Grupe}, D. and {Kiyota}, S. and {Masi}, G. and 
    {Nicholls}, B. and {Olivares}, F. and {Pignata}, G. and {Pojmanski}, G. and 
    {Simonian}, G.~V. and {Szczygiel}, D.~M. and {Wozniak}, P.~R.
    },
    title = "{ASASSN-15lh: A Highly Super-Luminous Supernova}",
  journal = {Science},
archivePrefix = "arXiv",
   eprint = {1507.03010},
 primaryClass = "astro-ph.HE",
     year = 2016,
    month = jan,
   volume = 351,
    pages = {257},
   adsurl = {http://adsabs.harvard.edu/abs/2015arXiv150703010D}
}

@ARTICLE{neustadt20,
       author = {{Neustadt}, J.~M.~M. and {Holoien}, T.~W. -S. and {Kochanek}, C.~S. and {Auchettl}, K. and {Brown}, J.~S. and {Shappee}, B.~J. and {Pogge}, R.~W. and {Dong}, Subo and {Stanek}, K.~Z. and {Tucker}, M.~A. and {Bose}, S. and {Chen}, Ping and {Ricci}, C. and {Vallely}, P.~J. and {Prieto}, J.~L. and {Thompson}, T.~A. and {Coulter}, D.~A. and {Drout}, M.~R. and {Foley}, R.~J. and {Kilpatrick}, C.~D. and {Piro}, A.~L. and {Rojas-Bravo}, C. and {Buckley}, D.~A.~H. and {Gromadzki}, M. and {Dimitriadis}, G. and {Siebert}, M.~R. and {Do}, A. and {Huber}, M.~E. and {Payne}, A.~V.},
        title = "{To TDE or not to TDE: the luminous transient ASASSN-18jd with TDE-like and AGN-like qualities}",
      journal = {\mnras},
         year = 2020,
        month = may,
       volume = {494},
       number = {2},
        pages = {2538-2560},
          doi = {10.1093/mnras/staa859},
archivePrefix = {arXiv},
       eprint = {1910.01142},
 primaryClass = {astro-ph.HE},
       adsurl = {https://ui.adsabs.harvard.edu/abs/2020MNRAS.494.2538N}
}

@ARTICLE{fausnaugh16,
   author = {{Fausnaugh}, M.~M. and {Grier}, C.~J. and {Bentz}, M.~C. and 
	{Denney}, K.~D. and {De Rosa}, G. and {Peterson}, B.~M. and 
	{Kochanek}, C.~S. and {Pogge}, R.~W. and {Adams}, S.~M. and 
	{Barth}, A.~J. and {Beatty}, T.~G. and {Bhattacharjee}, A. and 
	{Borman}, G.~A. and {Boroson}, T.~A. and {Bottorff}, M.~C. and 
	{Brown}, J.~E. and {Brown}, J.~S. and {Brotherton}, M.~S. and 
	{Coker}, C.~T. and {Crawford}, S.~M. and {Croxall}, K.~V. and 
	{Eftekharzadeh}, S. and {Eracleous}, M. and {Joner}, M.~D. and 
	{Henderson}, C.~B. and {Holoien}, T.~W.-S. and {Horne}, K. and 
	{Hutchison}, T. and {Kaspi}, S. and {Kim}, S. and {King}, A.~L. and 
	{Li}, M. and {Lochhaas}, C. and {Ma}, Z. and {MacInnis}, F. and 
	{Manne-Nicholas}, E.~R. and {Mason}, M. and {Montuori}, C. and 
	{Mosquera}, A. and {Mudd}, D. and {Musso}, R. and {Nazarov}, S.~V. and 
	{Nguyen}, M.~L. and {Okhmat}, D.~N. and {Onken}, C.~A. and {Ou-Yang}, B. and 
	{Pancoast}, A. and {Pei}, L. and {Penny}, M.~T. and {Poleski}, R. and 
	{Rafter}, S. and {Romero-Colmenero}, E. and {Runnoe}, J. and 
	{Sand}, D.~J. and {Schimoia}, J.~S. and {Sergeev}, S.~G. and 
	{Shappee}, B.~J. and {Simonian}, G.~V. and {Somers}, G. and 
	{Spencer}, M. and {Starkey}, D. and {Stevens}, D.~J. and {Tayar}, J. and 
	{Treu}, T. and {Valenti}, S. and {Van Saders}, J. and {Villanueva}, Jr., S. and 
	{Villforth}, C. and {Weiss}, Y. and {Winkler}, H. and {Zhu}, W.
	},
    title = "{Reverberation Mapping of Optical Emission Lines in Five Active Galaxies}",
  journal = {ArXiv e-prints},
archivePrefix = "arXiv",
   eprint = {1610.00008},
     year = 2016,
    month = sep,
   adsurl = {http://adsabs.harvard.edu/abs/2016arXiv161000008F}
}

@ARTICLE{blagorodnova17,
   author = {{Blagorodnova}, N. and {Gezari}, S. and {Hung}, T. and {Kulkarni}, S.~R. and 
	{Cenko}, S.~B. and {Pasham}, D.~R. and {Yan}, L. and {Arcavi}, I. and 
	{Ben-Ami}, S. and {Bue}, B.~D. and {Cantwell}, T. and {Cao}, Y. and 
	{Castro-Tirado}, A.~J. and {Fender}, R. and {Fremling}, C. and 
	{Gal-Yam}, A. and {Ho}, A.~Y.~Q. and {Horesh}, A. and {Hosseinzadeh}, G. and 
	{Kasliwal}, M.~M. and {Kong}, A.~K.~H. and {Laher}, R.~R. and 
	{Leloudas}, G. and {Lunnan}, R. and {Masci}, F.~J. and {Mooley}, K. and 
	{Neill}, J.~D. and {Nugent}, P. and {Powell}, M. and {Valeev}, A.~F. and 
	{Vreeswijk}, P.~M. and {Walters}, R. and {Wozniak}, P.},
    title = "{iPTF16fnl: A Faint and Fast Tidal Disruption Event in an E+A Galaxy}",
  journal = {\apj},
archivePrefix = "arXiv",
   eprint = {1703.00965},
 primaryClass = "astro-ph.HE",
     year = 2017,
    month = jul,
   volume = 844,
      eid = {46},
    pages = {46},
      doi = {10.3847/1538-4357/aa7579},
   adsurl = {http://adsabs.harvard.edu/abs/2017ApJ...844...46B}
}

@ARTICLE{bellm19,
   author = {{Bellm}, E.~C. and {Kulkarni}, S.~R. and {Graham}, M.~J. and 
	{Dekany}, R. and {Smith}, R.~M. and {Riddle}, R. and {Masci}, F.~J. and 
	{Helou}, G. and {Prince}, T.~A. and {Adams}, S.~M. and {Barbarino}, C. and 
	{Barlow}, T. and {Bauer}, J. and {Beck}, R. and {Belicki}, J. and 
	{Biswas}, R. and {Blagorodnova}, N. and {Bodewits}, D. and {Bolin}, B. and 
	{Brinnel}, V. and {Brooke}, T. and {Bue}, B. and {Bulla}, M. and 
	{Burruss}, R. and {Cenko}, S.~B. and {Chang}, C.-K. and {Connolly}, A. and 
	{Coughlin}, M. and {Cromer}, J. and {Cunningham}, V. and {De}, K. and 
	{Delacroix}, A. and {Desai}, V. and {Duev}, D.~A. and {Eadie}, G. and 
	{Farnham}, T.~L. and {Feeney}, M. and {Feindt}, U. and {Flynn}, D. and 
	{Franckowiak}, A. and {Frederick}, S. and {Fremling}, C. and 
	{Gal-Yam}, A. and {Gezari}, S. and {Giomi}, M. and {Goldstein}, D.~A. and 
	{Golkhou}, V.~Z. and {Goobar}, A. and {Groom}, S. and {Hacopians}, E. and 
	{Hale}, D. and {Henning}, J. and {Ho}, A.~Y.~Q. and {Hover}, D. and 
	{Howell}, J. and {Hung}, T. and {Huppenkothen}, D. and {Imel}, D. and 
	{Ip}, W.-H. and {Ivezi{\'c}}, {\v Z}. and {Jackson}, E. and 
	{Jones}, L. and {Juric}, M. and {Kasliwal}, M.~M. and {Kaspi}, S. and 
	{Kaye}, S. and {Kelley}, M.~S.~P. and {Kowalski}, M. and {Kramer}, E. and 
	{Kupfer}, T. and {Landry}, W. and {Laher}, R.~R. and {Lee}, C.-D. and 
	{Lin}, H.~W. and {Lin}, Z.-Y. and {Lunnan}, R. and {Giomi}, M. and 
	{Mahabal}, A. and {Mao}, P. and {Miller}, A.~A. and {Monkewitz}, S. and 
	{Murphy}, P. and {Ngeow}, C.-C. and {Nordin}, J. and {Nugent}, P. and 
	{Ofek}, E. and {Patterson}, M.~T. and {Penprase}, B. and {Porter}, M. and 
	{Rauch}, L. and {Rebbapragada}, U. and {Reiley}, D. and {Rigault}, M. and 
	{Rodriguez}, H. and {van Roestel}, J. and {Rusholme}, B. and 
	{van Santen}, J. and {Schulze}, S. and {Shupe}, D.~L. and {Singer}, L.~P. and 
	{Soumagnac}, M.~T. and {Stein}, R. and {Surace}, J. and {Sollerman}, J. and 
	{Szkody}, P. and {Taddia}, F. and {Terek}, S. and {Van Sistine}, A. and 
	{van Velzen}, S. and {Vestrand}, W.~T. and {Walters}, R. and 
	{Ward}, C. and {Ye}, Q.-Z. and {Yu}, P.-C. and {Yan}, L. and 
	{Zolkower}, J.},
    title = "{The Zwicky Transient Facility: System Overview, Performance, and First Results}",
  journal = {\pasp},
archivePrefix = "arXiv",
   eprint = {1902.01932},
 primaryClass = "astro-ph.IM",
     year = 2019,
    month = jan,
   volume = 131,
   number = 1,
    pages = {018002},
      doi = {10.1088/1538-3873/aaecbe},
   adsurl = {http://adsabs.harvard.edu/abs/2019PASP..131a8002B}
}

@ARTICLE{wevers19,
       author = {{Wevers}, T. and {Pasham}, D.~R. and {van Velzen}, S. and
         {Leloudas}, G. and {Schulze}, S. and {Miller-Jones}, J.~C.~A. and
         {Jonker}, P.~G. and {Gromadzki}, M. and {Kankare}, E. and
         {Hodgkin}, S.~T. and {Wyrzykowski}, {\L}. and
         {Kostrzewa-Rutkowska}, Z. and {Moran}, S. and {Berton}, M. and
         {Maguire}, K. and {Onori}, F. and {Mattila}, S. and {Nicholl}, M.},
        title = "{Evidence for rapid disc formation and reprocessing in the X-ray bright tidal disruption event candidate AT 2018fyk}",
      journal = {\mnras},
         year = "2019",
        month = "Oct",
       volume = {488},
       number = {4},
        pages = {4816-4830},
          doi = {10.1093/mnras/stz1976},
archivePrefix = {arXiv},
       eprint = {1903.12203},
 primaryClass = {astro-ph.HE},
       adsurl = {https://ui.adsabs.harvard.edu/abs/2019MNRAS.488.4816W}
}

@ARTICLE{chambers16,
   author = {{Chambers}, K.~C. and {Magnier}, E.~A. and {Metcalfe}, N. and 
	{Flewelling}, H.~A. and {Huber}, M.~E. and {Waters}, C.~Z. and 
	{Denneau}, L. and {Draper}, P.~W. and {Farrow}, D. and {Finkbeiner}, D.~P. and 
	{Holmberg}, C. and {Koppenhoefer}, J. and {Price}, P.~A. and 
	{Saglia}, R.~P. and {Schlafly}, E.~F. and {Smartt}, S.~J. and 
	{Sweeney}, W. and {Wainscoat}, R.~J. and {Burgett}, W.~S. and 
	{Grav}, T. and {Heasley}, J.~N. and {Hodapp}, K.~W. and {Jedicke}, R. and 
	{Kaiser}, N. and {Kudritzki}, R.-P. and {Luppino}, G.~A. and 
	{Lupton}, R.~H. and {Monet}, D.~G. and {Morgan}, J.~S. and {Onaka}, P.~M. and 
	{Stubbs}, C.~W. and {Tonry}, J.~L. and {Banados}, E. and {Bell}, E.~F. and 
	{Bender}, R. and {Bernard}, E.~J. and {Botticella}, M.~T. and 
	{Casertano}, S. and {Chastel}, S. and {Chen}, W.-P. and {Chen}, X. and 
	{Cole}, S. and {Deacon}, N. and {Frenk}, C. and {Fitzsimmons}, A. and 
	{Gezari}, S. and {Goessl}, C. and {Goggia}, T. and {Goldman}, B. and 
	{Grebel}, E.~K. and {Hambly}, N.~C. and {Hasinger}, G. and {Heavens}, A.~F. and 
	{Heckman}, T.~M. and {Henderson}, R. and {Henning}, T. and {Holman}, M. and 
	{Hopp}, U. and {Ip}, W.-H. and {Isani}, S. and {Keyes}, C.~D. and 
	{Koekemoer}, A. and {Kotak}, R. and {Long}, K.~S. and {Lucey}, J.~R and 
	{Liu}, M. and {Martin}, N.~F. and {McLean}, B. and {Morganson}, E. and 
	{Murphy}, D.~N.~A. and {Nieto-Santisteban}, M.~A. and {Norberg}, P. and 
	{Peacock}, J.~A. and {Pier}, E.~A. and {Postman}, M. and {Primak}, N. and 
	{Rae}, C. and {Rest}, A. and {Riess}, A. and {Riffeser}, A. and 
	{Rix}, H.~W. and {Roser}, S. and {Schilbach}, E. and {Schultz}, A.~S.~B. and 
	{Scolnic}, D. and {Szalay}, A. and {Seitz}, S. and {Shiao}, B. and 
	{Small}, E. and {Smith}, K.~W. and {Soderblom}, D. and {Taylor}, A.~N. and 
	{Thakar}, A.~R. and {Thiel}, J. and {Thilker}, D. and {Urata}, Y. and 
	{Valenti}, J. and {Walter}, F. and {Watters}, S.~P. and {Werner}, S. and 
	{White}, R. and {Wood-Vasey}, W.~M. and {Wyse}, R.},
    title = "{The Pan-STARRS1 Surveys}",
  journal = {ArXiv e-prints},
archivePrefix = "arXiv",
   eprint = {1612.05560},
 primaryClass = "astro-ph.IM",
     year = 2016,
    month = dec,
   adsurl = {http://adsabs.harvard.edu/abs/2016arXiv161205560C}
}

@ARTICLE{tonry18,
   author = {{Tonry}, J.~L. and {Denneau}, L. and {Heinze}, A.~N. and {Stalder}, B. and 
	{Smith}, K.~W. and {Smartt}, S.~J. and {Stubbs}, C.~W. and {Weiland}, H.~J. and 
	{Rest}, A.},
    title = "{ATLAS: A High-cadence All-sky Survey System}",
  journal = {\pasp},
archivePrefix = "arXiv",
   eprint = {1802.00879},
 primaryClass = "astro-ph.IM",
     year = 2018,
    month = jun,
   volume = 130,
   number = 6,
    pages = {064505},
      doi = {10.1088/1538-3873/aabadf},
   adsurl = {http://adsabs.harvard.edu/abs/2018PASP..130f4505T}
}

@INPROCEEDINGS{lantz04,
   author = {{Lantz}, B. and {Aldering}, G. and {Antilogus}, P. and {Bonnaud}, C. and 
	{Capoani}, L. and {Castera}, A. and {Copin}, Y. and {Dubet}, D. and 
	{Gangler}, E. and {Henault}, F. and {Lemonnier}, J.-P. and {Pain}, R. and 
	{Pecontal}, A. and {Pecontal}, E. and {Smadja}, G.},
    title = "{SNIFS: a wideband integral field spectrograph with microlens arrays}",
booktitle = {Optical Design and Engineering},
     year = 2004,
   series = {\procspie},
   volume = 5249,
   editor = {{Mazuray}, L. and {Rogers}, P.~J. and {Wartmann}, R.},
    month = feb,
    pages = {146-155},
      doi = {10.1117/12.512493},
   adsurl = {http://adsabs.harvard.edu/abs/2004SPIE.5249..146L}
}

@ARTICLE{roth16,
   author = {{Roth}, N. and {Kasen}, D. and {Guillochon}, J. and {Ramirez-Ruiz}, E.
	},
    title = "{The X-Ray through Optical Fluxes and Line Strengths of Tidal Disruption Events}",
  journal = {\apj},
archivePrefix = "arXiv",
   eprint = {1510.08454},
 primaryClass = "astro-ph.HE",
     year = 2016,
    month = aug,
   volume = 827,
      eid = {3},
    pages = {3},
      doi = {10.3847/0004-637X/827/1/3},
   adsurl = {http://adsabs.harvard.edu/abs/2016ApJ...827....3R}
}

@ARTICLE{vinko15,
   author = {{Vink{\'o}}, J. and {Yuan}, F. and {Quimby}, R.~M. and {Wheeler}, J.~C. and 
    {Ramirez-Ruiz}, E. and {Guillochon}, J. and {Chatzopoulos}, E. and 
    {Marion}, G.~H. and {Akerlof}, C.},
    title = "{A Luminous, Fast Rising UV-transient Discovered by ROTSE: A Tidal Disruption Event?}",
  journal = {\apj},
archivePrefix = "arXiv",
   eprint = {1410.6014},
 primaryClass = "astro-ph.HE",
     year = 2015,
    month = jan,
   volume = 798,
      eid = {12},
    pages = {12},
      doi = {10.1088/0004-637X/798/1/12},
   adsurl = {http://adsabs.harvard.edu/abs/2015ApJ...798...12V}
}

@ARTICLE{kara18,
       author = {{Kara}, E. and {Dai}, L. and {Reynolds}, C.~S. and {Kallman}, T.},
        title = "{Ultrafast outflow in tidal disruption event ASASSN-14li}",
      journal = {\mnras},
         year = "2018",
        month = "Mar",
       volume = {474},
       number = {3},
        pages = {3593-3598},
          doi = {10.1093/mnras/stx3004},
archivePrefix = {arXiv},
       eprint = {1711.06090},
 primaryClass = {astro-ph.HE},
       adsurl = {https://ui.adsabs.harvard.edu/abs/2018MNRAS.474.3593K}
}

@ARTICLE{baldwin81,
   author = {{Baldwin}, J.~A. and {Phillips}, M.~M. and {Terlevich}, R.},
    title = "{Classification parameters for the emission-line spectra of extragalactic objects}",
  journal = {\pasp},
     year = 1981,
    month = feb,
   volume = 93,
    pages = {5-19},
      doi = {10.1086/130766},
   adsurl = {http://adsabs.harvard.edu/abs/1981PASP...93....5B}
}

@ARTICLE{mcconnell13,
   author = {{McConnell}, N.~J. and {Ma}, C.-P.},
    title = "{Revisiting the Scaling Relations of Black Hole Masses and Host Galaxy Properties}",
  journal = {\apj},
archivePrefix = "arXiv",
   eprint = {1211.2816},
 primaryClass = "astro-ph.CO",
     year = 2013,
    month = feb,
   volume = 764,
      eid = {184},
    pages = {184},
      doi = {10.1088/0004-637X/764/2/184},
   adsurl = {http://adsabs.harvard.edu/abs/2013ApJ...764..184M}
}

@ARTICLE{arcavi14,
   author = {{Arcavi}, I. and {Gal-Yam}, A. and {Sullivan}, M. and {Pan}, Y.-C. and 
    {Cenko}, S.~B. and {Horesh}, A. and {Ofek}, E.~O. and {De Cia}, A. and 
    {Yan}, L. and {Yang}, C.-W. and {Howell}, D.~A. and {Tal}, D. and 
    {Kulkarni}, S.~R. and {Tendulkar}, S.~P. and {Tang}, S. and 
    {Xu}, D. and {Sternberg}, A. and {Cohen}, J.~G. and {Bloom}, J.~S. and 
    {Nugent}, P.~E. and {Kasliwal}, M.~M. and {Perley}, D.~A. and 
    {Quimby}, R.~M. and {Miller}, A.~A. and {Theissen}, C.~A. and 
    {Laher}, R.~R.},
    title = "{A Continuum of H- to He-rich Tidal Disruption Candidates With a Preference for E+A Galaxies}",
  journal = {\apj},
archivePrefix = "arXiv",
   eprint = {1405.1415},
 primaryClass = "astro-ph.HE",
     year = 2014,
    month = sep,
   volume = 793,
      eid = {38},
    pages = {38},
      doi = {10.1088/0004-637X/793/1/38},
   adsurl = {http://adsabs.harvard.edu/abs/2014ApJ...793...38A}
}

@ARTICLE{kauffmann03,
   author = {{Kauffmann}, G. and {Heckman}, T.~M. and {Tremonti}, C. and 
	{Brinchmann}, J. and {Charlot}, S. and {White}, S.~D.~M. and 
	{Ridgway}, S.~E. and {Brinkmann}, J. and {Fukugita}, M. and 
	{Hall}, P.~B. and {Ivezi{\'c}}, {\v Z}. and {Richards}, G.~T. and 
	{Schneider}, D.~P.},
    title = "{The host galaxies of active galactic nuclei}",
  journal = {\mnras},
   eprint = {astro-ph/0304239},
     year = 2003,
    month = dec,
   volume = 346,
    pages = {1055-1077},
      doi = {10.1111/j.1365-2966.2003.07154.x},
   adsurl = {http://adsabs.harvard.edu/abs/2003MNRAS.346.1055K}
}

@ARTICLE{kewley01,
   author = {{Kewley}, L.~J. and {Dopita}, M.~A. and {Sutherland}, R.~S. and 
	{Heisler}, C.~A. and {Trevena}, J.},
    title = "{Theoretical Modeling of Starburst Galaxies}",
  journal = {\apj},
   eprint = {astro-ph/0106324},
     year = 2001,
    month = jul,
   volume = 556,
    pages = {121-140},
      doi = {10.1086/321545},
   adsurl = {http://adsabs.harvard.edu/abs/2001ApJ...556..121K}
}

@ARTICLE{anderson14,
   author = {{Anderson}, J.~P. and {Gonz{\'a}lez-Gait{\'a}n}, S. and {Hamuy}, M. and 
    {Guti{\'e}rrez}, C.~P. and {Stritzinger}, M.~D. and {Olivares E.}, F. and 
    {Phillips}, M.~M. and {Schulze}, S. and {Antezana}, R. and {Bolt}, L. and 
    {Campillay}, A. and {Castell{\'o}n}, S. and {Contreras}, C. and 
    {de Jaeger}, T. and {Folatelli}, G. and {F{\"o}rster}, F. and 
    {Freedman}, W.~L. and {Gonz{\'a}lez}, L. and {Hsiao}, E. and 
    {Krzemi{\'n}ski}, W. and {Krisciunas}, K. and {Maza}, J. and 
    {McCarthy}, P. and {Morrell}, N.~I. and {Persson}, S.~E. and 
    {Roth}, M. and {Salgado}, F. and {Suntzeff}, N.~B. and {Thomas-Osip}, J.
    },
    title = "{Characterizing the V-band Light-curves of Hydrogen-rich Type II Supernovae}",
  journal = {\apj},
archivePrefix = "arXiv",
   eprint = {1403.7091},
 primaryClass = "astro-ph.HE",
     year = 2014,
    month = may,
   volume = 786,
      eid = {67},
    pages = {67},
      doi = {10.1088/0004-637X/786/1/67},
   adsurl = {http://adsabs.harvard.edu/abs/2014ApJ...786...67A}
}

@ARTICLE{phinney89,
   author = {{Phinney}, E.~S.},
    title = "{Cosmic merger mania}",
  journal = {\nat},
     year = 1989,
    month = aug,
   volume = 340,
    pages = {595-596},
      doi = {10.1038/340595a0},
   adsurl = {http://adsabs.harvard.edu/abs/1989Natur.340..595P}
}

@ARTICLE{rees88,
   author = {{Rees}, M.~J.},
    title = "{Tidal disruption of stars by black holes of 10 to the 6th-10 to the 8th solar masses in nearby galaxies}",
  journal = {\nat},
     year = 1988,
    month = jun,
   volume = 333,
    pages = {523-528},
      doi = {10.1038/333523a0},
   adsurl = {http://adsabs.harvard.edu/abs/1988Natur.333..523R}
}

@ARTICLE{evans89,
   author = {{Evans}, C.~R. and {Kochanek}, C.~S.},
    title = "{The tidal disruption of a star by a massive black hole}",
  journal = {\apjl},
     year = 1989,
    month = nov,
   volume = 346,
    pages = {L13-L16},
      doi = {10.1086/185567},
   adsurl = {http://adsabs.harvard.edu/abs/1989ApJ...346L..13E}
}

@ARTICLE{reines15,
   author = {{Reines}, A.~E. and {Volonteri}, M.},
    title = "{Relations between Central Black Hole Mass and Total Galaxy Stellar Mass in the Local Universe}",
  journal = {\apj},
archivePrefix = "arXiv",
   eprint = {1508.06274},
     year = 2015,
    month = nov,
   volume = 813,
      eid = {82},
    pages = {82},
      doi = {10.1088/0004-637X/813/2/82},
   adsurl = {http://adsabs.harvard.edu/abs/2015ApJ...813...82R}
}

@ARTICLE{dai18,
   author = {{Dai}, L. and {McKinney}, J.~C. and {Roth}, N. and {Ramirez-Ruiz}, E. and 
	{Miller}, M.~C.},
    title = "{A Unified Model for Tidal Disruption Events}",
  journal = {\apjl},
archivePrefix = "arXiv",
   eprint = {1803.03265},
 primaryClass = "astro-ph.HE",
     year = 2018,
    month = jun,
   volume = 859,
      eid = {L20},
    pages = {L20},
      doi = {10.3847/2041-8213/aab429},
   adsurl = {http://adsabs.harvard.edu/abs/2018ApJ...859L..20D}
}

@ARTICLE{kochanek17,
   author = {{Kochanek}, C.~S. and {Shappee}, B.~J. and {Stanek}, K.~Z. and 
	{Holoien}, T.~W.-S. and {Thompson}, T.~A. and {Prieto}, J.~L. and 
	{Dong}, S. and {Shields}, J.~V. and {Will}, D. and {Britt}, C. and 
	{Perzanowski}, D. and {Pojma{\'n}ski}, G.},
    title = "{The All-Sky Automated Survey for Supernovae (ASAS-SN) Light Curve Server v1.0}",
  journal = {\pasp},
archivePrefix = "arXiv",
   eprint = {1706.07060},
 primaryClass = "astro-ph.SR",
     year = 2017,
    month = oct,
   volume = 129,
   number = 10,
    pages = {104502},
      doi = {10.1088/1538-3873/aa80d9},
   adsurl = {http://adsabs.harvard.edu/abs/2017PASP..129j4502K}
}

@ARTICLE{kochanek16b,
   author = {{Kochanek}, C.~S.},
    title = "{Tidal disruption event demographics}",
  journal = {\mnras},
archivePrefix = "arXiv",
   eprint = {1601.06787},
 primaryClass = "astro-ph.HE",
     year = 2016,
    month = sep,
   volume = 461,
    pages = {371-384},
      doi = {10.1093/mnras/stw1290},
   adsurl = {http://adsabs.harvard.edu/abs/2016MNRAS.461..371K}
}

@ARTICLE{hung17,
   author = {{Hung}, T. and {Gezari}, S. and {Blagorodnova}, N. and {Roth}, N. and 
	{Cenko}, S.~B. and {Kulkarni}, S.~R. and {Horesh}, A. and {Arcavi}, I. and 
	{McCully}, C. and {Yan}, L. and {Lunnan}, R. and {Fremling}, C. and 
	{Cao}, Y. and {Nugent}, P.~E. and {Wozniak}, P.},
    title = "{Revisiting Optical Tidal Disruption Events with iPTF16axa}",
  journal = {\apj},
archivePrefix = "arXiv",
   eprint = {1703.01299},
 primaryClass = "astro-ph.HE",
     year = 2017,
    month = jun,
   volume = 842,
      eid = {29},
    pages = {29},
      doi = {10.3847/1538-4357/aa7337},
   adsurl = {http://adsabs.harvard.edu/abs/2017ApJ...842...29H}
}

@ARTICLE{cardelli89,
   author = {{Cardelli}, J.~A. and {Clayton}, G.~C. and {Mathis}, J.~S.},
    title = "{The relationship between infrared, optical, and ultraviolet extinction}",
  journal = {\apj},
     year = 1989,
    month = oct,
   volume = 345,
    pages = {245-256},
      doi = {10.1086/167900},
   adsurl = {http://adsabs.harvard.edu/abs/1989ApJ...345..245C}
}

@ARTICLE{oke95,
   author = {{Oke}, J.~B. and {Cohen}, J.~G. and {Carr}, M. and {Cromer}, J. and 
	{Dingizian}, A. and {Harris}, F.~H. and {Labrecque}, S. and 
	{Lucinio}, R. and {Schaal}, W. and {Epps}, H. and {Miller}, J.
	},
    title = "{The Keck Low-Resolution Imaging Spectrometer}",
  journal = {\pasp},
     year = 1995,
    month = apr,
   volume = 107,
    pages = {375},
      doi = {10.1086/133562},
   adsurl = {http://adsabs.harvard.edu/abs/1995PASP..107..375O}
}

@ARTICLE{bruzual03,
   author = {{Bruzual}, G. and {Charlot}, S.},
    title = "{Stellar population synthesis at the resolution of 2003}",
  journal = {\mnras},
   eprint = {astro-ph/0309134},
     year = 2003,
    month = oct,
   volume = 344,
    pages = {1000-1028},
      doi = {10.1046/j.1365-8711.2003.06897.x},
   adsurl = {http://adsabs.harvard.edu/abs/2003MNRAS.344.1000B}
}

@ARTICLE{wright10,
   author = {{Wright}, E.~L. and {Eisenhardt}, P.~R.~M. and {Mainzer}, A.~K. and 
    {Ressler}, M.~E. and {Cutri}, R.~M. and {Jarrett}, T. and {Kirkpatrick}, J.~D. and 
    {Padgett}, D. and {McMillan}, R.~S. and {Skrutskie}, M. and 
    {Stanford}, S.~A. and {Cohen}, M. and {Walker}, R.~G. and {Mather}, J.~C. and 
    {Leisawitz}, D. and {Gautier}, III, T.~N. and {McLean}, I. and 
    {Benford}, D. and {Lonsdale}, C.~J. and {Blain}, A. and {Mendez}, B. and 
    {Irace}, W.~R. and {Duval}, V. and {Liu}, F. and {Royer}, D. and 
    {Heinrichsen}, I. and {Howard}, J. and {Shannon}, M. and {Kendall}, M. and 
    {Walsh}, A.~L. and {Larsen}, M. and {Cardon}, J.~G. and {Schick}, S. and 
    {Schwalm}, M. and {Abid}, M. and {Fabinsky}, B. and {Naes}, L. and 
    {Tsai}, C.-W.},
    title = "{The Wide-field Infrared Survey Explorer (WISE): Mission Description and Initial On-orbit Performance}",
  journal = {\aj},
archivePrefix = "arXiv",
   eprint = {1008.0031},
 primaryClass = "astro-ph.IM",
     year = 2010,
    month = dec,
   volume = 140,
      eid = {1868},
    pages = {1868-1881},
      doi = {10.1088/0004-6256/140/6/1868},
   adsurl = {http://adsabs.harvard.edu/abs/2010AJ....140.1868W}
}

\end{document}